\documentclass[12pt]{article}
\pdfoutput=1
\usepackage{amsmath,amssymb,epsfig,amsfonts}
\usepackage{graphicx}
\usepackage{listings}
\usepackage[usenames, dvipsnames]{color}
\usepackage{multirow}
\usepackage[backref]{hyperref}
\usepackage[nosort]{cite}
\usepackage{lscape}
\usepackage{rotating}
\usepackage[normalem]{ulem} 
\usepackage{subcaption}
\usepackage{floatflt}
\usepackage{extarrows}

\usepackage{verbatim} 

\makeatletter

\newtheorem{Theorem}{Theorem}[section]
\newtheorem{fact}{Fact}[section]
\newtheorem{Corollary}[Theorem]{Corollary}
\newtheorem{definition}{Definition}[section]

\newcommand{\be}{\begin{equation}}
\newcommand{\ee}{\end{equation}}
\newcommand{\ba}{\begin{aligned}}
\newcommand{\ea}{\end{aligned}}

\newcommand{\bea}{\begin{eqnarray}}
\newcommand{\eea}{\end{eqnarray}}

\def\unit{{1\kern-.65ex {\rm l}}}
\def\1{{1\kern-.65ex {\rm l}}}

\newcommand{\bten}{\mathbf{10}}

\newcount\hour \newcount\minute
\hour=\time \divide \hour by 60
\minute=\time
\def\now{%
\ifnum \hour<13
  \ifnum \hour=0 \advance \hour by 12 \number\hour:\else \number\hour:\fi%
     \ifnum \minute<10 0\fi%
     \number\minute%
\ A.M.%
\else \advance \hour by -12 \number\hour:%
  \ifnum \minute<10 0\fi%
  \number\minute%
  \ P.M.%
\fi%
}

\makeatother

\begin{document}

\baselineskip=18pt  
\numberwithin{equation}{section}  
\allowdisplaybreaks  


%
%


\thispagestyle{empty}

\vspace*{-2cm} 
\begin{flushright}
{\tt KCL-MTH-15-03} 
\end{flushright}

\vspace*{0.8cm} 
\begin{center}

\bigskip

{\Huge F-theory and All Things Rational:}\\
\vspace{.5cm}
{\Large Surveying $U(1)$ Symmetries with Rational Sections}\\

 \vspace*{1.5cm}
{Craig Lawrie, Sakura Sch\"afer-Nameki, and Jin-Mann Wong}\\

 \vspace*{.8cm} 
 {\it Department of Mathematics, King's College London \\
 The Strand, London WC2R 2LS, England }\\
 {\tt {gmail:$\,$ craig.lawrie1729, sakura.schafer.nameki, jinmannwong}}\\

\vspace*{0.8cm}
\end{center}
\vspace*{.1cm}
\noindent 
We study elliptic fibrations for F-theory compactifications realizing 4d and 6d supersymmetric gauge theories with 
abelian gauge factors. In the fibration these $U(1)$ symmetries are realized in terms of additional rational sections. 
We obtain a universal characterization of all the possible $U(1)$ charges of matter fields by determining the corresponding codimension two fibers with rational sections.
In view of modelling supersymmetric Grand Unified Theories, one of the main examples that we analyze are $U(1)$ symmetries for  $SU(5)$ gauge theories with $\overline{\bf 5}$ and ${\bf 10}$ matter.
We use a combination of constraints on the normal 
 bundle of rational curves in Calabi--Yau three- and four-folds, as well as the splitting of rational curves in the fibers in codimension two,  to
determine the possible configurations of smooth rational sections. 
This analysis straightforwardly generalizes to multiple $U(1)$s. 
We study the flops of such fibers, as well as some of the  
Yukawa couplings in codimension three. Furthermore, we carry out a universal study of the $U(1)$-charged GUT singlets, including their KK-charges, and determine all realizations of singlet fibers.
By giving vacuum expectation values to these singlets, we propose a systematic way to analyze the Higgsing of  $U(1)$s to discrete gauge symmetries in F-theory.

\newpage

\tableofcontents



\section{Introduction}

Recent years have seen much progress towards refining F-theory compactifications, including the realization of symmetries
 of the low energy effective theory that allow more realistic model building. These developments have been fuelled 
 by increasingly sophisticated mathematical techniques 
  that are required to construct the geometries underlying such F-theory compactifications. 
In lockstep with this, there has been a definite trend towards characterizing universal aspects of 
string compactifications, with a view to going beyond an  example-driven approach. 
One of the areas where a universal characterization would be particularly bountiful is that of additional symmetries, such as abelian and discrete gauge symmetries, 
due to the direct phenomenological impact.
 
The main result of this paper is to provide such a universal characterization of possible $U(1)$ symmetries and associated matter charges  in F-theory.
Furthermore, we obtain a characterization of $U(1)$-charged singlets, which in turn can be used to Higgs abelian gauge groups to discrete symmetries.

The framework we are working within is F-theory compactifications on elliptically fibered Calabi--Yau three- and four-folds, where non-abelian gauge groups 
are modelled in terms of singularities above codimension one loci in the base of the fibration \cite{Vafa:1996xn}. Applications include the modelling of six-dimensional $N=(1,0)$ or four-dimensional $N=1$ supersymmetric gauge theories, 
whose gauge group is determined by the Kodaira type of the singularity \cite{Kodaira, Neron}. Matter is engineered from codimension two singularities, whose fibers are characterized in terms of representation theoretic data, associated
to the representation graph of the matter multiplet \cite{Hayashi:2014kca}. Abelian symmetries, which for instance are important model building tools for four-dimensional GUT models in F-theory \cite{Donagi:2008ca,Beasley:2008dc,  Beasley:2008kw}, are realized mathematically  in terms of rational sections of the elliptic fibrations, i.e. maps from the base  to the fiber \cite{Morrison:1996pp}. The rational sections, under the elliptic curve group law, form an abelian group, the Mordell--Weil group, $\mathbb{Z}^n \oplus \Gamma$, where $\Gamma$ is a discrete group, the origin of which is the zero-section $\sigma_0$. Such a rank $n$ Mordell--Weil group gives rise to $n$ abelian gauge factors in the low energy effective theory, by reducing the M-theory $C_3$-form upon the $(1,1)$-cycles that are dual to the rational sections. 

Numerous examples of F-theory compactifications with $U(1)$ symmetries are by now well-studied starting with the general theory of realizing the elliptic fiber with one \cite{Morrison:2012ei}, two \cite{Borchmann:2013jwa, Cvetic:2013nia, Borchmann:2013hta, Cvetic:2013jta} and three \cite{Cvetic:2013qsa} rational sections, toric constructions of various kinds \cite{Braun:2013yti, Braun:2013nqa, Braun:2014qka, Klevers:2014bqa}, models based on refined Weierstrass fibrations \cite{Grimm:2010ez, Braun:2011zm, Mayrhofer:2012zy, Morrison:2014era}, as well as a survey of all local spectral cover constructions \cite{Dolan:2011iu} or from Higgsing of $E_8$ \cite{Baume:2015wia}. 
Unfortunately, none of these approaches are both comprehensive, i.e. explore the complete set of possible $U(1)$ symmetries, and at the same time global (in the case of the spectral cover survey and $E_8$ embedding, which are general but only in terms of local models). 

Clearly it is highly desirable to determine the possible $U(1)$ symmetries in general, as these impose vital phenomenological input, and
can lead to potentially non-standard physics beyond the Standard Model (see e.g. \cite{Dolan:2011aq}). Furthermore, from a conceptual point of view, it is very appealing to be able to constrain these symmetries from the analysis of the fiber alone. 
One avenue that would lead in principle to such a general result is to determine the possible realizations of non-abelian gauge groups via Tate's algorithm \cite{Bershadsky:1996nh, Katz:2011qp} applied to the elliptic fibrations with extra sections in \cite{Morrison:2012ei, Cvetic:2013nia, Borchmann:2013hta}. This program was pursued in  \cite{Kuntzler:2014ila, Lawrie:2014uya}, resulting in a large class of new Tate-like models, however, in order to be able to carry out the algorithm, some technical simplifications had to be made, thus potentially jeopardizing the universality of the result. 

In this paper, we propose and provide a systematic analysis and universal characterization of such $U(1)$ symmetries in F-theory. 
Recall, that matter in a representation ${\bf R}$ of the gauge group, arises from wrapping M2-branes on irreducible components of the fiber in codimension two. 
The $U(1)$ charges of such matter multiplets are computed by intersecting the $U(1)$ generator, which is constructed from the rational sections, with these  fiber components. 
To classify the possible charges, one requires the following input: 
firstly, a complete understanding of the types of codimension two fibers that realize matter, which is now available in \cite{Hayashi:2014kca}, 
 and secondly, the possible configurations that the rational sections can take within these fibers. As we will demonstrate, the latter can be constrained in terms of general consistency requirements on
  $\mathbb{P}^1$s, i.e. rational curves, in Calabi--Yau varieties.

The possible codimension two fibers in an elliptic fibration with a holomorphic zero-section  can be characterized in terms of classical Coulomb phases of $d=5$ or $d=3$ $N=2$ supersymmetic gauge theories  \cite{Intriligator:1997pq, Aharony:1997bx, deBoer:1997kr, Diaconescu:1998ua, Grimm:2011fx, Hayashi:2013lra},  in terms of so-called box graphs \cite{Hayashi:2014kca}. In particular,  the box graphs characterize all possible splittings of the codimension one Kodaira fibers into codimension two fibers, which realize matter. In terms of the singular Weierstrass model, these characterize distinct small resolutions, which are connected by flop transitions. 
 
A rational section is characterized by the property that its intersection with the fiber is one. In codimension one, this implies that the section intersects a single rational curve in the Kodaira fiber transversally in a point\footnote{In principle, the section could contain codimension one fiber components, however, it would then not be irreducible. }. In codimension two, the section can again transversally intersect a single rational curve in the fiber, however, in addition, it can also contain components of the fiber. This effect has been referred to in the existing literature as the section \textit{wrapping} the fiber component.  This phenomenon was first observed in \cite{Morrison:2012ei}, where these fibers were shown to produce $U(1)$ charges distinct from fibers where both the zero-section and the additional section intersect transversally.  

For each section $\sigma$ there are two configurations that can occur in
codimension two. Either the section intersects a single component
transversally, or it contains (i.e. wraps) fiber components.  The wrapping is
highly constrained by the requirement that the intersection of $\sigma$ with
the fiber remains one, which we shall see translates into conditions on the
normal bundle degrees of the wrapped curves.  Concretely, we consider smooth
elliptic Calabi--Yau varieties $Y$ of dimension three and four and, subject to
the following constraint, we determine the possible section configurations:
intersections  of $\sigma$ with fiber components in codimension one are
preserved in codimension two, in particular, they are consistent with the
splitting as dictated by the box graphs.

For purposes of F-theory model building our main focus will be on $SU(n)$
gauge theories with fundamental and anti-symmetric matter, and in fact large
parts of this paper will focus on $n=5$ with the view to realize $SU(5)$ GUT
models in F-theory with additional $U(1)$ symmetries.  We determine all
possible section configurations in codimension two fibers for these matter
representations, and thereby the $U(1)$ charges.  For $SU(5)$ with one $U(1)$
there are three distinct codimension one configurations of the zero-section
$\sigma_0$, relative to the additional rational section $\sigma_1$, where they
intersect transversally the same $\mathbb{P}^1$ $I_5^{(01)}$, nearest
$I_5^{(0|1)}$ and next to nearest $I_5^{(0||1)}$ neighbor $\mathbb{P}^1$s of
the $I_5$ Kodaira fiber (see figure \ref{fig:I5Codim1}).

We determine all
section configurations for $\overline{\bf 5}$ and ${\bf 10}$ matter, under the
assumption that the sections remain smooth divisors in the Calabi--Yau
geometry -- the precise setup that enters this discussion is summarized in section \ref{sec:Assumptions}.
The resulting charges are as follows: 
\begin{equation}\label{eqn:11}
  \begin{aligned}
    \hbox{$U(1)$ charges of ${\bar{\bf 5}}$ matter for} &\quad 
    \left\{ 
      \begin{aligned}
        I_5^{(01)} &\in \left\{-3,-2, - 1, 0, +1,+2, +3 \right\} \cr
        I_5^{(0|1)} &\in \left\{-14, -9,-4,+1, + 6,+ 11 \right\} \cr
        I_5^{(0||1)} &\in \left\{- 13,- 8, -3, +2, + 7, + 12 \right\}  
      \end{aligned}
    \right.
    \cr 
    \hbox{$U(1)$ charges of ${{\bf 10}}$ matter for}&\quad 
    \left\{ 
      \begin{aligned}
        I_5^{(01)} &\in \left\{- 3, - 2,- 1, 0,  + 1, + 2, + 3 \right\} \cr
        I_5^{(0|1)} &\in \left\{- 12, - 7, - 2, + 3, + 8,+ 13 \right\} \cr
        I_5^{(0||1)} &\in \left\{-9, - 4, + 1,+6,+ 11\right\}  \,.
      \end{aligned}
    \right.
  \end{aligned}
\end{equation}
This result holds for both three- and four-folds alike, which we will
carefully derive using the constraints on the normal bundles of rational
curves  in Calabi--Yau varieties.  For four-folds we also discuss some
extension to Yukawa couplings, which arise in codimension three, and show how
the box graph analysis generalizes as well as how the $U(1)$ charges of the
interacting matter representations are consistent with the section
configuration in codimension three fibers.

At this juncture we should clarify an important point regarding the
normalization of the charges.  The rational section, $\sigma_1$, gives rise
to a $\mathbb{Q}$-divisor that is suitably orthogonal to the divisors
associated to the $SU(5)$ singular fibers, using the homomorphism between the
Mordell--Weil group and the $\mathbb{Q}$-divisors written in
\cite{MR1030197}, $\phi(\sigma_1)$. The generator of a $U(1)$ symmetry is an
integral divisor and must be a multiple of the above $\mathbb{Q}$-divisor to
be orthogonal to the gauge group, that is, it must have the form
$m\phi(\sigma_1)$ where $m$ is such that the divisor is integral.
Normalization of the $U(1)$ charges fixes the multiplier: there must not
exist another integral divisor $D \in H^2(Y, \mathbb{Z})$ such that
$m\phi(\sigma_1) = m^\prime D$ for any non-unit $m^\prime \in \mathbb{Z}$.
With a $U(1)$ generator so defined and normalized the $U(1)$ charges will be
in the possibilities listed in (\ref{eqn:11}).

One key realization here is that the analysis of the section configuration
holds true for any rational section, and thereby models with multiple sections
and thus $U(1)^n$ additional gauge symmetry, can be obtained by combining the
configurations in our classification. We discuss several examples with
multiple $U(1)$s in section \ref{sec:multiU1s}.  All matter charges and fiber
types in codimension two known from explicit models in the literature with one
ore more $U(1)$ symmetries appear in our classification, however these form a
strict subset of possible charges, and it would indeed be very interesting to
construct explicit realizations for the new  fiber types. We also compare our
charges to the ones obtained from Higgsing $E_8$ in \cite{Baume:2015wia}, and
find that our class of models is strictly larger than the ones arising from
$E_8$. Regarding the singlets in \cite{Baume:2015wia}, we provide realizations
for all charges of singlets in terms of $I_2$ fibers with rational sections. A
detailed discussion of the comparison to $E_8$ can be found in appendix
\ref{app:E8}.

Furthermore, we are able to  determine the fiber configurations for singlets, i.e. enhancements from $I_1$ fibers in codimension one to $I_2$ fibers in codimension two.
Contrary to the remaining part of the paper, this analysis is general only for three-folds. One important criterion for determining the singlets is the contractibility of curves, which is known
for three-folds, but not to our knowledge, in the case of four-folds. However, we determine all possible codimension two $I_2$ fibers with rational sections, without imposing any constraints on the normal bundle degree. 
This result can be seen as a general study of singlets, and imposing further constraints on the normal bundle to impose contractibility should then reduce these to the set of singlets in four-folds. 
Finally, we discuss flops of fibers with rational sections. It appears that flops can map out of the class of fibers where the section remains a smooth divisor in the Calabi--Yau, and it would be particularly interesting to study such singular flops in the future. 

Finally, we discuss the possibility, based on the singlet curve classification, to study more general Higgsings of the $U(1)$ symmetry to discrete symmetries, by giving $U(1)$-charged singlets a vacuum expectation value (vev). 
The case of charge $q=2, 3$ singlets and the Higgsing to $\mathbb{Z}_q$ has recently appeared in \cite{Mayrhofer:2014haa,Mayrhofer:2014laa,Cvetic:2015moa}. 
We provide both singlet fibers for higher charges, as well as determine the realization of the various KK-charges, i.e. intersections with the zero-section. 

The plan of this paper is as follows.  
In section \ref{sec:Box} we summarize all the necessary information about codimension two fibers from \cite{Hayashi:2014kca}. 
Furthermore, we extend that analysis, and determine  the Coulomb phases for $SU(n)$ gauge theories with a general (not necessarily the one arising from $U(n)$) additional $U(1)$ symmetry. 
In section \ref{sec:NormalBundles}, we discuss rational curves in Calabi--Yau three- and four-folds, and determine constraints on their normal bundles. These results will be an important input and constraining factor in our analysis. 
We then argue at the beginning of section \ref{sec:SU5F} that the constraints on the rational curves contained in a rational section, turn out to be identical in elliptic three- and four-folds\footnote{This is true only in this specific context of elliptically fibered Calabi--Yau geometries and we make the complete setup clear in section \ref{sec:NormalBundles}. It is by for not true, for rational curves in general Calabi--Yau varieties.}, thus allowing us in the remainder of this section to 
perform full classification of the codimension two fibers for both dimensions simultaneously. The case of fundamental matter for $SU(n)$ is discussed in the second half of section \ref{sec:SU5F} and the anti-symmetric matter for $n=5$ is discussed in section \ref{sec:SU5A} and appendix \ref{app:I1sSplittings}. The latter can of course also be generalized to $n>5$, however we leave this for the enterprising reader.   
Flops among these fibers are discussed in section \ref{sec:flops}.  Singlets are discussed in section \ref{sec:singlets} and multiple $U(1)$s, as well as Higgsing to discrete subgroups are the subject of section \ref{sec:multiU1s}.
For four-folds we generalize our results to codimension three, and describe some of the Yukawa couplings and section compatibility conditions in section \ref{sec:CodimThree}.
We close with discussions and future directions in section \ref{sec:Disc}.

To summarize the applicability of our results to three- and four-folds: sections \ref{sec:SU5F} and \ref{sec:SU5A} on charges of fundamental and anti-symmetric matter apply to both three- and four-folds. The section on flops is applicable to three-folds, the section on singlets \ref{sec:SingThree} to three-folds and section  \ref{sec:SingletsCY4} to four-folds. Finally, the section on codimension three to four-folds, only.


\section{Coulomb Phases and Fibers}
\label{sec:Box}

Before discussing rational sections we will review the results in
\cite{Hayashi:2014kca}, which give a comprehensive characterization of the
singular fibers in codimension two of an elliptic fibration. The main idea is
that the classical Coulomb phases of a 5d or 3d $N=2$ supersymmetric gauge
theory with matter obtained by compactifying M-theory on an elliptically
fibered Calabi--Yau three- or four-fold, encode the information about the
structure of singular fibers in codimensions one, two, and three. Distinct
Coulomb phases, which are separated by walls characterized by additional light
matter, correspond to distinct smooth Calabi--Yau varieties, which are related
by flop transitions. 

For this paper, the main case of interest is $\mathfrak{su}(5)$\footnote{From
the point of view of the box graphs, and also the elliptic fibration, it is
more natural to consider the Lie algebra, rather than group.} and we shall
restrict our attention in section \ref{sec:Coulomb} to explaining the
correspondence between singular fibers, gauge theory phases, and box graphs to
the case of $\mathfrak{su}(5)$ with matter in the $\overline{\bf 5}$ and ${\bf
10}$ representations, respectively.  For more general results see
\cite{Hayashi:2014kca}.  In addition, in section \ref{sec:CoulombU1} we will
also extend the analysis of Coulomb phases to $\mathfrak{su}(5) \oplus
\mathfrak{u}(1)$. 


\subsection{Box Graphs and Coulomb Phases}
\label{sec:Coulomb}

Our main interest regarding the results in \cite{Hayashi:2014kca} is the
characterization of the fibers in codimension two in an elliptically fibered
Calabi--Yau variety of dimension three or four.  We will assume that any such
fibration has at least one section.  The generic codimension one fibers in
such a variety are either smooth elliptic curves, or singular fibers, which
are collections of rational curves, i.e. smooth $\mathbb{P}^1$s, intersecting
in an affine Dynkin diagram of an ADE Lie algebra $\mathfrak{g}$. This 
classification, due to Kodaira and N\'{e}ron \cite{Kodaira, Neron}, holds true in codimension one, however fibers in higher
codimension can deviate from this.  The main result in \cite{Hayashi:2014kca},
is to map the problem of determining the codimension two fibers to the problem
of characterizing the Coulomb branch phases of a 3d or 5d $N=2$ supersymmetric
gauge theory with matter in a representation ${\bf R}$ of the gauge algebra
$\mathfrak{g}$ \cite{Intriligator:1997pq, Aharony:1997bx, deBoer:1997kr,
Diaconescu:1998ua, Grimm:2011fx}.  

Let us first discuss briefly the connection between Coulomb phases and resolutions of singular elliptic Calabi--Yau varieties.
The topologically distinct crepant resolutions, i.e. resolutions preserve keep the Calabi--Yau condition, of a singular Calabi--Yau variety are parameterised
by the phases of the classical Coulomb branch of the 3d $N=2$ gauge theory\footnote{A
  similar statement is true for Calabi--Yau three-folds in terms of the phases
of the associated 5d gauge theory.} 
obtained from the compactification of M-theory on the four-fold\cite{Grimm:2011fx,Hayashi:2013lra,  Hayashi:2014kca}. 

The 3d $N=2$  vector multiplet $V$ in
the adjoint of the gauge algebra $\mathfrak{g}$ has bosonic components given by the 
vector potential $A$ and a real scalar $\phi$. We are interested in the theory with additional chirals $Q$, transforming in a 
representation ${\bf R}$ of $\mathfrak{g}$.  The classical Coulomb branch is characterized by 
giving the scalars $\phi$ a vacuum expectation value, which breaks the gauge algebra $\mathfrak{g}$ to the  Cartan subalgebra, where 
 $\phi$ is such that 
\begin{equation}
  \langle \phi, \alpha_k \rangle \geq 0 \,,
\end{equation}
and  $\alpha_k$ are the simple roots of $\mathfrak{g}$. The Coulomb branch is therefore characterized
by the Weyl chamber of the gauge algebra $\mathfrak{g}$.

The presence of the chiral multiplets $Q$ in a representation ${\bf R}$ of $\mathfrak{g}$ adds a substructure to the
Coulomb branch. The vevs of $\phi$ give rise to a real mass term for the
chiral multiplets, 
\be 
L \supset | \langle \phi , \lambda \rangle |^2 |Q|^2 \,,
\ee
where $\lambda$ is a weight of the representation ${\bf R}$.
The mass term vanishes along walls  
\begin{equation}
  \langle \phi, \lambda \rangle = 0 \,.
\end{equation}
A classical Coulomb phase of the 3d
gauge theory is then one of the subwedges of the Weyl chamber delineated by
the walls where chiral multiplets become massless. A phase associated to the 
 representation ${\bf R }$ is then specified by a map 
\be\label{epsdef}
\ba
\varepsilon: \, \quad {\bf R} &\ \rightarrow \  \{\pm 1\}\cr 
			\lambda &\  \mapsto\  \varepsilon(\lambda) \,,
\ea
\ee
such that $\langle \phi, \lambda \rangle$ has a definite sign $\varepsilon (\lambda)$, i.e.
\begin{equation} \label{SubwedgeCon}
 \varepsilon (\lambda) \langle \phi, \lambda \rangle > 0  \,.
\end{equation}
Solutions for $\phi$ will not exist for every possible sign assignment $\varepsilon$, i.e. the phases 
are the non-empty subwedges of the Weyl chamber satisfying (\ref{SubwedgeCon}).
In particular the condition (\ref{SubwedgeCon}) means that the weight $\varepsilon(\lambda) \lambda$ is in
this subwedge that characterizes the corresponding phase.
 In \cite{Hayashi:2014kca} the phases for $\mathfrak{g}$ of ADE type were determined with various representations ${\bf R}$, and 
 shown to be characterized in terms of sign-decorated representation graphs,
 so-called box graphs, of ${\bf R}$, which are essentially a graphical depiction
 of the maps $\varepsilon$. It was shown that there are  simple, combinatorial rules for determining the box graphs corresponding to non-empty subwedges, and that furthermore  these 
 encode vital  information about the elliptic Calabi--Yau geometry (the intersection ring and relative cone of effective curves in the elliptic fiber).

\begin{figure}
\centering
  \includegraphics[width=4cm]{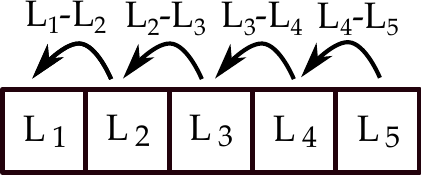} \qquad \qquad   \includegraphics[width=4cm]{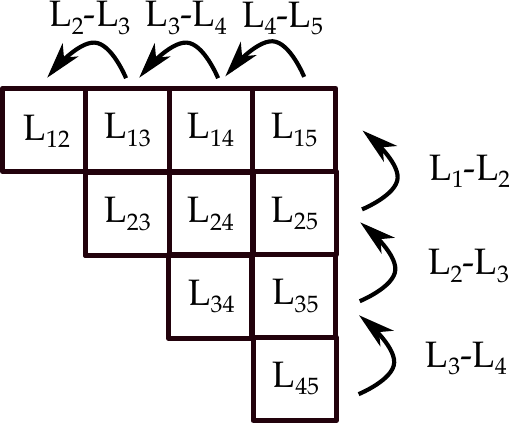} 
  \caption{The ${\bf 5}$  and ${\bf 10}$ representation of $SU(5)$. Each box
  represents a weight $L_i$ ($L_i+ L_j$) of the fundamental (anti-symmetric)
  representation and the walls inbetween each box correspond to the action of
  the simple roots $\alpha_k =L_k - L_{k+1}$ on the weights as indicated by
  the arrows. The direction of the arrow indicates the addition of the
  corresponding simple root. \label{fig:SU5Diagram5}}
\end{figure}

For our purposes $\mathfrak{g}=\mathfrak{su}(5)$ and ${\bf R} = {\bf 5}$ or ${\bf 10}$. 
We denote the weights of these representations in terms of the fundamental weights $L_i$ 
\be
{\bf 5}:\quad \lambda\in \{L_1, L_2, L_3, L_4, L_5\}\,, \qquad 
{\bf 10}: \quad \lambda \in \{L_i + L_j | \ i<j; \ i, j = 1, \cdots, 5\} \,,
\ee
where $\sum_i L_i =0$. The simple roots of $\mathfrak{su}(5)$ in this basis are
\be
\alpha_k = L_k- L_{k+1} \,.
\ee
The result of \cite{Hayashi:2014kca} applied to $\mathfrak{g}=\mathfrak{su}(5)$ with ${\bf R} = {\bf 5}$ can be summarized as follows:
each consistent phase $\Phi_\varepsilon$ is characterized by a map 
$\varepsilon$ as in (\ref{epsdef}),  subject to the constraint that it
satisfies
\be\label{Flow5}
\hbox{${\bf 5}$ flow rules}:\quad \left\{
\ba
\varepsilon (L_i) = +\quad \Rightarrow \quad \varepsilon (L_{j}) =+  \quad \hbox{for all }j <i\cr
\varepsilon (L_i) = -\quad \Rightarrow \quad \varepsilon (L_{j}) =-  \quad \hbox{for all }j >i 
\ea\right.
\ee
This results in phases that also include all $+$ or all $-$ sign assignments to the weights. These are in fact
phases of the $\mathfrak{su}(5)\oplus \mathfrak{u}(1)$ theory. The phases for the $\mathfrak{su}(5)$ theory need to satisfy an additional constraint, 
which ensures that the  sum of all the $L_i$  vanishes (trace condition) \cite{Hayashi:2014kca}. In this paper we are interested
in the phases for the theory with additional abelian factors. It is a priori not clear that all phases of any  $\mathfrak{su}(5)\oplus \mathfrak{u}(1)$ theory can be characterized 
in terms of the phases above, and we will prove this fact in section \ref{sec:CoulombU1}. 

Likewise, for ${\bf R}= {\bf 10}$ a sign assignment $\varepsilon$ gives rise to a phase, if and only if 
\be\label{Flow10}
\hbox{{\bf 10} flow rules}:\quad 
\left\{\ba
\varepsilon (L_i + L_j) = +\quad \Rightarrow \quad \varepsilon (L_k+L_l) =+  \quad \hbox{for all } (k,l), \quad k \leq i\,,\ l\leq j \cr
\varepsilon (L_i + L_j) = -\quad \Rightarrow \quad \varepsilon (L_k+L_l) =-  \quad \hbox{for all } (k,l), \quad k \geq i\,,\ l\geq j 
\ea\right.
\ee
Again for $\mathfrak{su}(5)$ there is an additional trace condition, which however we do not impose as we are interested in theories with $\mathfrak{u}(1)$ factors. 
The connection between Coulomb phases and box graphs is then formulated as follows (see \cite{Hayashi:2014kca} and section \ref{sec:CoulombU1}):
\begin{fact}\label{Fact1}
The classical Coulomb phases for 3d $N=2$ supersymmetric  $\mathfrak{su}(5)\oplus \mathfrak{u}(1)$ gauge theories with matter in the ${\bf R}= {\bf 5}$ or ${\bf 10}$ 
representation are in one-to-one correspondence with maps $\varepsilon$ as in
(\ref{epsdef}), satisfying the flow rules (\ref{Flow5}) or (\ref{Flow10}), respectively. We will denote these by $\Phi^{{\bf R}}_{\varepsilon}$.
 \end{fact}

Each phase $\Phi_\varepsilon^{\bf R}$ associated to such a map $\varepsilon$ can be represented graphically in terms of a box graph $\mathcal{B}_\varepsilon^{\bf R}$.
\begin{definition}
A box graph $\mathcal{B}_\varepsilon^{\bf R}$ for a Coulomb phase $\Phi^{\bf
R}_\varepsilon $ is given in terms of the representation graph of ${\bf R}$,
i.e. a graph where each weight  $\lambda$ of ${\bf R}$ is represented by a
box, and two weights are adjacent if they are mapped into each other by the action of a simple root, together with a sign assignment/coloring, given by $\varepsilon(\lambda)$.  
\end{definition}
Generically we will draw these by coloring $+$ as blue and $-$ as yellow.  The
representation graphs for ${\bf 5}$ and ${\bf 10}$ of $\mathfrak{su}(5)$ are
shown in figure \ref{fig:SU5Diagram5}. The phases/box graphs for
${\bf 5}$ are shown in figure \ref{fig:U5BoxGraphs5}, for ${\bf 10}$
in appendix \ref{app:I1sSplittings}.


\subsection{Box Graphs and Singular Fibers}

The Coulomb phases encode information about the effective curves of the elliptic fibration in codimension two. 
Let us begin with a few useful definitions. 
In the following $Y$ is a smooth elliptic Calabi--Yau variety  of dimension at least three with a section, which guarantees the existence of a Weierstrass model for this fibration. 
The information about the Coulomb phases can be reformulated in terms of the
geometric data of a certain relative subcone inside the cone of
effective curves. A curve is defined to be {\it effective} if it can be written in terms of a positive integral linear combination of integral curves (i.e. actual complex one-dimensional subspaces) of $Y$. 
The cone of effective curves in $Y$ is denoted by $NE(Y)$.\footnote{These are numerically effective curves, where we mod out by the equivalence that two curves are identified if they have the same intersections with all Cartier divisors. }
For an elliptic fibration, the notion of relative cone of curves is of
particular importance. Let $W$ be the singular Weierstrass model, associated
to $Y$. In fact, for a given singular Weierstrass model there are generically
several, topologically distinct smooth models, $Y_i$. The singular limit
corresponds,  in codimension one, to the maps
\be\label{piYiW}
\pi_i: \quad Y_i\quad  \rightarrow \quad W  \,,
\ee
such that all rational curves in the singular Kodaira fibers, which do not
meet the section, are contracted \cite{MR977771}. 
Associated to this, there is the notion of a relative cone of effective curves
(see e.g. \cite{MR1841091}):
\begin{definition}
The {relative cone of curves} $NE(\pi_i)$ of the morphism $\pi_i$ in (\ref{piYiW}) is the convex subcone of the cone of effective curves $NE(Y_i)$ generated by the curves that are contracted by $\pi_i$.
\end{definition}
The phases/box graphs are in one-to-one correspondence with pairs $(Y_i, \pi_i)$, specified in the following way:
Each fiber in codimension one is characterized by rational curves $F_k$ associated to the simple roots of the gauge group $G$. 
In codimension two some of the $F_k$ become reducible and split into a collection of rational curves 
\be\label{FkSplit}
F_k \quad \rightarrow \quad  C_1 + \cdots + C_\ell \,,
\ee
where each $C_j$ is associated to $\varepsilon(\lambda) \lambda$ for $\lambda$
a weight of the representation ${\bf R}$, or to a simple root. 
The main result in \cite{Hayashi:2014kca} can then be stated as follows:
\begin{fact}\label{Fact2}
There is a one-to-one correspondence between consistent phases or box graphs $\mathcal{B}^{\bf R}_{\varepsilon_i}$ characterized by the sign assignments $\varepsilon_i$ satisfying the conditions in Fact \ref{Fact1} 
and crepant resolution of $W$, $(Y_i, \pi_i)$. In particular, the box graphs determine the relative cone of effective curves for the maps $\pi_i$ as 
\be
NE (\pi_i) =\big\langle\  \{F_k \ | \   k= 0,\cdots, \hbox{rank}(\mathfrak{g})\} \ \cup\  \{C_{\varepsilon_i(\lambda) \lambda} \ | \ \lambda   \hbox{ weight  of }{\bf R} \}\ \big\rangle_{\mathbb{Z}^+}\,.
\ee
The extremal generators of this cone are 
\begin{enumerate}
\item The rational curves $F_k$, that remain irreducible in codimension two.
\item $C_{\varepsilon_i(\lambda)\lambda}$ is extremal if  there exists a $j$ such that $\mathcal{B}_{\varepsilon_j}^{\bf R} =\mathcal{B}_{\varepsilon_i}^{\bf R}|_{\varepsilon_j(\lambda) =-\varepsilon_i(\lambda)}$, i.e. there is another consistent box graph or phase, such that the only sign change occurs in the weight $\lambda$. 
\end{enumerate}
\end{fact}
From the box graphs we can determine which $F_k$ remain irreducible: 
$F_k$, associated to the simple roote $\alpha_k$, remains irreducible, if any
weight $\lambda$, for which $\lambda + \alpha_k$ is another weight in the
representation ${\bf R}$, the weight $\lambda + \alpha_k$  has the same sign assignment, i.e.\footnote{This condition is formulated in \cite{Hayashi:2014kca} as adding the simple root does not cross the anti-Dyck path that separates the + and - sign assigned weights in the box graph. } 
\be
\varepsilon(\lambda)= \varepsilon (\lambda + \alpha_k) \,.
\ee
\begin{fact}
Two crepant resolutions $(Y_i, \pi_i)$ and $(Y_j, \pi_j)$ of the singular Weierstrass model $W$ are related by a simple flop, if the corresponding box graphs are related by a single sign change
\be
\mathcal{B}_{\varepsilon_j}^{\bf R} =\mathcal{B}_{\varepsilon_i}^{\bf R}|_{\varepsilon_j(\lambda) =-\varepsilon_i(\lambda)}
\ee
for some weight $\lambda$. I.e. they correspond to single box changes of
signs, which map one extremal generator to minus itself. 
\end{fact}

\begin{figure}
  \centering
  \includegraphics[width=5cm]{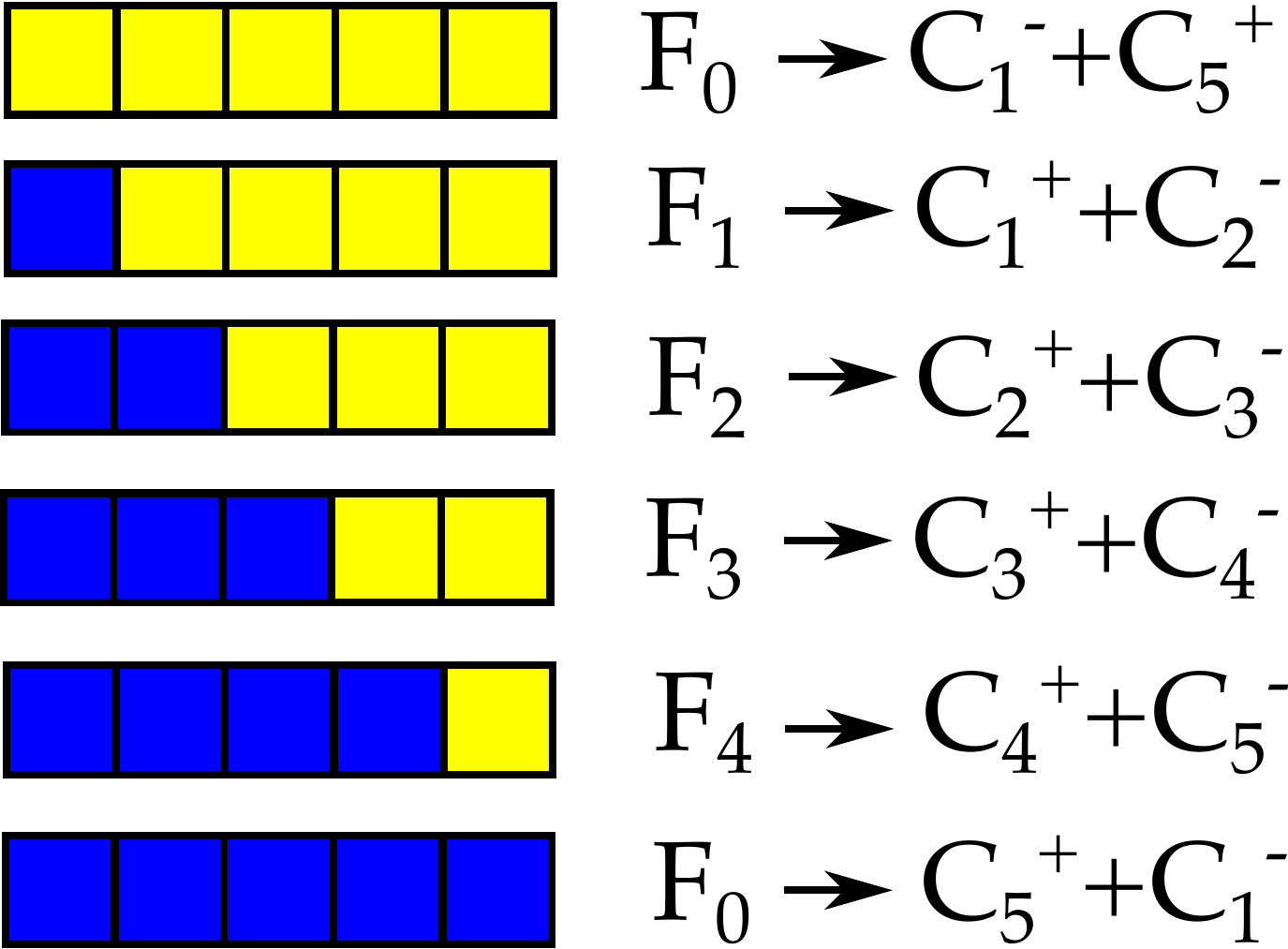} 
  \caption{Box graphs for $\mathfrak{u}(5)$ phases with ${\bf 5}$ matter. On the left are the splittings that occur over matter loci for the corresponding phase.\label{fig:U5BoxGraphs5}}
\end{figure}

In the remainder of this paper,  it will be very important to understand the degrees of normal bundles of curves in the fibers of elliptic Calabi--Yau varieties. 
The description of the codimension two fibers in terms of box graphs allows us to determine the intersections of the extremal generators with the so-called Cartan divisors,  $D_{F_k}$, which are $F_k$ fibered over 
the codimension one discriminant locus. 
They  are dual to the rational curves $F_k$,  with which they intersect in the Calabi--Yau $Y$ in the negative  Cartan matrix $-C_{kl}$ of the gauge algebra
\be\label{DFk}
D_{F_k} \cdot_Y F_l = - C_{kl} \,.
\ee
Consider now a codimension two fiber where $F_k$ splits as in (\ref{FkSplit}). Then  
\be
D_{F_{m}} \cdot_Y C_a = \varepsilon \left(\lambda^{(a)}\right) \lambda^{(a)}_m \,,\qquad m= 1, \cdots,  \hbox{rank} (\mathfrak{g}) \,,
\ee
 i.e. it intersects with the rational curves $C_a$ in a weight $\lambda^{(a)}$ 
of the representation ${\bf R}$. Which weight this is, i.e. the intersections of the fiber components with the Cartan divisors, and with which sign assignment it occurs can be determined from the 
box graphs. 
\begin{fact}
Let $C$ be an extremal generator of the cone $NE(\pi_i)$ for a pair  $(Y_i, \pi_i)$, associated to the box graph $\mathcal{B}^{\bf R}_{\varepsilon_i}$ as in Fact \ref{Fact2}, associated to a weight $\lambda$ of the representation ${\bf R}$. The Dynkin labels 
$\varepsilon_i(\lambda) \lambda_m = D_{F_m} \cdot_Y C$ can be computed from the box graph $\mathcal{B}^{\bf R}_{\varepsilon_i}$ as follows: 
If $\lambda \pm \alpha_m$ is not a weight in the representation then $D_{F_m} \cdot_Y C =0$. Else:
\begin{enumerate}
\item If  $\varepsilon_i(\lambda)= \varepsilon_i(\lambda \pm \alpha_m)$ then $D_{F_m} \cdot_Y C = +1 $.
\item If  $\varepsilon_i(\lambda)= -\varepsilon_i(\lambda \pm \alpha_m)$ then $D_{F_m} \cdot_Y C = -1$.
\end{enumerate}
\end{fact}
This fact together with $D_{F_m} \cdot_Y F_m = -2 $, will be used quite regularly in the analysis of the normal bundles in sections \ref{sec:SU5F} and \ref{sec:SU5A}.

Finally, let us note that the number $N^{{\bf R}_q}$ of phases, i.e. pairs $(Y_i, \pi_i)$, with matter in the representation ${\bf R}$ and $\mathfrak{u}(1)$ charge $q$ under the  gauge algebra $ \mathfrak{g} \oplus \mathfrak{u}(1)$ is given in terms of the quotiented Weyl group:
\begin{fact}\label{fact:Enumeration}
The number $N^{\bf R}_q$ of classical Coulomb phases for gauge algebras $\mathfrak{g} \oplus \mathfrak{u}(1)$ and representation ${\bf R}$ with $\mathfrak{u}(1)$ charge $q$ is 
\be
N^{{\bf R}_q} = \left| {W_{\widetilde{\mathfrak{g}}}\over W_{\mathfrak{g}}} \right| \,,
\ee
where $\widetilde{\mathfrak{g}}$ is the Lie algebra characterizing the local
enhancement in codimension two, i.e. decomposing its adjoint into
representations of the gauge algebra
contains the representation ${\bf R}_q$ and its conjugate as follows
\be
\ba
\widetilde{\mathfrak{g}} \quad& \rightarrow \quad \mathfrak{g} \oplus \mathfrak{u}(1) \cr
\hbox{Adj} \left( \widetilde{\mathfrak{g}}\right) \quad &\rightarrow \quad \hbox{Adj} \left(\mathfrak{g}\right)  \oplus \hbox{Adj} \left(\mathfrak{u}(1)\right)  \oplus {\bf R}_q \oplus \overline{\bf R}_{-q} \,.
\ea
\ee
\end{fact}
For $\mathfrak{g} =\mathfrak{su}(5)$ and ${\bf R} = {\bf 5}$ or ${\bf 10}$, $\widetilde{\mathfrak{g}} = \mathfrak{su}(6)$ or $\mathfrak{so}(10)$ and $N^{\bf 5} = 6$ and $N^{\bf 10} = 16$. 
For $\mathfrak{su}(5)$ with ${\bf 5}$ we summarized the phases in figure \ref{fig:U5BoxGraphs5}, including which of the $F_k$ split. 
The components into which they split are precisely those adjacent to the sign change, which is clear from the statements in Fact \ref{Fact2}. The curves $C^\pm_i$ correspond to the weights $\pm L_i$, which are generators of the cone defined by 
$\Phi^{\bf R}_\varepsilon$. Note that the ${\bf 5}$ representation can also arise from a higher rank enhancement e.g. to $\mathfrak{su}(n)$, $n>6$. Such enhancements when realized in the geometry would require very special tuning of the complex structure, with the fibers corresponding to monodromy-reduced $I_n$ fibers. These will not be considered here, but the reader is referred to \cite{CLSSN}.
The structure of splittings in codimension two for ${\bf 10}$ matter are listed in appendix \ref{app:I1sSplittings}, tables \ref{tab:10SplitPart1} and \ref{tab:10SplitPart2}, which include all the information about the splitting in codimension two, the extremal generators of the relative cone of effective curves, and  the associated box graphs.


\subsection{$U(1)$-Extended Coulomb Phases}
\label{sec:CoulombU1}

In \cite{Hayashi:2014kca} the phases for the $\mathfrak{su}(5) \oplus \mathfrak{u}(1)$ theory were
determined in the case where the $\mathfrak{u}(1)$ corresponds to $\sum_{i=1}^5 L_i$, 
where the $L_i$ are the fundamental weights introduced in the previous section, 
i.e. this $\mathfrak{u}(1)$ corresponds to the trace of the $\mathfrak{u}(5)$. 
In this section we show that the analysis there holds more generally for the classical Coulomb phases of $\mathfrak{su}(5)
\oplus \mathfrak{u}(1)$, where the $U(1)$ does not necessarily have this origin\footnote{There can corrections to the  classical Coulomb phase analysis 
with additional abelian factors, as discussed in $6d$ in \cite{Grimm:2013oga, Grimm:2015zea}, which will not play a role here.}. 
Note that the phases for the $\mathfrak{su}(5)\oplus \mathfrak{u}(1)$ theory
are one-to-one with the elements of the quotiented Weyl group
$W_{\widetilde{\mathfrak{g}}}/W_{\mathfrak{su}(5)}$, as summarized in Fact
\ref{fact:Enumeration}, which is strictly larger than the number of phases for
the theory without an abelian factor. 

Let ${\bf R}_q$ be a representation ${\bf R}$ of $\mathfrak{su}(5)$ with charge $q$ under the $\mathfrak{u}(1)$. 
Let us consider the maps $\varepsilon : {\bf R}_q \rightarrow \{\pm 1\}$
corresponding to a consistent, non-empty, subwedge of the fundamental Weyl
chamber. 
The walls of these subwedges are characterized by
\begin{equation}
  \langle \phi, (\lambda_i; q) \rangle \equiv \langle \phi, \lambda_i \rangle +
  q\phi_u = 0 \,,
\end{equation}
where $\phi_u$ is the  additional component of $\phi$ along the $\mathfrak{u}(1)$ generator. Consider the ${\bf
5}_q$ representation of $\mathfrak{su}(5) \oplus \mathfrak{u}(1)$. The
fundamental weights of
$\mathfrak{su}(5)$, the $L_i$, in the Cartan-Weyl basis take the form
\be
\ba
  \lambda_1 \,&:\quad (1,0,0,0) \cr
  \lambda_2 \,&:\quad (-1,1,0,0) \cr
  \lambda_3 \,&:\quad (0,-1,1,0) \cr
  \lambda_4 \,&:\quad (0,0,-1,1) \cr
  \lambda_5 \,&:\quad (0,0,0,-1) \,.
\ea
\ee
 In the same
basis the simple roots of the $\mathfrak{su}(5)$ are
\be
\ba
  \alpha_1 \,:\  (2,-1,0,0) \,, \quad 
  \alpha_2 \,:\  (-1,2,-1,0) \,,\quad 
  \alpha_3 \,:\  (0,-1,2,-1) \,,\quad 
  \alpha_4 \,:\  (0,0,-1,2) \,.
\ea\ee
To reiterate, to determine the maps $\varepsilon$ which correspond to non-empty
phases it is needed to find the maps $\varepsilon : {\bf 5}_q \rightarrow
\{\pm 1\}$
such that the inequalities
\begin{equation}\label{eqn:secondinequality}
\ba
  \langle \phi, \alpha_i \rangle & > 0  \cr 
  \varepsilon((\lambda_i; q))\langle \phi, (\lambda_i; q) \rangle & > 0
\ea
\end{equation}
have integral solutions for $\phi$. 

Similarly to the derivation of the flow rules alluded to in the earlier
parts of this section one can show that if $\varepsilon((\lambda_i; q)) = -1$ and
$\varepsilon((\lambda_{i+1}; q)) = +1$ then there would be no such solutions: for
such an $\varepsilon$ it would be the case that
\be
  \langle \phi, \lambda_{i+1} \rangle + q \phi_u 
  -\left(\langle \phi, \lambda_i \rangle + q \phi_u\right) > 0 
\quad \Leftrightarrow \quad 
  \langle \phi, \lambda_{i+1} - \lambda_i \rangle > 0 \,.
\ee
However, the simple roots are $ \alpha_i = \lambda_i - \lambda_{i+1}$
and the first of the inequalities in (\ref{eqn:secondinequality}) implies
\be
  \langle \phi, \lambda_i - \lambda_{i+1} \rangle > 0 \,.
\ee
Obviously there is no such $\phi$ which solves these inequalities: all subwedges of
the fundamental Weyl chamber defined by this map $\varepsilon$ are empty. 
This leads to the same flow rules as listed in (\ref{Flow5}).

Again
there are six phases, of which two have 
all positive or all negative signs, and are only non-empty in the theory with a
$\mathfrak{u}(1)$ symmetry in addition to the $\mathfrak{su}(5)$, indeed these extra phases occur
precisely for matter charged under the additional $\mathfrak{u}(1)$. 
Consider now the phase associated to the map $\varepsilon( (\lambda_i; q) ) =
{+1}$ for all $i$.  Then, using that $\sum \lambda_i = 0$, 
 as can be seen explicitly
above from the presentation in the Cartan--Weyl basis,
\be
  \sum_{i=1}^5 (\langle \phi, \lambda_i \rangle + q\phi_u) > 0 \quad \Leftrightarrow\quad 
  q\phi_u > 0 \,.
\ee
Such inequalities can only be solved if $q \neq 0$, and similiarly for the
all negative phase. These are the two additional phases for charged matter. 

One can also consider the ${\bf 10}_q$ representation of $\mathfrak{su}(5) \oplus \mathfrak{u}(1)$ in the same way. Similarly to the case when of the ${\bf 5}_q$
representation one finds an augmented set of maps $\varepsilon$ when $q$ is non-zero. There are sixteen phases when $q
\neq 0$ and eight when $q = 0$. These sets of phases correspond to  the
different sets of phases in \cite{Hayashi:2014kca},
except here there is no assumption that the generator of the $\mathfrak{u}(1)$ symmetry
is necessarily that in the $\mathfrak{u}(5)$.

To summarize if the matter is
charged under the $\mathfrak{u}(1)$ symmetry then there are additional phases of
the classical Coulomb branch for the $\mathfrak{su}(5) \oplus \mathfrak{u}(1)$ theory with
fundamental or anti-symmetric matter. The additional phases 
imply that there are additional distinct resolved geometries associated
to the singular Calabi--Yau four-fold, induced by the specialisation of
complex structure necessary to produce matter charged under the additional
$\mathfrak{u}(1)$, i.e. geometrically, the existence of additional rational sections.


\section{Rational Curves in Calabi--Yau Varieties}
\label{sec:NormalBundles}

The goal of this paper is to constrain the possible $U(1)$ charges of matter in 4d and 6d F-theory compactifications, by determining the possible codimension two fibers with rational sections. 
The relevant characteristic of the codimension two fibers that determine the $U(1)$ charge are the intersection numbers between the rational curves in the fiber and the section. 
We constrain these by combining the input from the box graphs on the codimension two fibers with general constraints on the normal bundles of rational curves in projective varieties. 
From section \ref{sec:Box} we obtain the information about the relative cone of effective curves $NE(\pi_i)$, for each resolution $(Y_i, \pi_i)$ of a singular Weierstrass model $W$. 
All curves in $NE(\pi_i)$ are {rational}, i.e. they are smooth $\mathbb{P}^1$s in $Y_i$.
In the following we will summarize several Theorems that we use in the later sections to  constain the fibers with rational sections for Calabi--Yau three- and four-folds. 
The protagonist in this discussion is the normal bundle of rational curves in Calabi--Yau varieties.


\subsection{Rational Curves and Normal Bundles}

In this section we collect useful results about rational curves in Calabi--Yau varieties,  in particular related to the normal bundle, which will allow us to constrain the fibers with rational sections. 
Unless otherwise stated $Y$ is a smooth Calabi--Yau variety. 

The first theorem constrains the degree of the normal bundle of a rational curve in a Calabi--Yau variety. 
\begin{Theorem}\label{thm:sum2}
  Let $Y$ be a smooth Calabi--Yau variety of dimension $n$ and $C$ a smooth
  rational curve in $Y$. Then the normal bundle of $C$ in $Y$,  ${N_{C/Y}}$,  is 
  \begin{equation*}
    N_{C/Y} = \bigoplus_{i=1}^{n-1} \mathcal{O}(a_i) \,, \quad \text{with}
    \quad\, \sum_{i=1}^{n-1} a_i = -2 \,.
  \end{equation*}
\end{Theorem}
{\it Proof:} E.g. for $n=3$ see \cite{MR1191424}.  Let $Y$ be of dimension $n$, then $N_{C/Y}$ is defined by the short exact sequence
\begin{equation}
  0 \rightarrow T_C \rightarrow T_{Y}|_C \rightarrow N_{C/Y} \rightarrow 0 \,,
\end{equation}
{where $T$ denotes the respective tangent bundles}. This  implies that $N_{C/Y}$ is a rank $n-1$ vector bundle on $C$ which, by
the Birkhoff-Grothendieck Theorem \cite{MR0087176}, can be written uniquely up
to permutations,  as a direct sum of line bundles on $C$,
\begin{equation*}
  N_{C/Y} = \bigoplus_{i=1}^{n-1} \mathcal{O}(a_i) \,.
\end{equation*}
By the Calabi--Yau condition on $Y$, the canonical bundle is trivial and thus, $c_1(T_Y|_C) = 0$. Combining this with
$c_1(T_C) = 2$ the exact sequence gives that $c_1(N_{C/Y}) = -2$. Thus $\sum
a_i = -2$. $\square$

In the following we will encounter rational curves which are contained within divisors, for instance, Cartan divisors associated to  the elliptic fibration, which we introduced in 
(\ref{DFk}). They are  ruled by 
the rational curves $F_k$ associated to simple roots of the gauge algebra, above the codimension one discriminant locus. Likewise we will see that 
the section, which we will assume to be a smooth divisor in the Calabi--Yau, can contain rational curves in the fiber that occur above codimension two. In all such 
instances it will be crucial to relate the normal bundle of the curve in the Calabi--Yau to the normal bundle in the divisor. This is achieved using the following exact sequence of normal bundles:
\begin{Theorem}\label{thm:NormalBundleSESGeneral}
  Let $Y$ be a smooth projective variety, $D$ a non-singular divisor in $Y$, and
  $C$ a smooth rational curve contained in $D$.  Then there is  a short exact sequence of normal bundles
  \be
    0 \rightarrow \ N_{C/D}  \rightarrow  N_{C/Y}  \rightarrow   \left.N_{D/Y}\right|_C  \rightarrow 0 \,.
    \ee
\end{Theorem}
{\it Proof:} \cite{MR0173675}, 19.1.5. $\square$\\

\noindent
One of the goals in later sections will be to determine the intersection of the rational section with various curves in the fiber. In particular, when these rational curves are 
contained in the section, this intersection is determined by the degree of the normal bundle of the divisor as follows -- 
{here $C$ does not necessarily have to be a rational curve}:

\begin{Theorem}\label{thm:adj}
Let $Y$ be a smooth projective variety, $D$ a divisor in $Y$ and $C$ a curve $C\subset D \subset Y$. 
Then 
\be 
D \cdot_Y C = \hbox{deg} \left(\left.N_{D/Y}\right|_{C} \right) 
\ee
\end{Theorem}
{\it Proof:} \cite{EisenbudHarris}, Theorem 15.1. $\square$\\

\noindent
Combining these properties, we can in fact relate the  
intersection of any non-singular divisor and a smooth
rational curve contained inside it in terms of the
degree of the normal bundle of the curve inside the divisor. 

\begin{Corollary}\label{cor:DC}
Let $Y$ be a smooth Calabi--Yau $n$-fold and $C$ a rational curve contained inside a smooth divisor $D$ in $Y$. 
Then 
\be
D\cdot_Y C = -2 - \hbox{deg} \left(N_{C/D}\right) \,.
\ee
\end{Corollary}
{\it Proof:} By Theorem \ref{thm:sum2} the degree of $N_{C/Y}$ is $-2$, which by Theorem \ref{thm:NormalBundleSESGeneral} has to be the sum of the degrees 
$-2 = \hbox{deg}( N_{C/D}) + \hbox{deg} (N_{D/Y}|_C) = \hbox{deg} (N_{C/D})+ D\cdot_Y C$  by  Theorem \ref{thm:adj}. $\square$\\

{With these general results we now turn to determining the possible degrees of normal bundles of rational curves in Calabi--Yau three-folds and four-folds in the next two sections, respectively. 
In particular we will constrain the normal bundles of rational curves in divisors, for instance rational sections, which by the above corollary will imply constraints on the intersections and thereby $U(1)$ charges.}


\subsection{Calabi--Yau Three-folds}

In this section, let  $Y$ be a smooth Calabi--Yau three-fold. 
Some results in rational curves in elliptically fibered three-folds (not necessarily Calabi--Yau varieties) can be found in Miranda \cite{Miranda}, which however does not discuss rational sections, or the generalization to higher dimensional varieties, which we will be important for us. 
Let $D$ be a smooth divisor in $Y$, and  $C$ a smooth rational curve contained in $D$. Then it follows directly from Corollary \ref{cor:DC} that\footnote{We will most of the time refrain from using $(C)^2_D = C\cdot_D C$ as this does not generalize to higher dimensional varieties. 
} 
  \begin{equation} 
    D \cdot_Y C = -2 - C \cdot_D C  \,.
  \end{equation}
We will often encounter the following situation: consider a rational curve $C$ in a smooth elliptic Calabi--Yau variety $Y$. From the box graph analysis, we know its normal bundle
in $Y$. We can then ask what normal bundles the curve can have in a divisor $D$ -- for instance the section. By the Corollary \ref{cor:DC}, the degree of the normal bundle $N_{C/D}$ is linked directly 
to the intersection in $Y$ of the divisor with the curve, which in the case when $D$ is a section determine the $U(1)$ charge. Thus, constraining the normal bundles of $C$ in the rational section results in constraints 
on the possible charges. 
The following theorem determines what the possible normal bundles of rational curves in divisors can be, given the normal bundle of the curve in $Y$. We furthermore summarize the bounds that are then implied 
upon the intersection of the divisor with the curve. 

\begin{Theorem}\label{thm:NormalBundleSES}
  Let $Y$ be a smooth Calabi--Yau three-fold, $D$ a non-singular divisor in $Y$, and
  $C$ a rational curve contained in $D$. 
  \begin{enumerate}
    \item[(i)] Let $(C)^2_D = \hbox{deg}(N_{C/D}) = k$. If $k\geq -1$ the short exact sequence of normal bundles in Theorem \ref{thm:NormalBundleSESGeneral} splits     
    and 
    \be
    N_{C/Y} = \mathcal{O}(k) \oplus
      \mathcal{O}(-2 - k) \,.
    \ee
     \item[(ii)] Let $N_{C/Y} = \mathcal{O}(-1) \oplus \mathcal{O}(-1)$. If $D$ is a smooth divisor containing $C$, then 
     \be 
     N_{C/D} = \mathcal{O}(k)\,,  \qquad k\leq -1 \,,
     \ee
     and there exists a non-trivial embedding
     \be
     \mathcal{O}(k) \  \hookrightarrow\  N_{C/Y} = \mathcal{O}(-1) \oplus \mathcal{O}(-1)  \,,
     \ee
     and 
     \be
     D\cdot_Y C = -2-k \geq -1 \,.
     \ee
     \item[(iii)] Let $N_{C/Y} = \mathcal{O} \oplus \mathcal{O}(-2)$. If $D$ is a smooth divisor containing $C$, then 
     \be 
     N_{C/D} = \mathcal{O}(k)\,, \qquad k=0 \quad \hbox{or} \quad k\leq -2  \,,
     \ee
     and there exists a non-trivial embedding
     \be
     \mathcal{O}(k) \  \hookrightarrow\  N_{C/Y} = \mathcal{O} \oplus \mathcal{O}(-2)  \,,
     \ee
     and 
     \be
     D\cdot_Y C = -2-k = \left\{\ba -2 & \quad\quad  k=0  \cr
     \geq 0 & \quad\quad  k\leq -2 
     \ea
     \right. \,.
     \ee

     \item[(iv)]  More generally, there is an embedding (without loss of generality $m\geq -1$)
     \be
     \mathcal{O}(k) \ \hookrightarrow \ \mathcal{O}(m) \oplus \mathcal{O}(-2-m)  \qquad \hbox { for } k= m \ \hbox{ or }\  k\leq  -2-m \,.
     \ee
  \end{enumerate}
\end{Theorem}
{\it Proof:} 
To show $(i)$ note that by Theorem \ref{thm:sum2} the degrees of the normal bundle have to sum to $-2$, so
$N_{C/Y} = \mathcal{O}(a) \oplus \mathcal{O}(-2-a)$, where without loss of generality  $a \leq -1$. By assumption $N_{C/D} = \mathcal{O}(k)$. 
The map $\mathcal{O}(k) \rightarrow \mathcal{O}(a)$ with $k\geq -1\geq a $  is trivial map,  unless $a=k$, in which case the Theorem follows. Else, if $a\not=k$ then $\mathcal{O}(k)$ needs to embed into $\mathcal{O}(-2-a)$ and therefore $k=-2-a$.   
Part (ii) follows by applying (i) which implies that if $k >-1$ then the normal bundle $N_{C/Y}$ cannot be $\mathcal{O}(-1) \oplus \mathcal{O}(-1)$. Thus $k\leq -1$, and there is an embedding of $\mathcal{O}(k)$ into $\mathcal{O}(-1) \oplus \mathcal{O}(-1)$. Similar arguments show parts (iii) and (iv). $\square$\\

Finally, the following theorem,  which we will only make use of in our analysis of singlets, determines the normal bundles of contractible curves in three-folds: 
\begin{Theorem}\label{thm:minus2}
Let $C$ be a smooth, rational curve that can be contracted in a smooth three-fold $Y$. Then the normal bundle is
\be
N_{C/Y} = \mathcal{O}(a) \oplus \mathcal{O}(b) \,,\qquad (a,b) = (-1,-1), \ (-2,0),  \hbox { or }\  (-3, 1) \,.
\ee
Such a curve is referred to as a $(-2)-$curve. 
\end{Theorem}
{\it Proof:} \cite{Reid, Laufer}.


\subsection{Calabi--Yau Four-folds}
\label{sec:four-folds}

For applications to 4d F-theory compactifications, including GUT model building, it is crucial to determine constraints for Calabi--Yau four-folds. 
In the following section, let $Y$ be a smooth Calabi--Yau four-fold, and $C$ a rational curve, contained in a smooth  divisor $D$. 
For elliptic fibrations, we will in fact be interested in a slightly more specialized situation, where inside the divisor $D$ there is a surface $S$ which is ruled by $C$. 
Specifically, we have in mind 
 what is usually referred to as matter surface,  which is a $\mathbb{P}^1$-fibration, i.e. a ruled surface,  over the matter curve (the codimension two locus in the base). 
These matter surfaces are contained within the Cartan divisors, which are dual to the rational cuves $F_i$ in the notation of section \ref{sec:Box}. 
In this setup, we will now show that  the classification for three-folds will in fact carry over directly to four-folds in codimension two. \footnote{It would appear that in fact it holds in codimension two for any elliptic Calabi--Yau $n$-fold. }

Again, the goal is to connect the intersection of divisors (in particular the section) with a rational curve $C$ in $Y$ to the degrees of the normal bundle of $C$ in $Y$. 
Recall the  short exact sequence of normal bundles from  Theorem \ref{thm:NormalBundleSESGeneral} \cite{MR0173675}
\be\label{ssNB}
  0 \ \rightarrow \  N_{C/D} \ \rightarrow\  N_{C/Y} \ \rightarrow \  N_{D/Y}|_C \ \rightarrow \ 0 \,.
\ee
By  Theorem \ref{thm:sum2}, the normal bundle is a direct sum of line bundles, where the sum of degrees needs to add up to $-2$
\be
N_{C/Y} = \mathcal{O} (a) \oplus \mathcal{O} (b) \oplus \mathcal{O}(-2 -a-b ) \,.
\ee
To determine the degrees $a$ and $b$, there are two cases of interest when $C$ is a rational curve in a codimension two fiber in an elliptic Calabi--Yau four-fold: 
either the rational curve $C$ corresponds to one of the curves that split in codimension two, or it remains irreducible. 
From the box graphs, we can determine the intersection of the Cartan divisors with the curves, $D\cdot_Y C$, which in turn by Theorem \ref{thm:adj}, constrain $N_{D/Y}|_C$. 
The following theorem determines the normal bundle $N_{C/Y}$ given the information about $N_{D/Y}|_C$:
\begin{Theorem} 
\label{thm:four-foldNB}
Let $C$ be a smooth rational curve, contained in a smooth divisor $D$ in a smooth Calabi--Yau four-fold $Y$. 
\begin{enumerate}
\item[(i)]  If $N_{D/Y}|_C = \mathcal{O}(-1)$ and $D$ contains a surface $S$, which is ruled by $C$,  then 
\be
N_{C/D} = \mathcal{O} \oplus \mathcal{O}(-1) \,,
\ee
and the short exact sequence (\ref{ssNB}) splits 
\be
N_{C/Y} =  \mathcal{O} \oplus \mathcal{O}(-1) \oplus \mathcal{O}(-1) \,. 
\ee
\item[(ii)] Likewise for $N_{D/Y}|_C = \mathcal{O}(-2)$ and $D$ is ruled by $C$ then 
\be
N_{C/D} = \mathcal{O} \oplus \mathcal{O} \,,
\ee
and 
\be
N_{C/Y} = \mathcal{O} \oplus \mathcal{O} \oplus \mathcal{O}(-2) \,.
\ee
\end{enumerate}
\end{Theorem}
{\it Proof:}
(i)  If there is a surface in $D$ which is ruled by $C$ then there is an embedding 
\be
\mathcal{O} \ \hookrightarrow\  N_{C/D} \,.
\ee
If $N_{D/Y}|_{C} =\mathcal{O}(-1)$ and given that the degrees in $N_{C/Y}$ sum to $-2$, it follows that 
\be
N_{C/D} = \mathcal{O}(m) \oplus \mathcal{O}(-1-m) \,.
\ee
As $\mathcal{O} = N_{C/S}$ needs to embed into $N_{C/D}$, it follows that $m=0$. 
 The extension group of $\mathcal{O} \oplus \mathcal{O}(-1)$ and $\mathcal{O}(-1)$ is trivial, and thereby the exact sequence splits. 
(ii) By similar arguments as in (i) $N_{C/D} = \mathcal{O}(m) \oplus \mathcal{O}(-m)$, and for $\mathcal{O}$ to embed into this $m=0$. 
Again the extension group is trivial and the normal bundle sequence splits. $\square$\\

For $\sigma$ a rational section, which contains curves in the fiber, we can now constrain the possible normal bundle degrees of $C$ in $\sigma$. 
The last theorem provides us with the information about the normal bundles $N_{C/Y}$. 
As in Theorem \ref{thm:NormalBundleSES}, we now determine the constraints on the intersection numbers $\sigma\cdot_Y C$ (where $\sigma$ will be now be a rational section) by constraining the degrees of the normal bundle
of $C$ in $\sigma$, which are related by Corollary \ref{cor:DC}. 
\begin{Theorem} \label{thm:abNB}
Let $\sigma$ be a smooth divisor in $Y$, a smooth Calabi--Yau four-fold,  and $C \subset \sigma$ a rational curve.
\begin{enumerate}
\item[(i)]
If  $N_{C/Y}= \mathcal{O} \oplus \mathcal{O} (-1 )\oplus \mathcal{O}(-1)$, then there is an embedding
\be
N_{C/\sigma}=  \mathcal{O}(a) \oplus \mathcal{O}(b) \ \hookrightarrow\ N_{C/Y}= \mathcal{O} \oplus \mathcal{O} (-1 )\oplus \mathcal{O}(-1) 
\ee
and 
\be
\sigma \cdot_Y C = -2 -a -b \,.
\ee
The values for $a$ and $b$ are constrained to be (wlog $a\geq b$)
\be
a\leq0 \,,\quad  b\leq -1 \,,\quad  a+b \leq -1 \,,
\ee
which implies that 
\be
\sigma \cdot_Y C \geq -1 \,.
\ee
\item[(ii)]
If  $N_{C/Y}= \mathcal{O} \oplus \mathcal{O} \oplus \mathcal{O}(-2)$, then there is an injection
\be
N_{C/\sigma}=  \mathcal{O}(a) \oplus \mathcal{O}(b) \ \hookrightarrow\ N_{C/Y}= \mathcal{O} \oplus \mathcal{O} \oplus \mathcal{O}(-2) 
\ee
and 
\be
\sigma \cdot_Y C = -2 -a -b \,.
\ee
The values for $a$ and $b$ are constrained to be  
\be
a=b=0 \qquad \hbox{ or } \qquad a\leq0 \,,\   b\leq 0 \,,\   a+b \leq -2 \,,
\ee
which implies that 
\be
\sigma \cdot_Y C  = \left\{  
\ba
-2 & \quad \quad a=b=0 \cr
\geq 0 & \quad \quad a+b \leq -2 
\ea \right. \,.
\ee

\end{enumerate}
\end{Theorem}
{\it Proof:} This follows directly from the short exact sequence (\ref{ssNB}) and Corollary \ref{cor:DC}. $\square$\\

{This concludes our summary of properties of rational curves. We now turn to combining these constraints on the intersection numbers and normal bundles, with the constraints from the box graphs 
that specify how codimension one fibers split in codimension two. The next two sections will discuss this in the case of $SU(n)$ with various matter representations. }


\section{$SU(5) \times U(1)$ with $\overline{\bf 5}$ Matter}
\label{sec:SU5F}

The ultimate physics application of our analysis of codimension two fibers is
the case of $SU(5)$ GUTs with additional $U(1)$ symmetries. The constraints on
the section and codimension two fiber structure provide a systematic way to
obtain a comprehensive list of all possible $U(1)$ charges for matter in the
$\overline{\bf 5}$ and ${\bf 10}$ representation of the GUT group $SU(5)$. In this
section we will first focus on  fundamental matter.

Throughout this section let $Y$ be an elliptically fibered Calabi--Yau
variety. The zero section of the fibration will be denoted by $\sigma_0$,
and the additional rational section needed for there to be a $U(1)$ symmetry
as $\sigma_1$. 


\subsection{Setup and Scope}
\label{sec:Assumptions}

There are a few assumptions that go into this analysis, and to make it clear what the scope of the results in this paper are, we will now list them. 

\begin{itemize}
\item[(1.)] We assume that each section in codimension one intersects exactly one fiber component transversally once, i.e. the sections do not contain components of codimension one fibers\footnote{This in fact seems to not be a real constraint, as wrapping in codimension one would imply that the section is either ruled by rational curves in the fiber (and thereby would contract to a curve in the singular limit) or not be irreducible.}. 
\item[(2.)] The rational sections, as divisors in $Y$,  will always be assumed to be smooth.
\item[(3.)] The codimension one locus in the base of the fibration, above which there are singular fibers $I_5$,  is smooth.
\item[(4.)] The $U(1)$ generator is an integral divisor normalized as
    described after (\ref{eq:shiodas}). 
\end{itemize}
Within the setup outlined above, the following can be regarded as complete classification of codimension two fibers for both Calabi--Yau three- and four-folds with one extra rational section, and thereby the possible matter charges. 


\subsection{Codimension one Fibers with Rational Sections}
\label{sec:SU5codim1}

The codimension one fibers for $SU(5)$ GUTs realized in F-theory are fibers of
Kodaira type $I_5$. These fibers consist of a ring of five smooth rational
curves, $F_i$ for $i = 0,\cdots,4$.

Further, as these curves are the
components of the fiber over generic points above a codimension one locus in the base, $S_{GUT}$, 
one can define divisors in $Y$, which are ruled by the curves $F_i$ over $S_{GUT}$.
 These divisors, $D_{F_i}$, are called
the Cartan divisors, and satisfy 
\begin{equation}
  D_{F_i} \cdot_Y F_j = -C_{ij} \,,
\end{equation}
where $C_{ij}$ is the Cartan matrix of affine $SU(5)$.

\begin{figure}
  \centering
  \includegraphics[width=9cm]{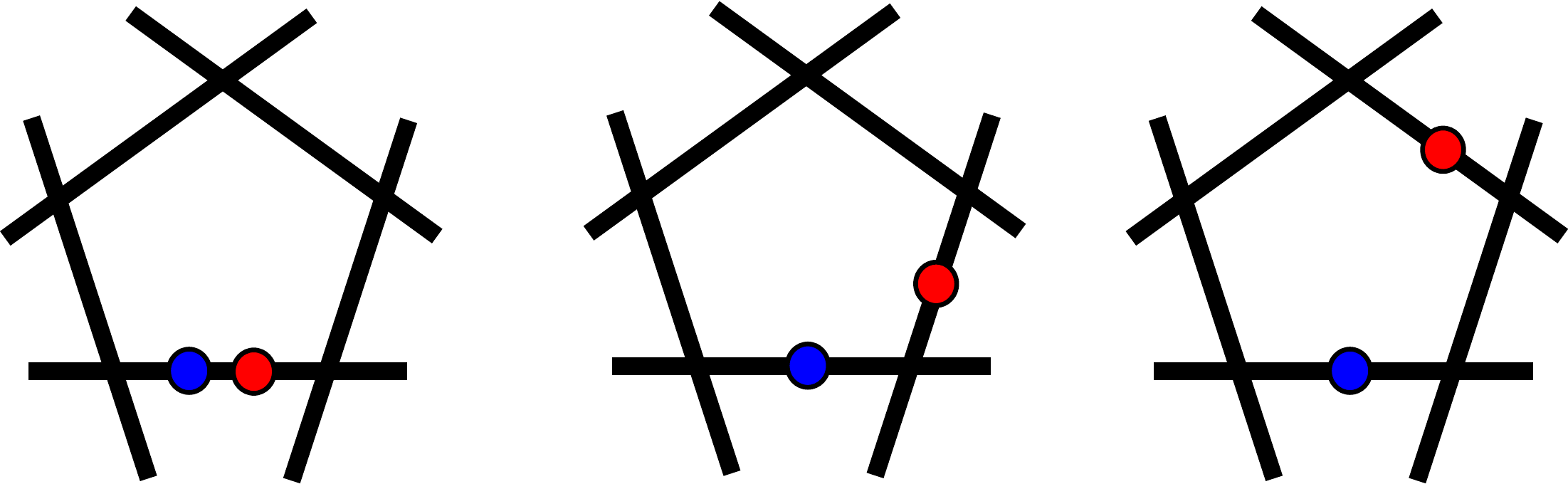} 
  \caption{Three types of codimension one $I_5$ fibers with sections $\sigma_0$ (blue) and $\sigma_1$ (red) distributed as $I_5^{(01)}$, $I_5^{(0|1)}$ and $I_5^{(0||1)}$, respectively. \label{fig:I5Codim1}}
\end{figure}

Let $\sigma$ be a rational section of the elliptic fibration, i.e. it has to satisfy 
\begin{equation}
  \sigma \cdot_Y \text{Fiber} = 1 \,.
\end{equation}
Throughout this paper it shall be assumed, see section \ref{sec:Assumptions}, that this condition is satisfied by
$\sigma$ having exactly one transversal intersection with one of the
components of the generic codimension one fiber and having no intersection with the other
components. The section thus intersects, say, the  $m$th component of the
fiber
\begin{equation}
  \sigma \cdot_Y F_i = \begin{cases}
    1 \quad i = m \cr
    0 \quad i \neq m \,. 
  \end{cases} 
\end{equation}
It shall always be supposed, without loss of generality, that one section, the
zero-section, shall intersect the component $F_0$. Up to inverting the order
of the simple roots there are three distinct codimension one fiber types once
this information about the additional rational section is included. These
are, using the notation introduced in \cite{Kuntzler:2014ila}, 
\begin{equation}\label{I5fibertypes}
  \begin{aligned}
    I_5 ^{(01)}: &\qquad  \sigma_0 \cdot_Y F_0 = \sigma_1 \cdot_Y F_0 =1 \cr
    I_5 ^{(0|1)}: &\qquad  \sigma_0 \cdot_Y F_0 = \sigma_1 \cdot_Y F_1 =1 \cr
    I_5 ^{(0||1)}: &\qquad  \sigma_0 \cdot_Y F_0 = \sigma_1 \cdot_Y F_2 =1 \,,
  \end{aligned}
\end{equation}
corresponding to the three configurations shown in figure \ref{fig:I5Codim1}.

The $U(1)$ generator comes from the Shioda map as applied to the extra
rational section, $\sigma_1$. The Shioda map associates to a rational section $\sigma_1$ an element $\mathcal{S}(\sigma_1)$ in 
$H_{2d-2}(Y, \mathbb{Z})$, where $d$ is the complex dimension of
$Y$,  which is perpendicular to all horizontal divisors (i.e. divisors pulled back from the base), the zero section as well as 
 the Cartan divisors, associated to the $F_i$, which ensures that the non-abelian $SU(5)$ gauge bosons are uncharged under the $U(1)$ \cite{Morrison:2012ei}.
In order to compute $U(1)$ charges of matter, we are interested in the intersection of the Shioda map with curves in the fiber, for which the subtractions from contributions of horizontal divisors are not relevant, and we therefore define 
$S(\sigma_1)$ to be such that 
 \be
 S(\sigma_1)  \cdot_Y C = q(C) \,, 
 \ee
 the charge under the $U(1)$. 
In this way the Shioda map
is specified by the codimension one data of the fibration. For $SU(5)$ with
Mordell--Weil group rank one the Shioda divisors are
\begin{equation}\label{eq:shiodas}
  \begin{aligned}
    I_5 ^{(01)}: &\qquad S(\sigma_1)= \sigma_1 -  \sigma_0 \cr
    I_5 ^{(0|1)}: &\qquad S(\sigma_1)= 5 \sigma_1 - 5\sigma_0 + 4 D_{F_1} + 3 D_{F_2} +
    2 D_{F_3} + D_{F_4} \cr
    I_5 ^{(0||1)}: &\qquad S(\sigma_1) =  5 \sigma_1 - 5\sigma_0 + 3 D_{F_1} + 6 D_{F_2}
    + 4 D_{F_3} + 2 D_{F_4} \,.
  \end{aligned}
\end{equation}
To arrive at the specific forms above some further assumptions need to be
made for the divisor $S(\sigma_1)$ that generates the $U(1)$ symmetry from
the Shioda map. Imposing orthogonality to the $SU(5)$ Cartan divisors
specifies the above up to a multiplicative constant. This constant is fixed
by the requirement that $S(\sigma_1)$ should be integral, and that there
should be no other integral divisor $D$ such that $S(\sigma_1) = m^\prime D$
for some $|m^\prime| > 1$. The last condition is required for the $U(1)$
symmetry to be normalized appropriately. Assumption (4.) in section
\ref{sec:Assumptions} is precisely that there does not exist such an
integral divisor $D$.

\subsection{Normal Bundles in Elliptic Calabi--Yau Varieties}
\label{sec:NBinCY}

\begin{figure}
  \centering
  \includegraphics[width=17cm]{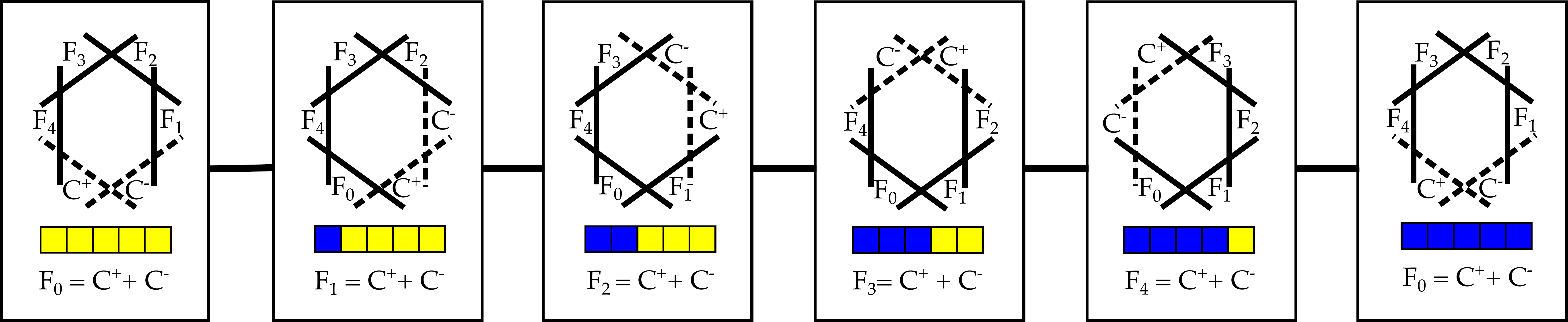} 
  \caption{Box graphs and codimension two fibers where the $F_j$ that split into $C^\pm$ in codimension two are shown with dashed lines, for the  $\mathfrak{su}(5)\oplus \mathfrak{u}(1)$ theory with matter in the fundamental representation. \label{fig:SU5BoxGraphs5}}
\end{figure}

We start with an $I_5$ fiber, with components $F_i$, intersecting in the affine Dynkin diagram of $SU(5)$. 
Along codimension two enhancement loci, some fiber components become reducible. 
The resulting codimension two fibers, which give rise to matter in the fundamental representation, were determined in section
\ref{sec:Box}, from the Coulomb phases/box graphs, where one of the $F_j$ curves splits as follows
\begin{equation}
  F_j \rightarrow C^+ + C^- \,.
\end{equation}
In the case of $SU(5)$ with $\overline{\bf 5}$ these are shown in  figure 
\ref{fig:SU5BoxGraphs5}, including the fibers that split, shown as dashed lines. 

In this analysis we allow for a non-holomorphic zero section \cite{Cvetic:2013nia, Braun:2013yti}  which means that over codimension two $\sigma_0$ can also contain curves in the fiber. 
Let $\sigma$ denote either $\sigma_0$ or $\sigma_1$. 
We will now determine the fibers including the rational sections in codimension two. 
In addition to intersecting the components of the codimension two fiber transversally, the section can contain entire fiber components $C \subset \sigma$, which in the existing literature is refered to as \textit{wrapping}. 
In addition to consistency of the embedding of the rational curves into the divisors $\sigma$, we will use two constraints to determine all possible fibers:
\begin{enumerate}
  \item If $\sigma \cdot_Y F_i =0$ or $1$, then this
    holds also in codimension two, in particular when the curve $F_i$ splits
    it is necessary that the sum of the two curves, $C^+$ and $C^-$,
    intersects with the section as $F_i$ did.
  \item $\sigma \cdot_Y \text{Fiber} = 1$.
\end{enumerate}

Denote by $F_p$ the 
codimension one fiber component that splits 
\be
F_p \ \rightarrow\ C^+ + C^- \,.
\ee
 From the box graph analysis it is known that the intersection with $D_{F_p}$ of these curves is
\be\label{DFCpm}
      D_{F_p} \cdot_Y C^\pm = -1 \,.
   \ee
For the case where a curve $F_i$ in the fiber remains irreducible, again from the box graph analysis, we have that
\be \label{DFFi}
      D_{F_i} \cdot_Y F_i = -2 \,.
\ee
We will now determine, using \eqref{DFCpm} and \eqref{DFFi}, the normal bundles of the curves $C^{\pm}$ and $F_i$ in $Y$, which will in turn fix the possible intersection of these curves with the section.

\subsubsection{Three-folds}
   
First consider the case where $Y$ is a Calabi--Yau three-fold. 
Then by Theorem \ref{thm:NormalBundleSES} (i),  (\ref{DFCpm}) fixes the normal bundles to be 
    \begin{equation}
      N_{C^{\pm}/Y} = \mathcal{O}(-1) \oplus \mathcal{O}(-1) \,.
    \end{equation}
    If a curve $C=C^\pm$ is contained in the divisor $\sigma$, $C\subset \sigma$, then from Theorem  \ref{thm:NormalBundleSES}  (ii) it follows that     \begin{equation}
      N_{C /\sigma} = \mathcal{O} (k)\,,\quad k\leq -1 \,,
    \end{equation}
    and this in turn bounds the intesection of the curve with the section
        \be\label{eqn:sigmaCeqm1}
    \sigma \cdot_Y C = -2 - k  \geq -1 \,.
    \ee
    On the other hand, if $\sigma$ does not contain one of the curves $C= C^\pm$, then  $\sigma \cdot_Y C \geq 0$. 
    In summary we can conclude that the intersection number of $\sigma$ with the two curves $C^\pm$ is always bounded below as follows
    \be\label{sigmaCbound}
    \sigma \cdot_Y C^\pm \geq -1 \,.
    \ee
If $F_i$ is irreducible and $F_i\subset \sigma$ then its normal bundle in $Y$ is given by
\be
N_{C^{\pm}/Y} = \mathcal{O}\oplus \mathcal{O}(-2) \,,
\ee
and by \eqref{DFFi}  and Theorem \ref{thm:NormalBundleSES} (iii)
    \be\label{FisigmaNB}
    N_{F_i/\sigma} = \mathcal{O}(k),\qquad k=0 \quad \hbox{or} \quad k\leq -2 \,,
    \ee    
    and 
    \be
    \sigma \cdot_Y F_i =  \left\{ \ba -2 & \qquad k=0 \cr \geq 0 & \qquad k\leq -2 
\ea \right. \,.
    \ee
    
    \subsubsection{Four-folds}
    \label{sec:four-foldNB}
    
 Likewise we can consider the case when $Y$ is a smooth Calabi--Yau four-fold. We will now show that the constraints on the intersections of the section with the fiber components
 in this case are the same as the ones we derived for three-folds. 
 In section \ref{sec:NBinCY} we started by considering a rational $F_p$ in the fiber, which in codimension two splits and  
 \be
D_{F_p} \cdot_Y C^\pm =-1 \,.
\ee
Let $S^{\pm}$ be the surfaces ruled by $C^\pm$ over the codimension two locus in the base.
Then $S^\pm \subset D_{F_p}$  which implies by  Theorem \ref{thm:four-foldNB} (i), that 
\be
N_{C^\pm/D_{F_p}} =  \mathcal{O} \oplus \mathcal{O}(-1) \,. 
\ee
and that the normal bundle to these curves in the four-fold is 
\be
N_{C^{\pm}/Y} = \mathcal{O} \oplus \mathcal{O}(-1) \oplus \mathcal{O}(-1) \,.
\ee 
Consider now the situation that $S=S^\pm$ is contained in $\sigma$, and thereby $C= C^\pm \subset \sigma$. 
There is a normal bundle exact sequence 
\be\label{NCSsigma}
0 \rightarrow N_{C/S} \rightarrow N_{C/\sigma} \rightarrow N_{S/\sigma}|_{C} \rightarrow 0 \,.
\ee
As $S$ is ruled by $C$ we know that $N_{C/S} = \mathcal{O}$. On the other hand, we know that by the normal bundle 
exact sequence for $C\subset \sigma \subset Y$ 
\be
 0\rightarrow N_{C/\sigma} \rightarrow \ N_{C/Y} = \mathcal{O} \oplus  \mathcal{O}(-1) \oplus \mathcal{O}(-1)  \ \rightarrow N_{\sigma/Y}|_C \rightarrow 0 \,,
\ee
thus writing $ N_{C/\sigma} = \mathcal{O}(a) \oplus \mathcal{O}(b)$ Theorem \ref{thm:abNB} (i) states that $a\leq0$, $b\leq -1$ and $a+b \leq -1$. However, from 
(\ref{NCSsigma}), we know that $\mathcal{O} \hookrightarrow \mathcal{O}(a) \oplus \mathcal{O}(b)$, therefore we must have $a= 0$ and $b\leq -1$, i.e.
\be
N_{C^{\pm}/\sigma} = \mathcal{O} \oplus \mathcal{O} (k) \,,\qquad k\leq -1 \,.
\ee
This proves that the conditions on the normal bundle degrees of $N_{C/\sigma}$ for four-folds are exactly the same as the ones we derived in the case of three-folds (\ref{eqn:sigmaCeqm1})
resulting in the same bounds on $\sigma\cdot_Y C^\pm$ as in (\ref{sigmaCbound}). 

Likewise, when $F_i\subset S_i$ is contained in the section, where $S_i$ is the surface ruled by $F_i$ over the codimension two locus in the base, 
then $D_{F_i} \cdot_Y F_i=-2$  and by  Theorem \ref{thm:four-foldNB} (ii)
\be
N_{F_i/Y} = \mathcal{O} \oplus \mathcal{O} \oplus \mathcal{O} (-2) \,.
\ee
Again applying the normal bundle exact sequences to $F_i \subset S_i \subset \sigma$ as well as $F_i\subset \sigma \subset Y$ 
we infer from \ref{thm:abNB} (ii) that 
\be
N_{F_i /\sigma} = \mathcal{O} \oplus \mathcal{O}(k) \,,\qquad k =0 \quad \hbox{or} \quad k\leq -2 \,,
\ee
which again is identical to the constraints that we had on the normal bundle degree for $F_i \subset \sigma$ in the three-fold case in (\ref{FisigmaNB})
and thus the bound on $\sigma \cdot_Y F_i$ is also identical to that case and depends only on $k$.

It seems that similar arguments will hold for elliptic Calabi--Yau $n$-folds in codimension two, quite generally for $n\geq3$, where instead of a ruled surface $S^\pm$, there is a ruled  
$n-2$ dimensional sub-variety, which is ruled by the rational curves in the fiber. This seems to only add additional $\mathcal{O}$ summands to the normal bundle, and the constraints on the intersections would appear to be the same as the ones we derived for $n=3$ and $n=4$.


\subsection{Codimension two Fibers with Rational Sections}
\label{sec:Codim2Analysis}

In the last section we have shown that the conditions on the normal bundle degrees for rational curves in the elliptic fibration 
which are contained in the section, are characterized, for both three- and four-folds by one integer, namely, the degree of the normal bundle 
$N_{C/\sigma} = \mathcal{O}(k)$ for three-folds, and  $N_{C/\sigma} = \mathcal{O} \oplus \mathcal{O}(k)$, for four-folds, respectively, where $k$ is bounded as described in the previous section. 
The happy fact, that the degrees in three-and four-folds (in this specifc context), are constrained in the same way, allows us to carry out a full classification simultaneously for both cases. 
The only important input is the degree of the normal bundles deg$(N_{C/\sigma}) =k$, upon which the charges will depend.
One last word of caution before we start our analysis: in the case of four-folds, whenever a rational curve $C$ in the fiber is contained in $\sigma$, we mean this to imply always, that 
there is a surface $S$, which is ruled by $C$ over the codimension two locus, which is also contained in $\sigma$ (i.e. in compliance with the general discussion in section \ref{sec:four-foldNB}).

The two cases to consider now separately are
\be
\sigma \cdot_Y F_p = \sigma \cdot_Y (C^+ + C^- ) = \left\{ \ba 0  & \quad \hbox{Case (a)} \cr  1  &\quad \hbox{Case (b)} \ea \right.\,.
\ee
\begin{enumerate}
\item[(a)] $\sigma \cdot_Y F_p =0$:\\
From (\ref{eqn:sigmaCeqm1}) it follows that $\sigma \cdot_Y C^\pm \geq -1$. 
There are three solutions to $\sigma \cdot_Y F_p =0$:
\be
(\sigma \cdot_Y C^{+}, \sigma \cdot_Y C^{-}) = (-1, 1)\, , \ (0, 0)\  \hbox{ and }  \ (1,-1) \,.
\ee
There are several ways that each of these intersections can be realized: 
$\sigma \cdot_Y C^{+}=-1$ implies $C^+ \subset \sigma$ and the degree of the normal bundle of $C^+$ in $\sigma$ is $\hbox{deg}(N_{C^+/\sigma}) =-1$. Likewise, $\sigma \cdot_Y C^{+}=0$ implies $C^+ \subset \sigma$ and $\hbox{deg}(N_{C^+/\sigma}) = -2$ or $C^+\not\subset\sigma$ with no transverse intersection. 
On the other hand the intersections for $C^-$ can be realized as follows: $\sigma \cdot_Y C^{-}=1$ implies either, that $C^- \not\subset \sigma$, and intersects $\sigma$ transversally once, or $C^- \subset \sigma$ and $\hbox{deg}(N_{C^-/\sigma}) =-3$. The case for $\sigma \cdot_Y C^+=1$ proceeds in the same fashion, by swapping $C^+$ and $C^-$.
The intersection $\sigma \cdot_Y C^{-}=0$ implies either, that $C^- \not\subset \sigma$, and does not intersects $\sigma$, or $C^- \subset \sigma$ and $\hbox{deg}(N_{C^-/\sigma})=-2$. 

In the last case, it is important to note that by the structure of the codimension two fiber the two curves $C^\pm$, which are both contained in the divisor $D_{F_p}$, intersect
\be\label{CpCm}
C^+ \cdot_{D_{F_p}} S^- = 1 \,,
\ee
where $S^-$ is the matter surface, which is ruled by $C^-$ in the case of four-folds, and is equal to $C^-$ for three-folds.
I.e. if one of the curves is contained in the section, then the other curve will automatically acquire an intersection with the section. Thus the combinations $C^+ \subset \sigma$, $\hbox{deg}(N_{C^+/\sigma}) =-2$  and $C^- \not\subset \sigma \,,\  \sigma\cdot_Y C^- =0$ do not have any solution in an $I_6$ fiber. 

In summary we obtain the following configurations:
\be\label{sigmaFp0}
\begin{array}{c|c||l|l}
\sigma\cdot_Y C^+ & \sigma\cdot_Y C^- & \quad C^+ \hbox { configuration  }& \quad C^- \hbox { configuration  }\cr\hline\hline
-1 	&  1  &  	C^+ \subset \sigma \,,\ \hbox{deg}(N_{C^+/\sigma})=-1 & C^- \not\subset \sigma \,,\  \sigma\cdot_Y C^- =1\cr
    	&    	&   	C^+ \subset \sigma \,,\ \hbox{deg}(N_{C^+/\sigma})=-1 & C^- \subset \sigma \,,\  \hbox{deg}(N_{C^-/\sigma})=-3\cr\hline
0   	&  0 	&    	C^+ \subset \sigma \,,\ \hbox{deg}(N_{C^+/\sigma})=-2 & C^- \subset \sigma \,,\  \hbox{deg}(N_{C^-/\sigma})=-2\cr 
	&	&	C^+ \not\subset \sigma \,,\  \sigma\cdot_Y C^+ =0&C^- \not\subset \sigma \,,\  \sigma\cdot_Y C^- =0 \cr\hline
1 	&  -1 &    	C^+ \not\subset \sigma \,,\  \sigma\cdot_Y C^+ =1 & C^- \subset \sigma \,,\ \hbox{deg}(N_{C^-/\sigma})=-1\cr
     	&    &    	C^+ \subset \sigma \,,\  \hbox{deg}(N_{C^+/\sigma})=-3 & C^- \subset \sigma \,,\ \hbox{deg}(N_{C^-/\sigma})=-1 
\end{array}
\ee

\item[(b)] $\sigma \cdot_Y F_p =1$:\\
Making use again of the bound (\ref{eqn:sigmaCeqm1}), the solutions to $\sigma \cdot_Y (C^+ + C^-) =1$ are 
\be
(\sigma \cdot_Y C^{+}, \sigma \cdot_Y C^{-}) = (-1, 2)\,, \  (0,1) \,,\  (1, 0)\ \hbox{ and } \ (2,-1) \,.
\ee
The only new configuration that has not already appeared in case (a) is $\sigma \cdot_Y C^{-} =2$. One configuration that realizes this is $C^- \not\subset \sigma$, but $C^-$ has two transverse intersection points with $\sigma$. Note that in this case $C^+$ is contained in $\sigma$, and thus contributes an intersection by (\ref{CpCm}). 
If $C^- \subset \sigma$ then $\hbox{deg}(N_{C^-/\sigma})=-4$. The complete set of section configurations in this case are summarized in the following table\footnote{We will see that the intersection configurations with $(*)$ in fact do not have a realization in an $I_6$ fiber. }: 
\be\label{sigmaFp1}
\begin{array}{c|c||l|l}
\sigma\cdot_Y C^+ & \sigma\cdot_Y C^- & \quad C^+ \hbox { configuration  }& \quad C^- \hbox { configuration  }\cr\hline\hline
-1 	&  2 	&  	C^+ \subset \sigma \,,\ \hbox{deg}(N_{C^+/\sigma})  =-1 		& C^- \not\subset \sigma \,,\  \sigma\cdot_Y C^- =2\cr
     	&    	&   	C^+ \subset \sigma \,,\ \hbox{deg}(N_{C^+/\sigma})  =-1 		& C^- \subset \sigma \,,\  \hbox{deg}(N_{C^-/\sigma})  =-4\cr\hline
0   	&  1 	&    	C^+ \subset \sigma \,,\ \hbox{deg}(N_{C^+/\sigma})  =-2 		& C^- \not\subset \sigma \,,\  \sigma\cdot_Y C^- =1 \  (*)\cr 
	& 	& 	C^+ \subset \sigma \,,\ \hbox{deg}(N_{C^+/\sigma})  =-2 		& C^-  \subset \sigma \,,\   \hbox{deg}(N_{C^-/\sigma})  =-3 \cr 
	&	&	C^+ \not\subset \sigma \,,\  \sigma\cdot_Y C^+ =0	&C^- \not\subset \sigma \,,\  \sigma\cdot_Y C^- =1 \cr\hline
1  	&  0 	&   	C^+ \not\subset \sigma \,,\  \sigma\cdot_Y C^+ =1  	&  C^- \subset \sigma \,,\ \hbox{deg}(N_{C^-/\sigma})  =-2\  (*)\cr 
	&   	&  	C^+  \subset \sigma \,,\   \hbox{deg}(N_{C^+/\sigma})  =-3		& C^- \subset \sigma \,,\ \hbox{deg}(N_{C^-/\sigma})  =-2 \cr 
	&	&	C^+ \not\subset \sigma \,,\  \sigma\cdot_Y C^+ =1	&C^- \not\subset \sigma \,,\  \sigma\cdot_Y C^- =0 \cr\hline
2 	&  -1 &  	C^+ \not\subset \sigma \,,\  \sigma\cdot_Y C^+ =2 	& C^- \subset \sigma \,,\ \hbox{deg}(N_{C^-/\sigma})  =-1\cr
     	&    	&   	C^+\subset \sigma \,,\  \hbox{deg}(N_{C^+/\sigma})  =-4		&C^- \subset \sigma \,,\ \hbox{deg}(N_{C^-/\sigma})  =-1\cr
\end{array}
\ee
\end{enumerate}

Note that for each value of $\sigma \cdot_Y C^\pm$ there are two realizations in terms of different configurations, and in the following we will only consider one of these. 

Furthermore, we need to discuss the remaining fiber components. From the box graphs, we know that the intersection of rational curves in the fiber in codimension two is that of an $I_6$ Kodaira fiber. Thus, if a component $C^\pm$ is contained in $\sigma$ it induces intersections of the section with the adjacent fiber components.  
Depending on the position of the section in codimension one, there are two cases again to consider: 
let $F_q$ be such that it remains an irreducible fiber component in codimension two. Then 
\begin{enumerate}
\item[(a)] $\sigma \cdot_Y F_q= 0$: \\
Either $F_q\not\subset \sigma$ and has no transverse intersections, or $F_q \subset \sigma$ then $\hbox{deg}(N_{F_q /\sigma}) =-2 $.
\item[(b)] $\sigma \cdot_Y F_q= 1$: \\
Either $F_q\not \subset \sigma$ and has one transverse intersection, or $F_q\subset\sigma$ then $\hbox{deg}(N_{F_q /\sigma}) =-3$. 
\end{enumerate}

We can now determine the complete set of fibers in codimension two with a rational section $\sigma$. Again, $F_p \rightarrow C^+ + C^-$ is the rational curve that becomes reducible in codimension two: 

\begin{enumerate}
  \item[(i)] $C^+, C^- \not\subset\sigma$:  
       \begin{enumerate}
      \item $\sigma \cdot_Y F_p = 0$  and $\sigma \cdot_Y F_m=1$, $p\not=m$:\\
        It follows from table  \ref{sigmaFp0} that the only configuration is 
        \be
        C^+, C^-\not\subset \sigma \,,\quad \sigma\cdot_Y C^{\pm} =0 \,.
        \ee
        The section does not intersect either of the split components, indeed
        it must merely remain on the component that it originally intersected in
        codimension one, $F_m$. Figures \ref{fig:WrapEx} and \ref{fig:InFib1} (i) represent this
        configuration.
      \item $\sigma \cdot_Y F_p = 1$: \\
        From table  \ref{sigmaFp1} the only two solutions are 
        \be
        C^+, C^-  \not\subset \sigma \,,\quad \sigma\cdot_Y C^{\pm} =1 \,,\quad  \sigma\cdot_Y C^{\mp} =0 \,.
        \ee
	In this case the section intersects one of the split components transversally, and does not contain any curves in the fiber. This is 
	 shown in figure \ref{fig:WrapEx}, and more generally, in figures \ref{fig:InFib2}, (i) and (ii),
        respectively.
   \end{enumerate}
   
\begin{figure}
  \centering
  \includegraphics[width=16cm]{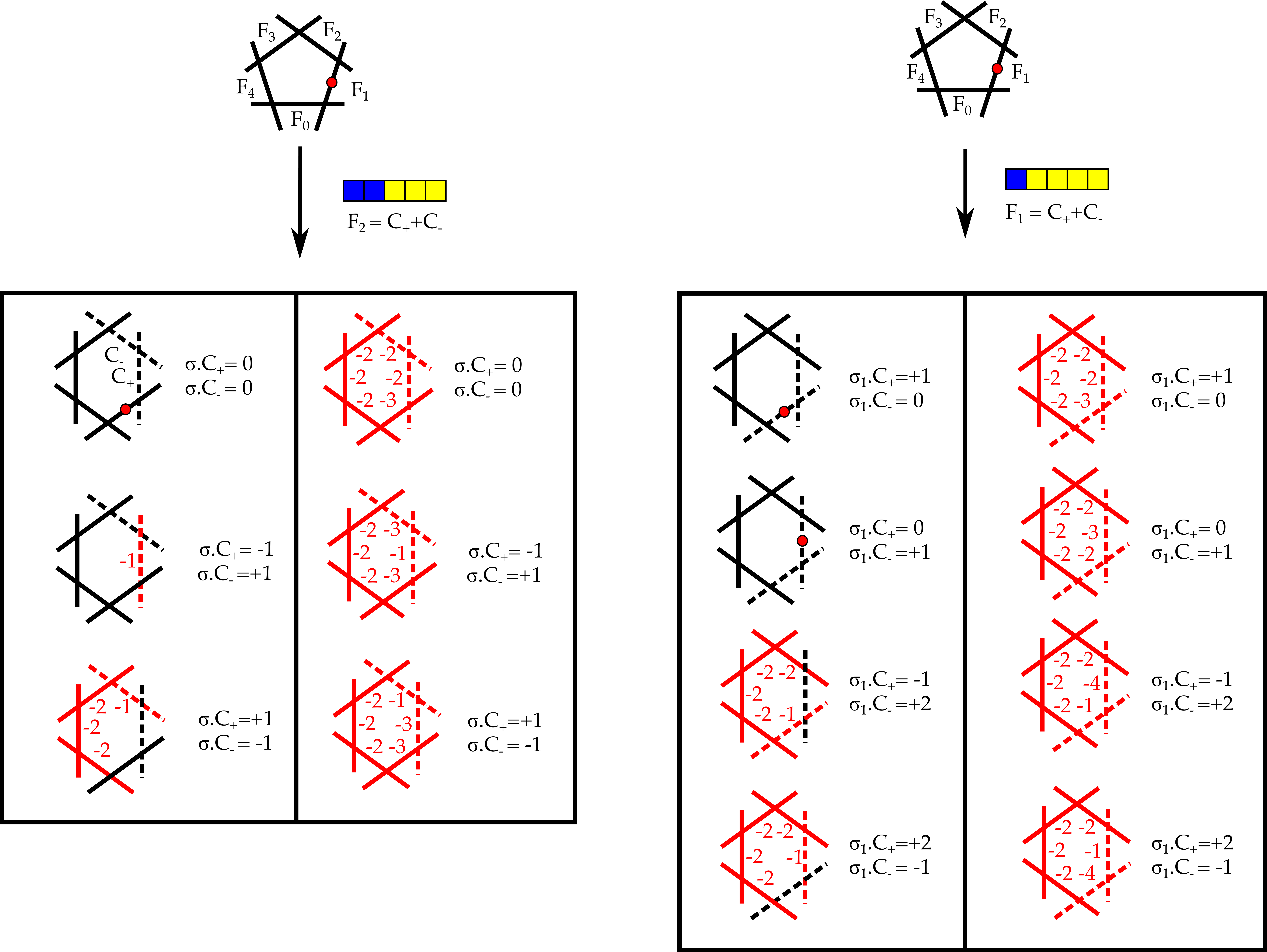} 
  \caption{$I_5$ fiber with rational section $\sigma$, shown intersecting $F_1$ in codimension one. The left hand side shows the case $F_2\rightarrow C^+ + C^-$ in codimension two and all the the section configurations that are consistent, which correspond to all case (a) in the main text. 
  The fiber components that are contained in $\sigma$ are colored red, and the numbers next to it refer to the degree of the normal bundle of the curves inside $\sigma$. Furthermore, in each row the two configurations give rise to the same intersection of $\sigma\cdot_Y C^\pm$, and are thus, from the point of view of $U(1)$ charges, identical. Note that for one of these configurations the entire fiber is contained in the section.
    The right hand side shows the case when the fiber component $F_1$, which intersects the section in codimension one, becomes reducible in codimension two. Again, for each pair $(\sigma\cdot_Y C^+, \sigma \cdot_Y C^-)$ there are two configurations realizing those intersection numbers. \label{fig:WrapEx}}
\end{figure}
   
  \item[(ii)] $C^+ \subset \sigma$, $C^- \not\subset \sigma$:
      \begin{enumerate}
      \item $\sigma \cdot_Y F_p = 0$  and $\sigma \cdot_Y F_m=1$, $p\not=m$:\\
       The configuration from table \ref{sigmaFp0} is
       \be
       \ba
       &C^+ \subset \sigma\,,\ \hbox{deg}(N_{C^+/\sigma})  =-1 \cr
       &C^- \not\subset \sigma \,, \ \sigma\cdot_Y C^- =1 \,.
       \ea
       \ee 
        The positive intersection of $\sigma$ with $C^-$ arises from the
        single point of intersection between the curves $C^+$ and $C^-$. 
        Any fiber components, $F_i$, which are positioned in the ring between
        $C^+$ and $F_m$ must also be contained in $\sigma$, so that $\sigma \cdot_Y F_i = 0$.
        This can be seen by considering first the intersection point of $C^+$ with
        the curve $F_i$, which is adjacent to it in the ring. Clearly this
        would have $\sigma \cdot_Y F_i = 1$, which would be inconsistent with
        codimension one unless $i = m$. Therefore $F_i$ must be contained in
        $\sigma$, with $F_i \cdot_\sigma D_{F_i} = -2$, so that it has zero intersection number in $Y$.
        This is consistent with Theorems \ref{thm:NormalBundleSES} and \ref{thm:abNB}.
         Identically, such wrapping must
        continue until the section meets the fiber component that it
        intersects in codimension one. This configuration is depicted in
        figure \ref{fig:WrapEx} and, more generally, for $I_n$, in figure \ref{fig:InFib1} (ii).
        
      \item $\sigma \cdot_Y F_p = 1$: \\
      There are two solutions in this case from table \ref{sigmaFp1}, however we will see only the following gives rise to a
      consistent fiber:
       \be\label{GoodSolsFp1}
       \ba
       &C^+ \subset \sigma\,,\ \hbox{deg}(N_{C^+/\sigma})  =-1 \cr
       &C^- \not\subset \sigma \,, \ \sigma\cdot_Y C^- =2 \,.
       \ea
       \ee
       The second solution       
        characterized by $C^+ \subset \sigma\,,\ \hbox{deg}(N_{C^+/\sigma})  =-2$ and 
       $C^- \not\subset \sigma \,, \ \sigma\cdot_Y C^- =1$ would imply that the section wraps $C^+$, and thus by the argument
       in the last paragraph, would gain a non-trivial intersection with all $F_i$ between $C^+$ and $C^-$ unless, all of these curves
       are contained in $\sigma$ with normal bundle degree $-2$, so that $\sigma \cdot_Y F_i =0$. However, then $C^-$ would be the only 
       not contained fiber component, and would have intersection 2 with the section, which would be in contradiction. 
       Thus we are left with the only configuration (\ref{GoodSolsFp1}).
        Again, by the same arguments as given in the previous paragraph the section
        must contain all the $F_i$ between $C^+$ and $C^-$. If there were to
        be some $F_i$ which was not contained in $\sigma$ then it would have a
        strictly positive intersection number with $\sigma$ from its neighbour
        in the ring, contradicting codimension one. $C^-$ then has one
        intersection point with $\sigma$ from the intersection with $C^+$ and
        one from the intersection with the $F_i$ on its other side, giving the
        required intersection number of $+2$. The fiber is represented in figure \ref{fig:WrapEx} and for $I_n$ in 
        figure \ref{fig:InFib2} (iv).
    \end{enumerate}
  \item[(iii)] $C^- \subset \sigma$, $C^+ \not\subset \sigma$: \\
    The analysis in the case is essentially identical to the analysis in case (ii), by exchanging the roles of $C^+$ and $C^-$,  
      and we do not repeat it here.
    \begin{enumerate}
      \item $\sigma \cdot_Y F_p = 0$: \\
        See figure \ref{fig:WrapEx}  and figure \ref{fig:InFib1} (iii).
      \item $\sigma \cdot_Y F_p = 1$: \\
        See figure \ref{fig:WrapEx}  and figure \ref{fig:InFib2} (iii).
    \end{enumerate}
  \item[(iv)] $C^+, C^- \subset \sigma$: 
   \begin{enumerate}
      \item  $\sigma \cdot_Y F_p = 0$  and $\sigma \cdot_Y F_m=1$, $p\not=m$: \\
       From table \ref{sigmaFp0} there are three configurations, corresponding to degree of the normal bundle of the curves in $\sigma$ 
       \be
       \left( \hbox{deg}(N_{C^+/\sigma})  ,\ \hbox{deg}(N_{C^-/\sigma})  \right) = (-1, -3)\,,\ (-2,-2)\,,\ (-3, -1) \,.
       \ee
	In all of these cases, all $F_i$ need to be contained in $\sigma$, which again follows by noting that if only $C^\pm$ were contained in $\sigma$, 
	then both $F_{p-1}$ and $F_{p+1}$ gain an intersection from the wrapping of $C^\pm$. Thus in order for all 
	but $F_m$ to have zero intersection with $\sigma$, the entire fiber needs to be contained in $\sigma$ with 
	\be
	\hbox{deg}(N_{F_m /\sigma})  = -3 \,,\qquad \hbox{deg}(N_{F_i /\sigma})  = -2 \,,\quad i\not=m, p  \,.
	\ee 
	The degree of $\hbox{deg}(N_{F_m/\sigma})$ ensures that this component has, consistently with codimension one, intersection $+1$ with $\sigma$. 
        See figure \ref{fig:WrapEx}  and figure \ref{fig:InFib1} parts (iv)-(vi).
              \item $\sigma \cdot_Y F_p = 1$: \\
       Table \ref{sigmaFp1} implies there are four configurations of this type:
       \be
       \left( \hbox{deg}(N_{C^+/\sigma})  ,\ \hbox{deg}(N_{C^-/\sigma})  \right) = (-1, -4)\,,\ (-2,-3)\,,\ (-3, -2)\,,\ (-4, -1) \,.
       \ee
       Again, just as in the last paragraph, the entire fiber needs to be contained in $\sigma$ with 
       \be
       \hbox{deg}(N_{F_i/\sigma})  =-2 \,,\qquad i\not= p \,.
       \ee
        See figure \ref{fig:WrapEx} and figure \ref{fig:InFib2} parts (v)-(viii).
            \end{enumerate}
\end{enumerate}
This completes the analysis of what fiber configurations in codimension two are possible with one rational section. 

\begin{sidewaysfigure}
  \centering
  \includegraphics[scale=0.1]{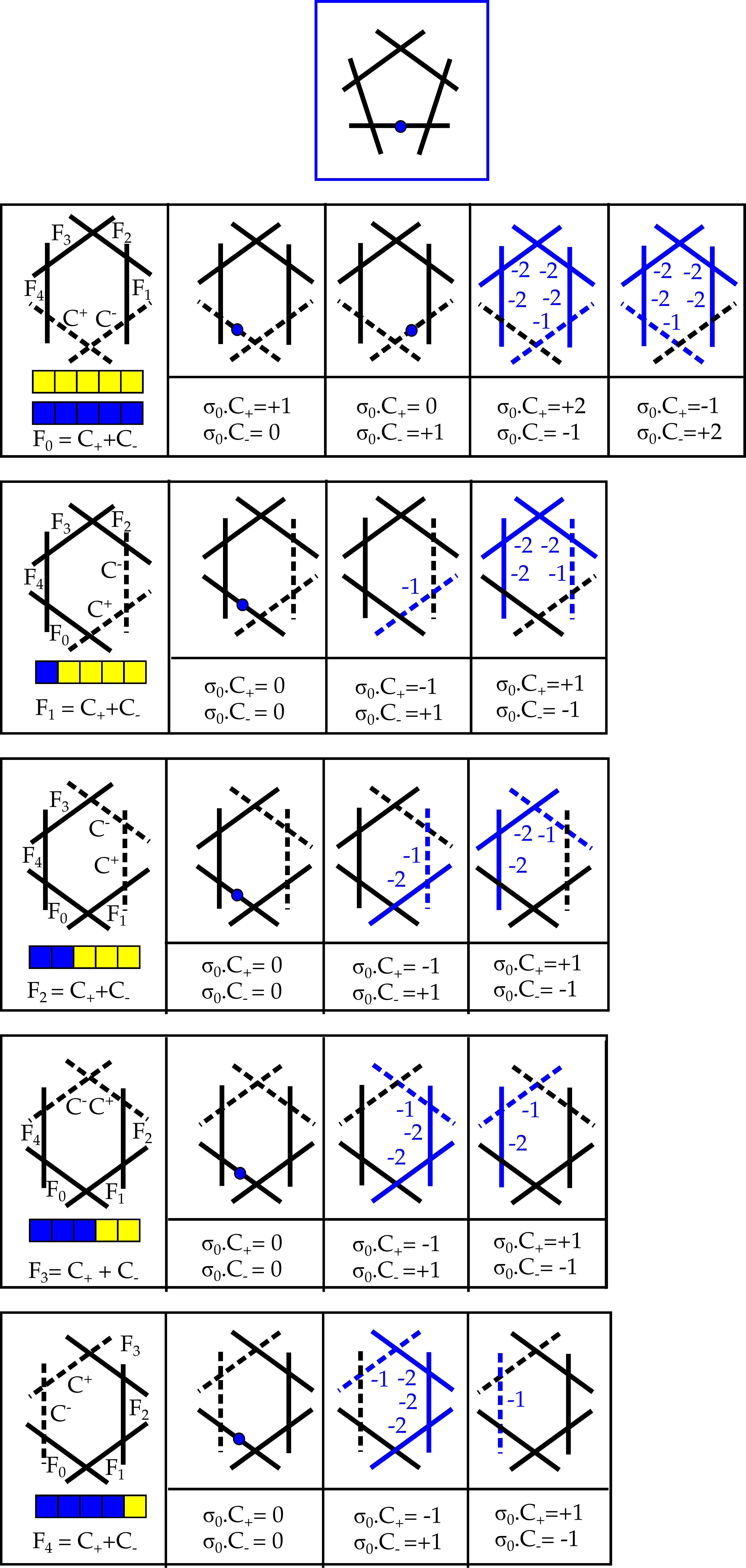} \qquad
  \includegraphics[scale=0.1]{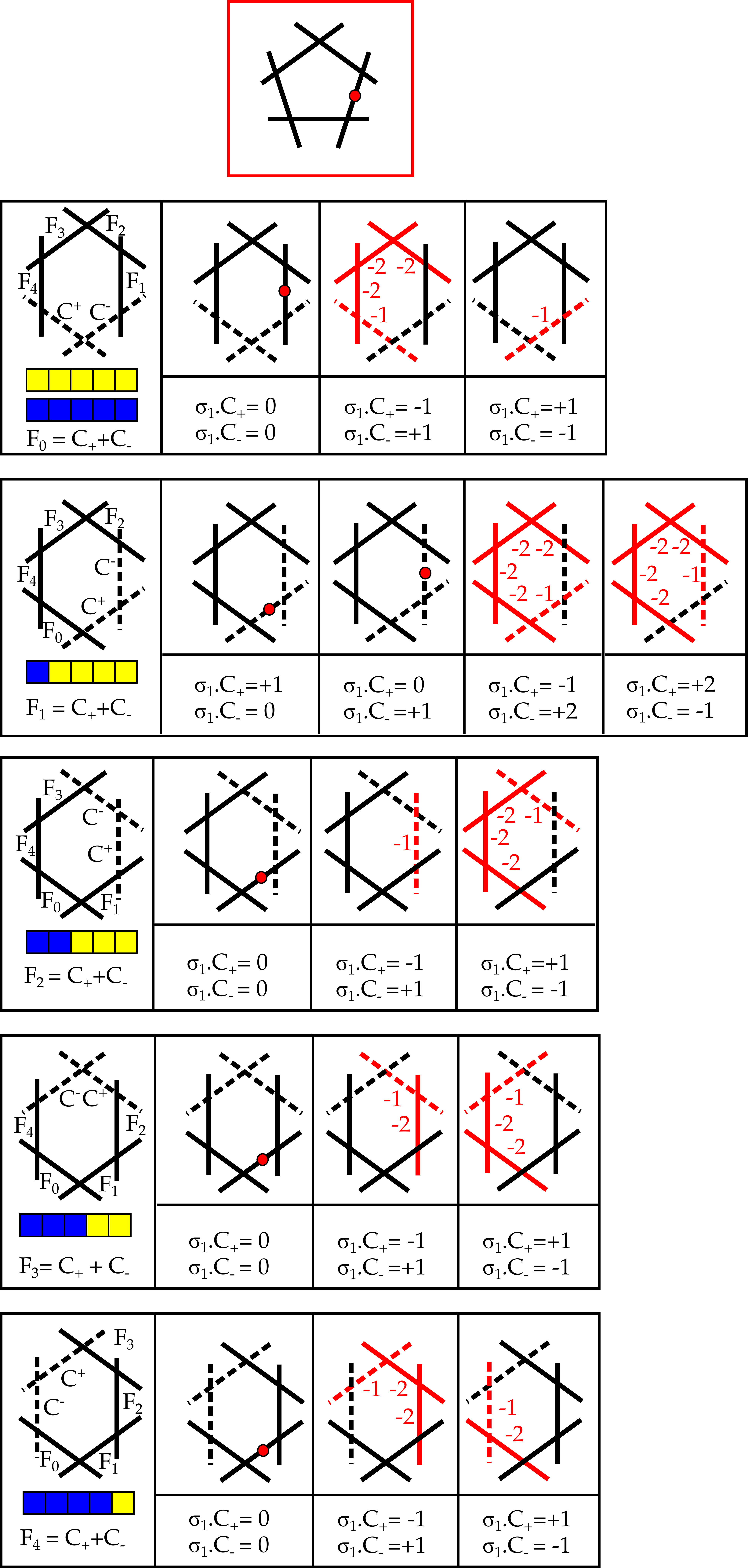} \qquad
  \includegraphics[scale=0.1]{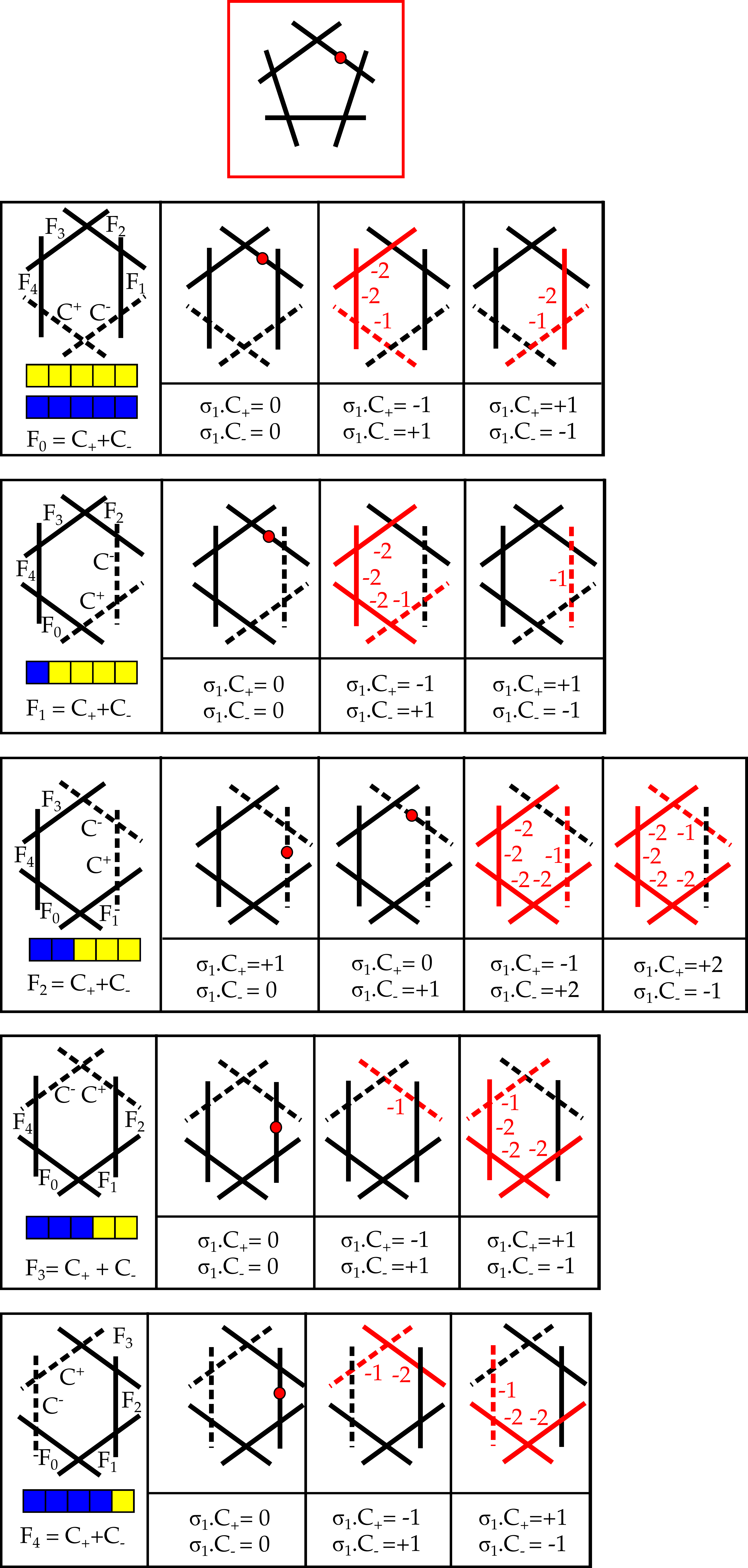} 
  \caption{For each phase/box graph we show the full set of codimension two $I_6$ fibers with rational sections, for the codimension one fiber where the section intersects $F_0$, $F_1$ and $F_2$, respectively, shown at the top. 
  The components that split are shown by dashed lines, and colored (either blue or red) components correspond to rational curves that are contained in the section, with the numbers indicating the degrees of the normal bundle in the section. Dots indicate transverse intersections of the section with the fiber components. 
  We list the intersection numbers $\sigma \cdot_Y C^\pm$. More details can be found in the main text. }
  \label{fig:F0SigmaCodim2Fibs}
\end{sidewaysfigure}


\subsection{Compilation of  Fibers}

The analysis in the last section allows us now to characterize all possible fibers in codimension two for an $SU(5)$ model with one rational section.
There are in total three distinct codimension one configurations for the section, up to inverting the order of the 
curves $F_i$ in codimension one. For each of these, we now determine the fibers with rational section in codimension two. 
As shown in tables \ref{sigmaFp0} and \ref{sigmaFp1}, for each value of $(\sigma\cdot_Y C^+, \sigma\cdot_Y C^-)$ there are two realizations in terms of fibers, see e.g.  figure \ref{fig:WrapEx}. As these are indistinguishable from the point of view of $U(1)$ charges, in the following, we will only consider the fibers with minimal wrapping. 
The different configurations are drawn for each phase of each codimension one
fiber type in figure \ref{fig:F0SigmaCodim2Fibs}. These tables
contain information about
\begin{itemize}
\item Phase: given in terms of the box graph as well as the splitting $F_i  = C^+ + C^-$ for each phase. 
\item Codimension two fiber: in the present case for fundamental matter, the enhancement is to an $I_6$ fiber, i.e. $SU(6)$. The intersection of the exceptional $\mathbb{P}^1$s  is shown, including the curves $C^\pm$ that arise from the splitting are marked by dashed lines. 
\item All possible codimension two fibers with section: a dot on one of the
  $\mathbb{P}^1$s corresponds to a section intersecting the fiber component transversally in  $+1$. If a fiber component is contained in the section $\sigma$, then it is
  colored (blue or red). The ``wrapped'' components carry a numerical label, which indicates the normal bundle degree of the curve inside the section $\sigma$. 
\item Matter intersections: finally, the table contains the information about the intersection of the section $\sigma$ with the curves $C^\pm$, which will then be used to compute the $U(1)$ charges. 
\end{itemize}

Knowing the various configurations one can read off the values of $\sigma
\cdot_Y C^\pm$ in each case. It is these values which determine the $U(1)$
charges, after the application of the Shioda map, as shall be seen in the
subsequent section. In the phase where the codimension one component $F_p$
splits the possible values of $\sigma \cdot_Y C^\pm$ are
\begin{enumerate}
  \item[(a)]  $\sigma \cdot_Y F_p = 0$
    \begin{equation}\label{eqn:wrapinteroptA}
      \sigma \cdot_Y C^\pm \in \{-1,0,1\} \,. 
    \end{equation}
  \item[(b)] $\sigma \cdot_Y F_p = 1$ 
    \begin{equation}\label{eqn:wrapinteroptB}
      \sigma \cdot_Y C^\pm \in \{-1,0,1,2\} \,.
    \end{equation}
\end{enumerate}
These values are the contributions to the $U(1)$ charges from the rational
sections. One sees that there is an additional value for $\sigma \cdot_Y C$
when the codimension one curve that splits, $F_p$, had the rational section
intersecting it in codimension one. We should then anticipate seeing
additional $U(1)$ charges in those phases where such a component of the
$I_5$ fiber splits. Indeed we will see this in the next section.


\subsection{$U(1)$ Charges}

The $U(1)$ charges of the curves $C^\pm$, which are labelled by the weights of the
fundamental representation, are obtained by intersecting them with the Shioda map of
the section $\sigma_1$
\be\label{ShiodaS} 
  S(\sigma_1) = 5(\sigma_1 - \sigma_0) + S_f \,,
\ee 
where $\sigma_0$ is the zero-section. Here, $S_f$ depends on the codimension one
fibers and is
determined by requiring that for all $i$ 
\be 
S(\sigma_1)\cdot_Y  F_i=0 \,.  
\ee 
In
particular, if $F_i \rightarrow C^+ + C^-$ splits then $(C^+ + C^-) \cdot_Y S(\sigma_1)=0$
is required. The $U(1)$ charges of $C^+$ and $C^-$ is given by $S(\sigma_1) \cdot_Y  C^+$ and
$S(\sigma_1) \cdot_Y C^-$ respectively, and  are always conjugate.
 For $I_5^{(01)}$, $S_f$ is trivial, and for the 
 remaining codimension one fiber types they are listed in tables
\ref{table:sfI50s1fives} and  \ref{table:sfI50ss1fives}.

\begin{table}
\centering
\begin{tabular}{|c|c|c|c|}
\hline 
Phase & $S_f$ & $S_f \cdot_Y C^+$ & $S_f \cdot_Y C^-$ \\\hline\hline
\multirow{2}{*}{\includegraphics[width=1.5cm]{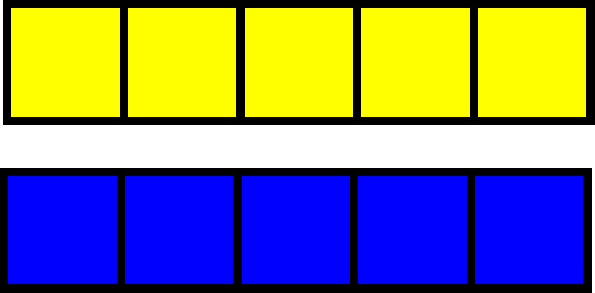}} & \multirow{6}{*}{$4D_{F_1} + 3D_{F_2} + 2D_{F_3} + D_{F_4}$}
& \multirow{2}{*}{$+1$} & \multirow{2}{*}{$+4$} \\
&&& \\
\includegraphics[width=1.5cm]{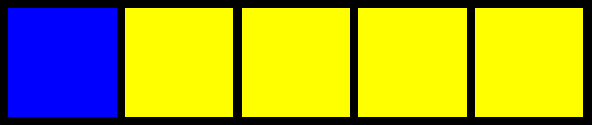} & 
& $-4$ & $-1$ \\[2pt]
\includegraphics[width=1.5cm]{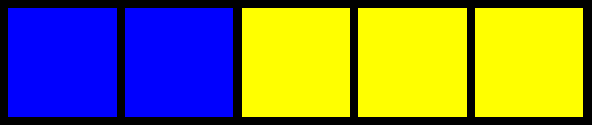} & 
& $+1$ & $-1$ \\[2pt]
\includegraphics[width=1.5cm]{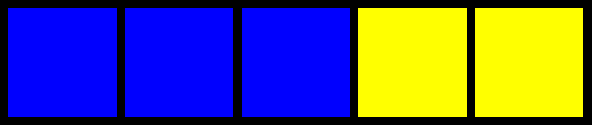} & 
& $+1$ & $-1$ \\[2pt]
\includegraphics[width=1.5cm]{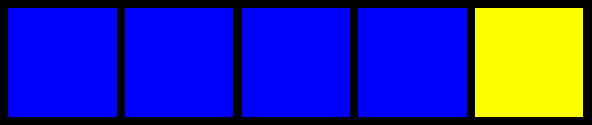} & 
& $+1$ &  $-1$ \cr\hline
\end{tabular}
\caption{Values for $S_f \cdot_Y C^{\pm}$ for $I_5^{(0|1)}$ local enhancement to $I_6$.}
\label{table:sfI50s1fives}
\end{table}

\begin{table}
\centering
\begin{tabular}{|c|c|c|c|}
\hline 
Phase & $S_f$ & $S_f \cdot_Y C^+$ & $S_f \cdot_Y C^-$ \\\hline\hline
\multirow{2}{*}{\includegraphics[width=1.5cm]{I6phase1.pdf}} & \multirow{6}{*}{$3D_{F_1} + 6D_{F_2} + 4D_{F_3} + 2D_{F_4}$}
& \multirow{2}{*}{$+2$} & \multirow{2}{*}{$+3$} \\
&&& \\
\includegraphics[width=1.5cm]{I6phase2.pdf} & 
& $-3$ & $+3$ \\[2pt]
\includegraphics[width=1.5cm]{I6phase3.pdf} & 
& $-3$ & $-2$ \\[2pt]
\includegraphics[width=1.5cm]{I6phase4.pdf} & 
& $+2$ & $-2$ \\[2pt]
\includegraphics[width=1.5cm]{I6phase5.pdf} & 
& $+2$ &  $-2$ \cr\hline
\end{tabular}
\caption{Values for $S_f \cdot_Y C^\pm$ for $I_5^{(0||1)}$ local enhancement to $I_6$.}
\label{table:sfI50ss1fives}
\end{table}

In the section \ref{sec:Codim2Analysis} we determined a comprehensive list of possible fibers in codimension two, given that a rational section $\sigma$ intersects either $F_0$, $F_1$, or $F_2$ in codimension one, respectively. In a model with one $U(1)$, we apply this analysis to the zero-section $\sigma_0$ and additional section $\sigma_1$. Without loss of generality, $\sigma_0\cdot_Y F_0=1$, and thus the possible codimension two fibers are listed in figure \ref{fig:F0SigmaCodim2Fibs}.   
Depending on which codimension one fiber type \eqref{I5fibertypes} we start with, in addition the section $\sigma_1$ can be in one of the configurations in figures \ref{fig:F0SigmaCodim2Fibs}. Obviously, only fiber types in the same phase can be combined. 

\begin{figure}
  \centering
  \includegraphics[width=14cm]{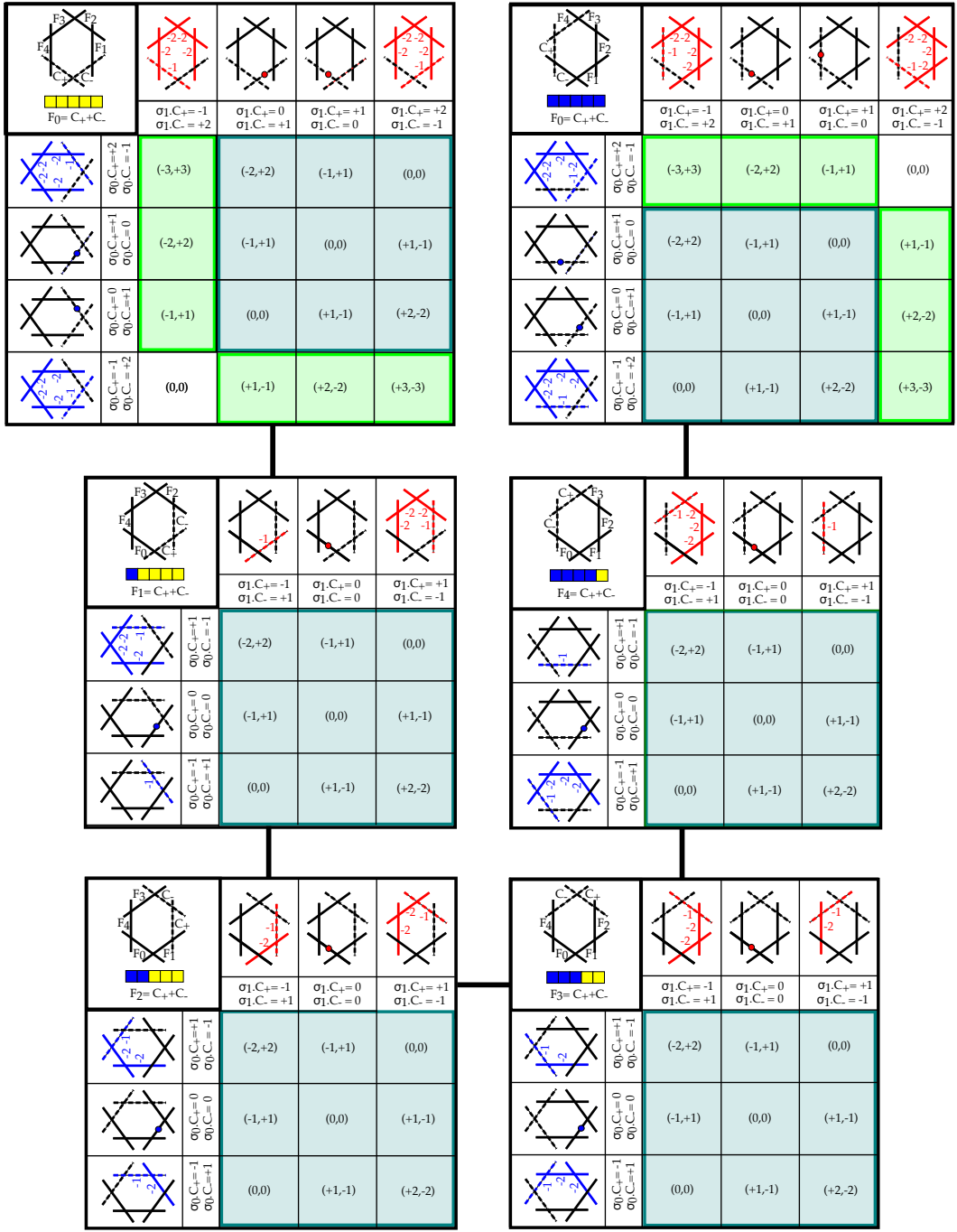} 
 \caption{Codimension two fibers and charges for ${\overline{\bf 5}}$ matter for $I_5^{(01)}$ models. For details see  (\ref{figlab}).}
 \label{fig:01Charges}
\end{figure}

\begin{figure}
  \centering
  \includegraphics[width=14cm]{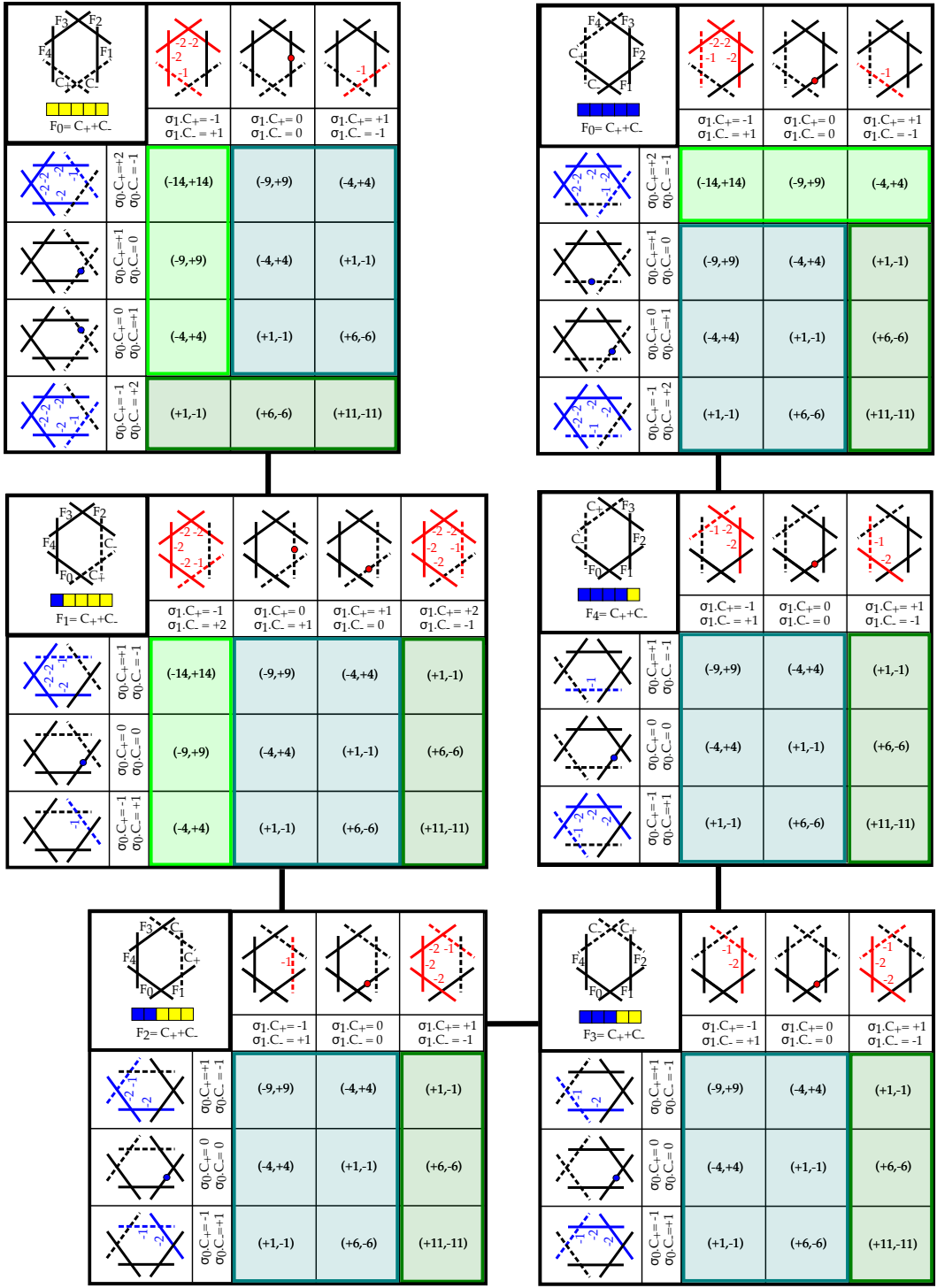} 
 \caption{Codimension two fibers and charges for ${\overline{\bf 5}}$ matter for $I_5^{(0|1)}$ models.  For details see  (\ref{figlab}).
 }
 \label{fig:0s1Charges}
\end{figure}

\begin{figure}
  \centering
  \includegraphics[width=14cm]{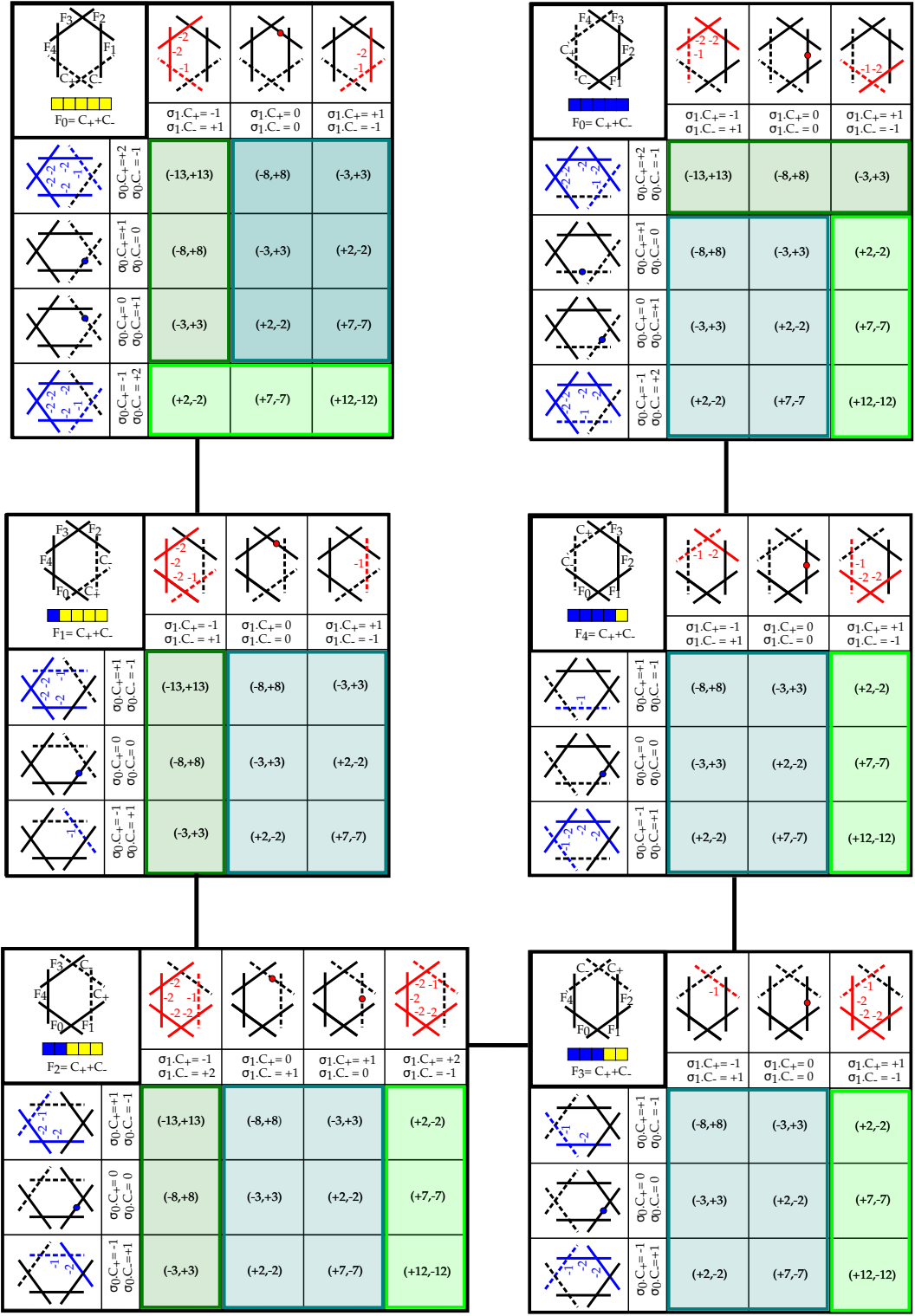} 
 \caption{Codimension two fibers and charges for ${\overline{\bf 5}}$ matter for $I_5^{(0||1)}$ models.  For details see  (\ref{figlab}).}
 \label{fig:0ss1Charges}
\end{figure}

The charge is computed by intersecting the Shioda map $S(\sigma_1)$ (\ref{ShiodaS}) with
the split curves $C^+$ and $C^-$. The result is shown for all codimension one
fiber types in figures \ref{fig:01Charges}, \ref{fig:0s1Charges}, and
\ref{fig:0ss1Charges}. Each of the figures contains the information
\be\label{figlab}
{\hbox{Caption for Figures \ref{fig:01Charges}, \ref{fig:0s1Charges}, and
\ref{fig:0ss1Charges}:}}
\ee
\begin{itemize}
  \item The phase, specified by the box graph, and the fiber in codimension
    two that results, {without the section information}.
  \item The horizontal (vertical) axis shows the different
    configurations for curves of the fiber in the section $\sigma_1$ ($\sigma_0$).
  \item The entries of the tables contain the $U(1)$ charges $(a, -a)$
    determined by $S(\sigma_1) \cdot_Y C^+$ and $S(\sigma_1) \cdot_Y C^-$ respectively. 
  \item The lines between the phases, that is, connecting the six large boxes,
    denote that there exist flop transitions between those linked phases.\footnote{These are the flops that exist generically, as explained in
      \cite{Hayashi:2014kca}.  This will be
    discussed later on.} The coloring of the charges is related these flops and will be discussed later.
\end{itemize}

In summary the charges for $\bar{\bf 5}$ (and negative of these for the conjugate ${\bf 5}$) that we find are:
\be  \label{FundamentalCharges}
\ba
\hbox{$U(1)$ charges of ${\bar{\bf 5}}$ matter for} &\quad 
\left\{ 
\ba
I_5^{(01)} &\in \left\{-3,-2, - 1, 0, +1,+2, +3 \right\} \cr
I_5^{(0|1)} &\in \left\{-14, -9,-4,+1, + 6,+ 11 \right\} \cr
I_5^{(0||1)} &\in \left\{- 13,- 8, -3, +2, + 7, + 12 \right\} \,.
\ea
\right.
\ea
\ee
This concludes the analysis of possible $U(1)$ charges for an $SU(5)$ gauge theory in F-theory with fundamental matter, for one additional abelian gauge factor. Note that all known charges from explicit realizations of the fiber in various toric tops as well as Tate models, including the individual $U(1)$ charges from models with multiple $U(1)$ factors, are a  (strict) subset. We discuss the relation to the embedding into $E_8$, as discussed in \cite{Baume:2015wia}, in appendix \ref{app:E8}.

\subsection{$SU(n)\times U(1)$ with Fundamental Matter}

\begin{sidewaysfigure}
 \centering
 \begin{minipage}[t]{0.45\linewidth}
   \centering
  \begin{subfigure}[b]{0.2\textwidth}
    \includegraphics[width=\textwidth]{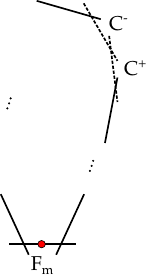}
    \caption{}
  \end{subfigure} \quad\quad 
  \begin{subfigure}[b]{0.2\textwidth}
    \includegraphics[width=\textwidth]{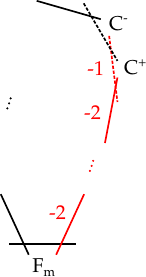}
   \caption{}
  \end{subfigure} \quad 
  \begin{subfigure}[b]{0.2\textwidth}
   \includegraphics[width=\textwidth]{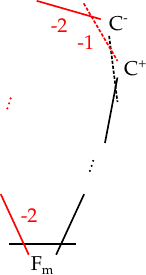}
    \caption{}
      \end{subfigure}\\
  \vspace{0.3cm}
     \begin{subfigure}[b]{0.2\textwidth}
   \includegraphics[width=\textwidth]{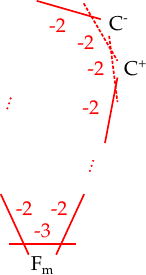}
    \caption{}
  \end{subfigure}\quad \quad
       \begin{subfigure}[b]{0.2\textwidth}
   \includegraphics[width=\textwidth]{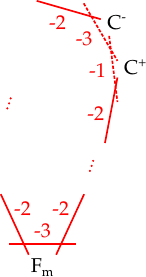}
    \caption{}
  \end{subfigure}\quad \quad
       \begin{subfigure}[b]{0.2\textwidth}
   \includegraphics[width=\textwidth]{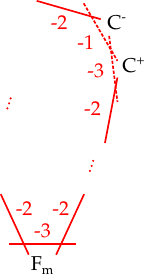}
    \caption{}
  \end{subfigure}
  \caption{The six consistent wrapping configurations for $I_n \rightarrow
    I_{n+1}$ in the phases where the component $F_p$, which the section
    intersects in codimension one, does not split. Components which are
    coloured red are wrapped by the section, and the red numbers indicate the normal bundle degree of that curve inside the divisor $\sigma$. A red node indicates that the section intersects that component transversally.}
    \label{fig:InFib1}
  \end{minipage}
  \qquad
 \begin{minipage}[t]{0.45\linewidth}
 \centering
  \begin{subfigure}[b]{0.2\textwidth}
    \includegraphics[width=\textwidth]{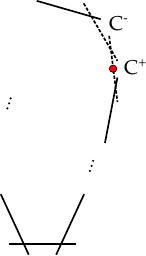}
    \caption{}
    \label{fig:InFib2_1}
  \end{subfigure} \quad
  \begin{subfigure}[b]{0.2\textwidth}
    \includegraphics[width=\textwidth]{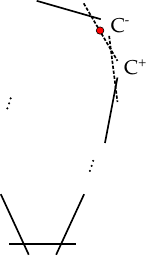}
    \caption{}
    \label{fig:InFib2_2}
  \end{subfigure} \quad 
  \begin{subfigure}[b]{0.2\textwidth}
    \includegraphics[width=\textwidth]{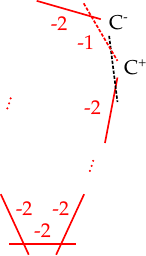}
    \caption{}
  \end{subfigure} \quad 
  \begin{subfigure}[b]{0.2\textwidth}
    \includegraphics[width=\textwidth]{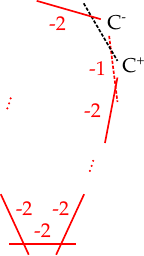}
    \caption{} 
      \end{subfigure}\\
  \vspace{0.3cm}
     \begin{subfigure}[b]{0.2\textwidth}
   \includegraphics[width=\textwidth]{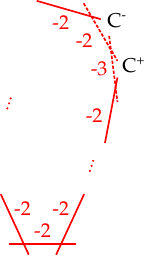}
    \caption{}
  \end{subfigure}\quad
       \begin{subfigure}[b]{0.2\textwidth}
   \includegraphics[width=\textwidth]{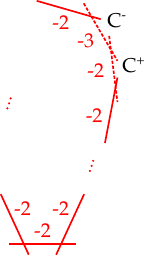}
    \caption{}
  \end{subfigure}\quad
       \begin{subfigure}[b]{0.2\textwidth}
   \includegraphics[width=\textwidth]{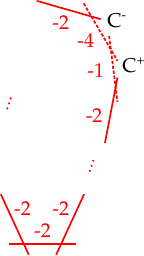}
    \caption{}
  \end{subfigure}\quad
       \begin{subfigure}[b]{0.2\textwidth}
   \includegraphics[width=\textwidth]{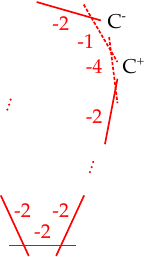}
    \caption{}
   \end{subfigure}
  \caption{The eight consistent wrapping configurations for $I_n \rightarrow
    I_{n+1}$ in the phases where the component $F_p$ which the section
    intersects in codimension one, $\sigma \cdot_Y F_p =1$, splits: $F_p \rightarrow C^+ + C^-$. Components which are
    coloured red are wrapped by the section, and the red numbers
  indicate the normal bundle degree of that curves inside the divisor $\sigma$. A red node indicates that the
    section intersects that component transversally.}
    \label{fig:InFib2}
  \end{minipage}
\end{sidewaysfigure}

In our discussion of fiber configurations in section  \ref{sec:Codim2Analysis} it was in fact of no particular importance that we started with 
an $I_n$ fiber with $n=5$.  Indeed the situation is very similar and easily generalizes, 
to $SU(n)$ with fundamental (i.e. the  $\overline{\bf n}$ representation) matter, where the fiber enhances from an
$I_n$ to an $I_{n+1}$. Each section in codimension one intersects one of the rational curves $F_i$, $i=0,1, \cdots, n-1$, which intersect in an affine $SU(n)$ Dynkin diagram. 
In codimension two, one of the $F_i$ splits, as shown in  \cite{Hayashi:2014kca}.  For an elliptic fibration with sections $\sigma_0$ and $\sigma_1$, we again use the notation
\be
I_n^{(0|^m 1)} :\qquad \sigma_0 \cdot_Y F_0 =1 \,,\qquad \sigma_1 \cdot_Y F_m =1 \,.
\ee
Let $F_p$ be the component that splits in codimension two. Then there are two cases to consider: either $\sigma \cdot_Y F_p=0$ or $1$, which are
shown in figures \ref{fig:InFib1} and \ref{fig:InFib2}, respectively. 
The reasoning is entirely as in section \ref{sec:Codim2Analysis}, with the only difference being the length of the chain of rational curves $F_i$ that are located between $C^+$ and $C^-$.
The distinct cases of intersections $(\sigma \cdot_Y C^+, \sigma \cdot_Y C^-)$ are also analogous to the $SU(5)$ case.

The Shioda
map can be constructed for an $I_n^{(0|^m1)}$ fiber and the $U(1)$
charges of a fibration with a specified wrapping configuration can be
written in terms of $m$ and $n$. The Shioda map for an $I_n$ fiber with
separation $m$ between the sections is
determined by the $m$th row of the inverse Cartan matrix associated to the
codimension one singularity type \cite{Morrison:2012ei}. The inverse Cartan matrix of $SU(n)$ is an
$(n - 1) \times (n-1)$ matrix with elements
\begin{equation}\label{eqn:inverscartan}
  C_{mc} = \frac{1}{n}
  \begin{cases}
    c(n-m) \quad c \leq m \cr
    m(n-c) \quad m < c \,. 
  \end{cases}
\end{equation}
The Shioda map for an $I_n^{(0|^m1)}$ fiber is then of the form
\begin{equation}
  S(\sigma_1) = n(\sigma_1 - \sigma_0) + \sum_{i=1}^{n-1}C_{mi}D_{F_i} \,,
\end{equation}
ignoring contributions from the base. For ease
of notation we will allow $c_p$ to denote the coefficient of the term $D_{F_p}$ in
the Shioda map, that is $C_{mp}$. The Shioda map excepting the term
$n(\sigma_1 - \sigma_0)$ will be denoted by $S_f$ as before. The conjugate $U(1)$ charges
are obtained from the intersection numbers
\begin{equation}
  S(\sigma_1) \cdot_{Y} C^\pm \,.
\end{equation}
Such an intersection can be broken into two parts, contributions from
$(\sigma_1 - \sigma_0) \cdot_Y C^\pm$, which were enumerated for each section in
(\ref{eqn:wrapinteroptA}, \ref{eqn:wrapinteroptB}), and
contributions from $S_f \cdot_Y C^\pm$, which are determined here. Let us consider
the phase where $F_p \rightarrow C^+ + C^-$, and we shall content
ourselves with only obtaining the $U(1)$ charge of $C^+$, as the
charge for  $C^-$ is simply its negative. From the resulting fiber it
is observed that the only contributions from $S_f \cdot_Y C^+$ come from $c_p$
and $c_{p - 1}$, as these are the coefficients in the Shioda map of the divisors  $D_{F_i}$, 
which $C^+$ intersects, i.e. 
\begin{equation}
  S_f \cdot_{Y} C^+ = c_{p - 1} - c_p \,.
\end{equation}
Given (\ref{eqn:inverscartan}) this can be expanded explicitly in terms of $m$
and $n$ (importantly the dependence on the phase is minimal)
\begin{equation}\label{eqn:sfplus}
  S_f \cdot_{Y} C^+ = 
  \begin{cases}
    (m - n) \quad p \leq m \cr
    m \quad\quad\quad\,\, m < p \,.
  \end{cases}
\end{equation}

In the above we considered only the so-called $SU(n)$-phases, where $p =
1,\cdots, n-1$. What remains is to consider the phases with an additional $U(1)$, where $F_0
\rightarrow C^+ + C^-$. In this case the only contribution to $S_f \cdot_Y
C^+$ comes from $c_{n-1}$, which is $m$. In the previous section
the possible values of $\sigma_i \cdot_Y C^+$ were determined
from the possible consistent wrapping scenarios to be such that 
\begin{equation}
  \sigma_i \cdot_Y C^+ \in \{-1, 0, 1, 2\} \,.
\end{equation}
Combining this information with (\ref{eqn:sfplus}) tables can be
constructed for all possible charges in each phase. The two tables which cover
all the phases for $I_n^{(0|^m1)}$ are given in table \ref{tbl:chargesNoZero}. It can be seen
that the possible charges are
\begin{equation}
  S(\sigma_1) \cdot_Y C^+ = m - 3n \,,\, m - 2n \,,\, \cdots \,,\, m + 2n \,.
\end{equation}
The subset of charges that exist in {\it every} phase is
\begin{equation}
  S(\sigma_1) \cdot_Y C^+ = m - 2n \,,\, m - n \,,\, \cdots \,,\, m + n \,.
\end{equation}

\begin{table}
  \centering
\begin{tabular}{|cc||c|c|c|c|}
  \hline
  &  & \multicolumn{4}{|c|}{$\sigma_1 \cdot_Y C^+$} \cr\cline{3-6}
   &  & $-1$ & $0$ & $1$ & $2$ \cr\hline\hline
   \multirow{3}{*}{\rotatebox{90}{$\sigma_0 \cdot_Y C^+$}}
   & \multicolumn{1}{|c||}{$-1$} & $m-n$ & $m$ & $m+n$ & $m+2n$ \cr\cline{2-6}
   & \multicolumn{1}{|c||}{$0$} & $m-2n$ & $m-n$ & $m$ & $m+n$ \cr\cline{2-6}
   & \multicolumn{1}{|c||}{$1$} & $m-3n$ & $m-2n$ & $m-n$ & $m$ \cr\hline
\end{tabular}
\quad
\begin{tabular}{|cc||c|c|c|}
  \hline
  &  & \multicolumn{3}{|c|}{$\sigma_1 \cdot_Y C^+$} \cr\cline{3-5}
   &  & $-1$ & $0$ & $1$ \cr\hline\hline
   \multirow{4}{*}{\rotatebox{90}{$\ \sigma_0 \cdot_Y C^+$}} 
   & \multicolumn{1}{|c||}{$-1$} & $m$ & $m+n$ & $m+2n$ \cr\cline{2-5}
   & \multicolumn{1}{|c||}{$0$} & $m-n$ & $m$ & $m+n$ \cr\cline{2-5}
   & \multicolumn{1}{|c||}{$1$} & $m-2n$ & $m-n$ & $m$\cr\cline{2-5}
   & \multicolumn{1}{|c||}{$2$} & $m-3n$ & $m-2n$ & $m-n$ \cr\hline
\end{tabular}
  \caption{The $U(1)$ charges of all the possible wrapping combinations of the
    codimension one $I_n^{(0|^m1)}$ fiber enhancing to an $I_{n+1}$ fiber.  On
    the left are the charges in phase where $F_p$ splits for $p = 1,\cdots,m$,
    and on the right are the charges for the phases where $p = m+1,\cdots,n-1$
    or $p=0$. In each configuration, the cases $\sigma \cdot_Y C^+ =2$ only appear in the $p=m$ or $p=0$ phases.}
  \label{tbl:chargesNoZero}
\end{table}

While these are the charges that appear in every phase for every $m$, there
are some special end-point values of $m$ for which extra charges appear in all
phases. When $m = 1$ or $m = n-1$ then charges $m+2n$ and $m-3n$ respectively
appear in all phases. In addition, when $m = 0$ the tables degenerate on top
of each other and the charge $m + 2n$ appears in all phases. In the phase
where $F_0$ splits there is a new charge $m + 3n$ from $\sigma_1 \cdot_Y C^+ = 2$ and
$\sigma_0 \cdot_Y C^+ = -1$.

There are charges, which do not appear in every phase within the framework of fibers satisfying the setup outlined in 
section \ref{sec:Assumptions}. This has in particular to do with the flops of configurations of the type shown 
in (iii) and (iv) of figure \ref{fig:InFib2}, which we will elaborate on in 
section \ref{sec:flops}. 

%



\section{$SU(5)\times U(1)$ with {\bf 10} Matter}
\label{sec:SU5A}

In this section we find the possible charges for $\bf 10$ matter by analysing how the sections can behave under an $I_5$ to $I_1^*$ enhancement. The codimension one $I_5$ fibers and Shioda maps are the same as those given in section \ref{sec:SU5codim1}.

\subsection{Codimension two Fibers with Rational Sections}\label{sec:I1sconfigs}

The fibers of the  ${\bf 10}$ representation are obtained from the box graphs
in tables \ref{tab:10SplitPart1} and \ref{tab:10SplitPart2} in appendix
\ref{app:I1sSplittings}. The resulting fibers are all $I_1^*$, consistent with
the local enhancement to $\mathfrak{so}(10)$, with the correct multiplicities.
To find the charges of the ${\bf 10}$ representation we employ the same method
as before, solving for the possible configurations under the constraints of
consistency with codimension one, $\sigma \cdot_Y \text{Fiber} = 1$. The
multiplicity of each component in the $I_1^*$ fiber must be taken into account
when imposing the latter condition.

\begin{figure}
  \centering
  \includegraphics[width=15cm]{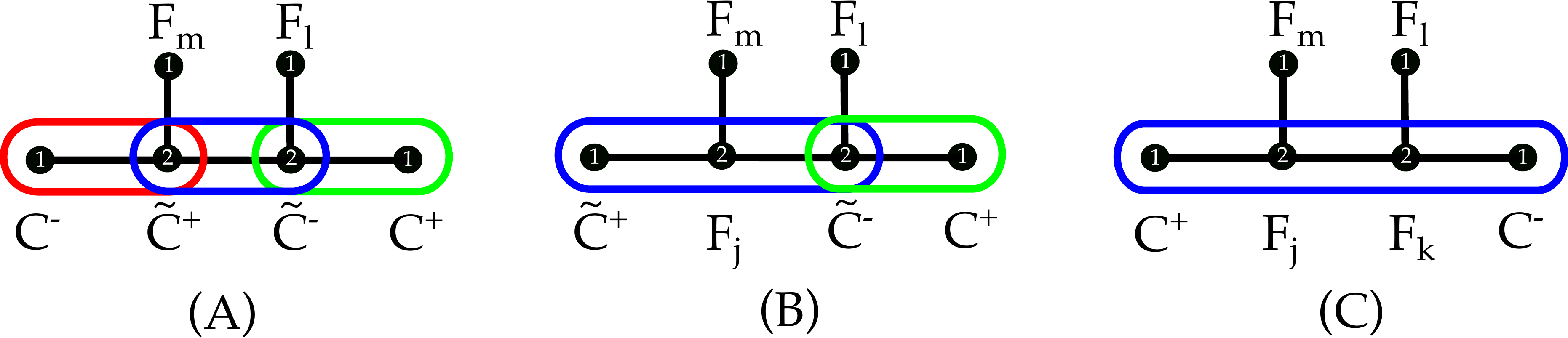} 
  \caption{The three abstract splittings for $I_5$ to $I_1^*$ enhancements. The colored loops indicate that there exists a root that splits into the encircled curves in codimension two.}
\label{fig:I1SplitTypes}
\end{figure}

There are three classes of splitting types that can occur in the enhancement to
$I_1^*$, shown in figure \ref{fig:I1SplitTypes}. They are one of the following,
\begin{enumerate}
\item[(A)] $F_i  \rightarrow C^+ + \tilde{C}^- $, $F_j \rightarrow \tilde{C}^+ + \tilde{C}^- $, $F_k \rightarrow \tilde{C}^+ + C^- $
\item[(B)] $F_i  \rightarrow \tilde{C}^\pm + F_j + \tilde{C}^\mp$, $F_k
  \rightarrow C^\pm + \tilde{C}^\mp$
\item[(C)] $F_i  \rightarrow C^+ + F_j + F_k + C^- , \, j \neq k \hbox{ and }
  j,k \neq i $ .
\end{enumerate}

In each of the three cases there are different subcases to consider depending
on which of the components of the fiber the section intersects in codimension
one. There are five different options corresponding to the number of
components in codimension one, however the reflection symmetry of the
intersection graphs allows one to consider only eleven different
configurations, instead of fifteen. The configurations will be termed the
``splitting types'' and will be denoted as 

\vspace{0.1cm}
\begin{minipage}[t]{.33\textwidth}
\begin{enumerate}
  \item[A.1:] $\sigma \cdot_Y F_l = 1$
  \item[A.2:] $\sigma \cdot_Y F_i = 1$
  \item[A.3:] $\sigma \cdot_Y F_j = 1$
\end{enumerate}
\end{minipage}
\begin{minipage}[t]{.33\textwidth}
\begin{enumerate}
  \item[B.1:] $\sigma \cdot_Y F_l = 1$
  \item[B.2:] $\sigma \cdot_Y F_k = 1$
  \item[B.3:] $\sigma \cdot_Y F_j = 1$
  \item[B.4:] $\sigma \cdot_Y F_m = 1$
  \item[B.5:] $\sigma \cdot_Y F_i = 1$
\end{enumerate}
\end{minipage}
\begin{minipage}[t]{.33\textwidth}
\begin{enumerate}
  \item[C.1:] $\sigma \cdot_Y F_l = 1$
  \item[C.2:] $\sigma \cdot_Y F_k = 1$
  \item[C.3:] $\sigma \cdot_Y F_i = 1$ .
\end{enumerate}
\end{minipage}\vspace{0.3cm}

For each splitting type one can determine the values of the intersection
numbers, from the intersection of the section with the split curves, that are
consistent with the constraints from codimension one and the requirement
that the normal bundles of subspaces embed as subbundles
of the total normal bundle. Each possible set of intersection numbers may have
multiple realizations in terms of configurations of the curves inside the
section. The intersection numbers with $\sigma$ are all that is  necessary to
determine $U(1)$ charges via the Shioda map. In this section splitting type
A.2 will be detailed explicitly and the tables of results for all the other
ten splitting types will be relegated to appendix \ref{app:I1sSplittings}.

\begin{figure}
  \centering
    \includegraphics[scale=2]{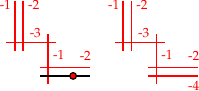}\vspace{0.2cm}
    \includegraphics[scale=2]{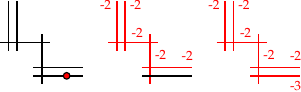}\vspace{0.2cm}
    \includegraphics[scale=2]{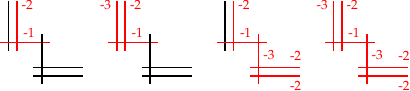}
    \caption{The different realizations of the intersection number solutions $(i)$ (top row),
    $(ii)$ (middle row), and $(iii)$ (bottom row) for splitting type A.2. The
  red integers are the degree of the normal bundles of each curve inside the
section.}
    \label{fig:A2configs}
\end{figure}

Consider then splitting type A.2, defined as the splitting
\begin{equation}
  \begin{aligned}
    F_i &\rightarrow \tilde{C}^+ + C^- \cr
    F_j &\rightarrow \tilde{C}^+ + \tilde{C}^- \cr
    F_k &\rightarrow C^+ + \tilde{C}^- \,,
  \end{aligned}
\end{equation}
with $\sigma \cdot_Y F_i = 1$, and the intersection of the section with all
other codimension one fiber components being zero. As such the constraints
from the split curves become
\begin{equation}\label{eqn:a2const}
  \begin{aligned}
    \sigma \cdot_Y (\tilde{C}^+ + C^-) &= 1 \cr
    \sigma \cdot_Y (\tilde{C}^+ + \tilde{C}^-) &= 0 \cr
    \sigma \cdot_Y (C^+ + \tilde{C}^-) &= 0 \,.
  \end{aligned}
\end{equation}
Any one of the intersection numbers $\sigma \cdot_Y C$ for any curve $C$
determines all the other intersection numbers with the $C$s. As the normal
bundle to the curves $C$ that come from the splitting of the curves $F_i$ in
codimension two is $\mathcal{O}(-1) \oplus \mathcal{O}(-1)$ for three-folds and
$\mathcal{O} \oplus \mathcal{O}(-1) \oplus \mathcal{O}(-1)$ for four-folds it
is known by Theorems \ref{thm:NormalBundleSES} and \ref{thm:abNB} that $\sigma \cdot_Y C \geq -1$ for all
such $C$. Solving the constraints (\ref{eqn:a2const}) subject to these inequalities
leads to the three solutions
\begin{align}
  (i) \quad &\sigma \cdot_Y C^- = 2 \,,\quad \sigma \cdot_Y \tilde{C}^+ = \sigma \cdot_Y
  C^+ = -1 \,,\quad \sigma \cdot_Y \tilde{C}^- = 1 \cr
  (ii) \quad &\sigma \cdot_Y C^- = 1 \,,\quad \sigma \cdot_Y \tilde{C}^+ = \sigma \cdot_Y
  \tilde{C}^- = \sigma \cdot_Y C^+ = 0 \cr
  (iii) \quad &\sigma \cdot_Y C^- = 0 \,,\quad \sigma \cdot_Y \tilde{C}^+ = \sigma \cdot_Y
  C^+ = 1 \,,\quad \sigma \cdot_Y \tilde{C}^- = -1 \,.
\end{align}
Each of these solutions has in addition that $\sigma \cdot_Y F_l = \sigma
\cdot_Y F_m = 0$ from consistency of the curves which do not split with codimension
one. It remains to ask whether there are any possible realizations of these
intersection numbers. All the configurations realizing each of these three
solutions are shown in figure \ref{fig:A2configs}. If a curve is such that $\sigma
\cdot_Y C = -1$ then it must be contained in $\sigma$ with
$\text{deg}(N_{C/\sigma}) = -1$, else if a curve is such that $\sigma \cdot_Y C = k \geq 0$
then the curve is either not contained in $\sigma$ and has $k$ transverse
intersections with $\sigma$, or it is contained in $\sigma$ with
$\text{deg}(N_{C/\sigma})
= -k-2$. In this way configurations of curves inside the section with
particular intersection numbers can be constructed.

\subsection{$U(1)$ charges}

\begin{table}
  \centering
  \begin{tabular}{|c|c|c|c|}
    \hline
    Phase & $I_5^{(01)}$ charges & $I_5^{(0|1)}$ charges & $I_5^{(0||1)}$
    charges \cr\hline
    1 & $-3,-2,-1,0,+1,+2,+3$ & $-12,-7,-2,+3,+8,+13$ & $-9,-4,+1,+6,+11$ \cr
    2 & $-2,-1,0,+1,+2$ & $-12,-7,-2,+3,+8\phantom{,+13}$ & $-9,-4,+1,+6\phantom{,+11}$ \cr
    3 & $-2,-1,0,+1,+2$ & $-12,-7,-2,+3,+8\phantom{,+13}$ & $-9,-4,+1,+6\phantom{,+11}$ \cr
    4 & $-2,-1,0,+1,+2$ & $-12,-7,-2,+3,+8\phantom{,+13}$ & $-9,-4,+1,+6,+11$ \cr
    5 & $-2,-1,0,+1,+2$ & $-12,-7,-2,+3,+8\phantom{,+13}$ & $-9,-4,+1,+6\phantom{,+11}$ \cr
    6 & $-2,-1,0,+1,+2$ & $\phantom{-12}-7,-2,+3,+8\phantom{,+13}$ & $-9,-4,+1,+6\phantom{,+11}$ \cr
    7 & $-2,-1,0,+1,+2$ & $-12,-7,-2,+3,+8\phantom{,+13}$ & $-9,-4,+1,+6,+11$ \cr
    8 & $-2,-1,0,+1,+2$ & $\phantom{-12}-7,-2,+3,+8\phantom{,+13}$ & $-9,-4,+1,+6,+11$ \cr
    9 & $-2,-1,0,+1,+2$ & $-12,-7,-2,+3,+8\phantom{,+13}$ & $-9,-4,+1,+6,+11$ \cr
    10 & $-2,-1,0,+1,+2$ & $\phantom{-12}-7,-2,+3,+8\phantom{,+13}$ & $-9,-4,+1,+6,+11$ \cr
    11 & $-2,-1,0,+1,+2$ & $-12,-7,-2,+3,+8,+13$ & $-9,-4,+1,+6,+11$ \cr
    12 & $-2,-1,0,+1,+2$ & $\phantom{-12}-7,-2,+3,+8,+13$ & $-9,-4,+1,+6,+11$ \cr
    13 & $-2,-1,0,+1,+2$ & $\phantom{-12}-7,-2,+3,+8\phantom{,+13}$ & $-9,-4,+1,+6,+11$ \cr
    14 & $-2,-1,0,+1,+2$ & $\phantom{-12}-7,-2,+3,+8,+13$ & $-9,-4,+1,+6,+11$ \cr
    15 & $-2,-1,0,+1,+2$ & $\phantom{-12}-7,-2,+3,+8,+13$ & $-9,-4,+1,+6,+11$ \cr
    16 & $-3,-2,-1,0,+1,+2,+3$ & $-12,-7,-2,+3,+8,+13$ & $-9,-4,+1,+6,+11$ \cr\hline
  \end{tabular}
  \caption{The range of possible $U(1)$ charges for each codimension one
    fiber type. The phases are those listed in tables \ref{tab:10SplitPart1}
    and \ref{tab:10SplitPart2} in appendix \ref{app:I1sSplittings}.}
  \label{tbl:I1sCharges}
\end{table}

The possible codimension two fibers are obtained by combining the $\sigma_0$ and $\sigma_1$
configurations appearing in the same phase. The $U(1)$ charges of the $\bten$ representation for each such combined configuration 
are determined from the  $C^+/ C^-$ intersections with the sections
listed in the figures and the appropriate Shioda map \eqref{ShiodaS}. The
results are 
shown in table \ref{tbl:I1sCharges}. Each entry in the table lists
the possible  charges in each phase for a particular codimension one
fiber type, and is summarized in terms of the following set of possible charges:
\be\label{10MatterCharges}
\ba
\hbox{$U(1)$ charges of ${{\bf 10}}$ matter for}&\quad 
\left\{ 
\ba
I_5^{(01)} &\in \left\{- 15, - 10,- 5, 0,  + 5, + 10, + 15 \right\} \cr
I_5^{(0|1)} &\in \left\{- 12, - 7, - 2, + 3, + 8,+ 13 \right\} \cr
I_5^{(0||1)} &\in \left\{-9, - 4, + 1,+6,+ 11\right\}  \,.
\ea\right.
\ea
\ee
Again, like for the case of fundamental matter, the known charges that occur in concrete realizations of elliptic fibrations of $SU(5)$ GUTs are a strict
subset of these. The comparison to the embedding into $E_8$ can be found in appendix \ref{app:E8}.


\section{Flops and Rational Sections}
\label{sec:flops}

Flops between distinct resolutions of singular elliptic Calabi--Yau fibrations have been discussed in terms of the Coulomb phases, or box graphs, in \cite{Hayashi:2014kca}, and 
realized in terms of explicit elliptic fibrations (based on Tate models) in \cite{Hayashi:2013lra, Braun:2014kla, Esole:2014hya, ABSSN}. In this section, we will 
study the flops for codimension two fibers with sections wrapping fiber components. For simplicity we consider here three-folds, however we expect all of the flops to generalize quite straightforwardly 
to four-fold flops, e.g. as discussed in \cite{Matsuki, MatsukiWeyl}.

\subsection{Flops and Intersections}
\label{sec:FlopsIntro}

The small resolutions of the singular fibers are related by flops along curves in the fiber in codimension two. 
To determine how the flops change the normal bundle degrees of $C \subset D$, which in the three-fold case is given by the self-intersections of the curves in $D$, it is useful to recapitulate some of the mathematical results on this for three-folds. 
The first important notion is that of a $(-2)$-curve as introduced in Theorem \ref{thm:minus2} (see \cite{Reid} for more details).
Recall that the normal bundle of the curves $F_i$, which remain irreducible in codimension two, are
\be
N_{F_i/Y} = \mathcal{O} \oplus \mathcal{O}(-2) \,,
\ee
whereas if $F_p \rightarrow C^+ + C^-$ becomes reducible in codimension two, then each of the irreducible components $C^\pm$ have normal bundle in $Y$
\be
N_{C^\pm/Y} = \mathcal{O}(-1) \oplus \mathcal{O}(-1) \,.
\ee

\begin{figure}
\centering
\includegraphics[width=9cm]{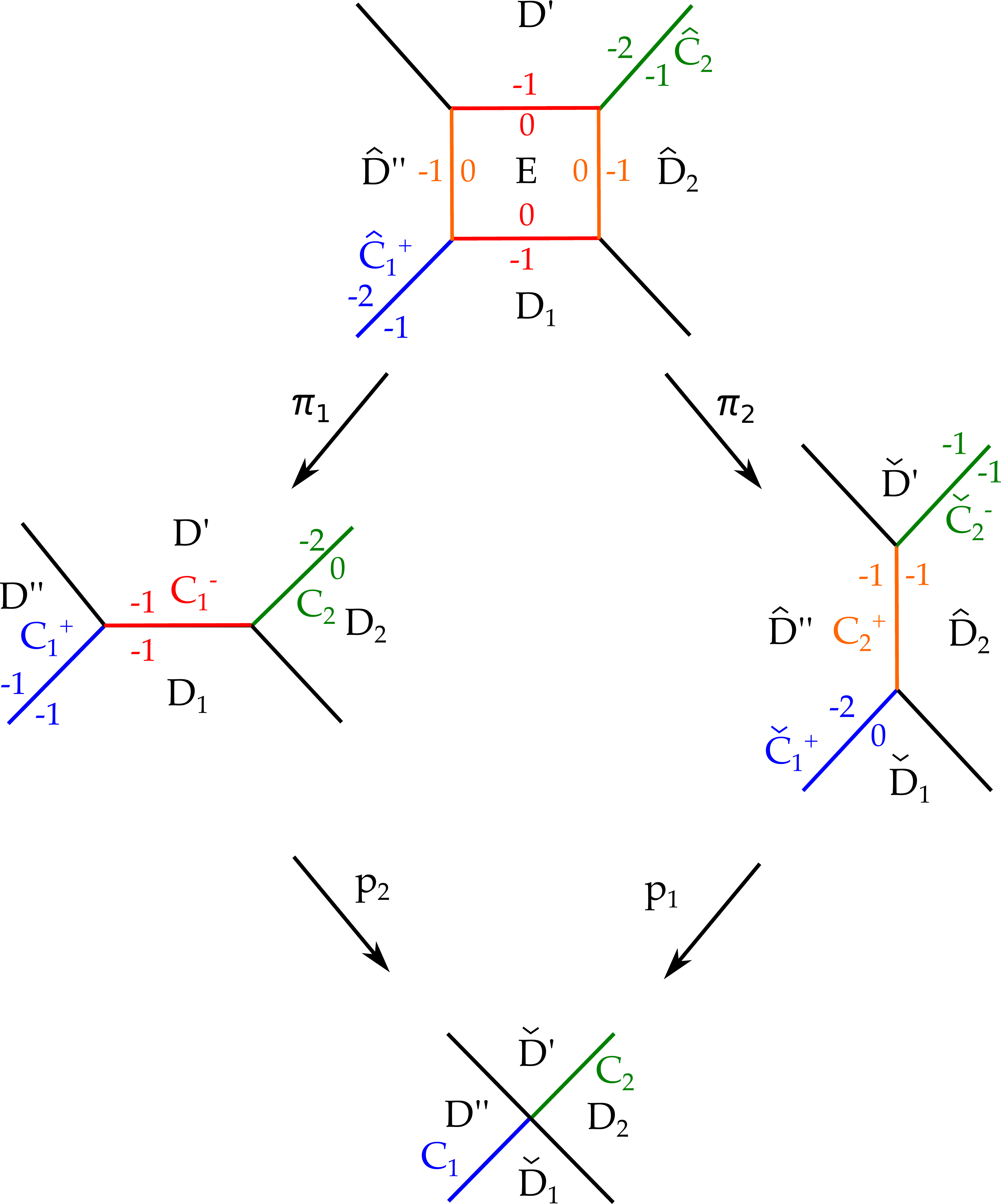}
\caption{Flop of the curve $C_1^-$ into $C_2^+$. $D$'s are divisors, $C$ the curves at their intersections, and the small numbers indicate the degree of the normal bundles of the curves inside the divisors. The exceptional divisor, $E = \mathbb{P}^1 \times \mathbb{P}^1$, is introduced in the blow up as an intermediate stage. Alternatively one can blow down to the singular configuration at the bottom of the picture. }
\label{fig:FlopDiags}
\end{figure}

Consider the situation shown in figure \ref{fig:FlopDiags}, starting with the configuration in the lower left hand side. 
The curves $C_1^{\pm}$ both have normal bundles of degree $(-1, -1)$, the curve $C_2$ has normal bundle $(-2,0)$ (i.e. it is, in our standard notation, one of the $F_i$). 
Consider blowing up along the curve $C_1^-$. 

Let $D$ and $\hat{D}$ be divisors and $\pi_1: \hat{D} \rightarrow {D}$ the blow up of a curve $C$. The canonical class changes as
\be\label{Kchangebu}
K_{\hat{D}} = \pi_1^* K_{{D}} + C \,.
\ee
Here the blow up affects the two divisors $D_2$ and $D''$, in particular under $\pi_1:  \hat{D}_2 \rightarrow D_2$ the canonical class changes by the new curve, ${C}_2^+$,
\be
K_{\hat{D}_2} = \pi_1^* K_{D_2} + {C}_2^+ \,,\qquad K_{\hat{D}''} = \pi_1^* K_{D''} + {C}_2^+  \,.
\ee  
The curves ${C}_2$ and ${C}_1^+$, are contained within these two divisors, and their normal bundles change in the blow up. Denoting
their images under the blow up by $\hat{C}$, the normal bundle degrees are (using adjunction that $K_D \cdot_D C  = - (C)^2_D -2$)
\be\ba
\hbox{deg} (N_{\hat{C}_2/\hat{D}_2})
 &= (\hat{C}_2)^2_{\hat{D}_2}  = - K_{\hat{D}_2 } \cdot_{\hat{D}_2} \hat{C}_2  -2 \cr
& = -( \pi_1^* K_{D_2 } + \hat{C}_2^+) \cdot_{\hat{D}_2} \hat{C}_2-2  = -(-2+ 1) -2 = -1  \cr
\hbox{deg} (N_{\hat{C}_1^+/\hat{D}''})
 &= (\hat{C}_1^+)^2_{\hat{D}''}  = - K_{\hat{D}'' } \cdot_{\hat{D}''} \hat{C}_1^+  -2 \cr
& = -( \pi_1^* K_{D''} + \hat{C}_2^+) \cdot_{\hat{D}''} \hat{C}_1^+ -2 = -(-1+ 1) -2 = -2 \,. 
\ea\ee
The normal bundles of $\hat{C}_2^-$, $\hat{C}_1^+$ in the divisors ${D}'$, ${D}_1$ respectively, are unchanged as the canonical class of these divisors remains the same under the blow up. 
The resulting configuration is shown on the top of figure \ref{fig:FlopDiags}.

The flop is completed by blowing down the curve $\hat{C}_1^-$. The canonical classes change again as in  (\ref{Kchangebu}) for the two divisors, which contain this curve, i.e.
${D}_{1}$ and $D'$ under the blow down $\pi_2: D \rightarrow \check{D}$
\be
K_{D_1} = \pi_2^* K_{\check{D}_1} + \hat{C}_1^- \,,\qquad K_{D'} = \pi_2^* K_{\check{D}'} + \hat{C}_1^- \,.
\ee
After the blow down, denote the curve corresponding to $\hat{C}_2$ and $\hat{C}_1^+$ by $\check{C}_2^-$ and $\check{C}_1$, respectively. Then the normal bundles change as follows
\be
\ba
\hbox{deg} (N_{\check{C}_2^-/\check{D}_1}) & = (\check{C}_2^-)_{\check{D}'}^2 = - K_{\check{D}'} \cdot_{\check{D}'} \check{C}_2^- -2   \cr
&= - (K_{D'} - \hat{C}_1^-) \cdot_{D'} \hat{C}_2  -2 = - (0 -1) -2 =-1\cr
\hbox{deg} (N_{\check{C}_1/\check{D}_1}) & = (\check{C}_1)_{\check{D}_1}^2 = - K_{\check{D}_1} \cdot_{\check{D}_1} \check{C}_1 -2   \cr
&= - (K_{D_1} - \hat{C}_1^-) \cdot_{D_1} \hat{C}_1^+   -2= -(-1 -1) -2=0\,.
\ea
\ee
On the other hand, $\hat{C}_1^-$ is not in $\hat{D}_2$ or $\hat{D}''$, so the blow down does not affect the normal bundle of $\check{C}_2^-$ in $\hat{D}_2$ or of $\check{C}_1^+$ in $\hat{D}''$. 
Thus the flop of $C_1^-$, which was previously the intersection of $D'$ and $D_1$, produces a new curve $\hat{C}_2^+$ which is no longer contained inside either $D'$ or $D_1$ but instead intersects them in a point.

Alternatively, one can consider first blowing down with $p_2$ in figure \ref{fig:FlopDiags}, and then blowing up. The advantage of the process we described here, is that the geometry in every step is
smooth, whereas the lower, singular configuration would require particular care in applying the intersection calculus. 

The prior analysis can now be applied to the case of $SU(5)$ models with e.g. fundamental matter. Taking one of the divisors $D'$ or $D_1$ above to be one of the rational sections we see that, under a flop, a curve contained inside the section is flopped to one that intersects the section in a point and vice versa. Consider a configuration in figure \ref{fig:F0SigmaCodim2Fibs}, for example where $\sigma \cdot_Y F_1=1$ in codimension one, then 
the generic flops for fibers studied in \cite{Hayashi:2014kca} dictate how the configurations flop into each other. However for fibers with rational sections,  not every configuration appears to have a flop image in the category of fiber configurations that satisfy our initial setup.  This is indicated in the shading of the charges in figures \ref{fig:01Charges}$-$\ref{fig:0ss1Charges}, showing which charges flop into each other. The charges in blue appear in every phase whereas the charges highlighted in green only appear in certain phases. The flop of the configurations, which do not appear in all phases will be discussed in section \ref{sec:FlopToSingularSection}.


\subsection{An $I_1^*$ Flop}

\begin{figure}
  \centering
  \includegraphics[width=8cm]{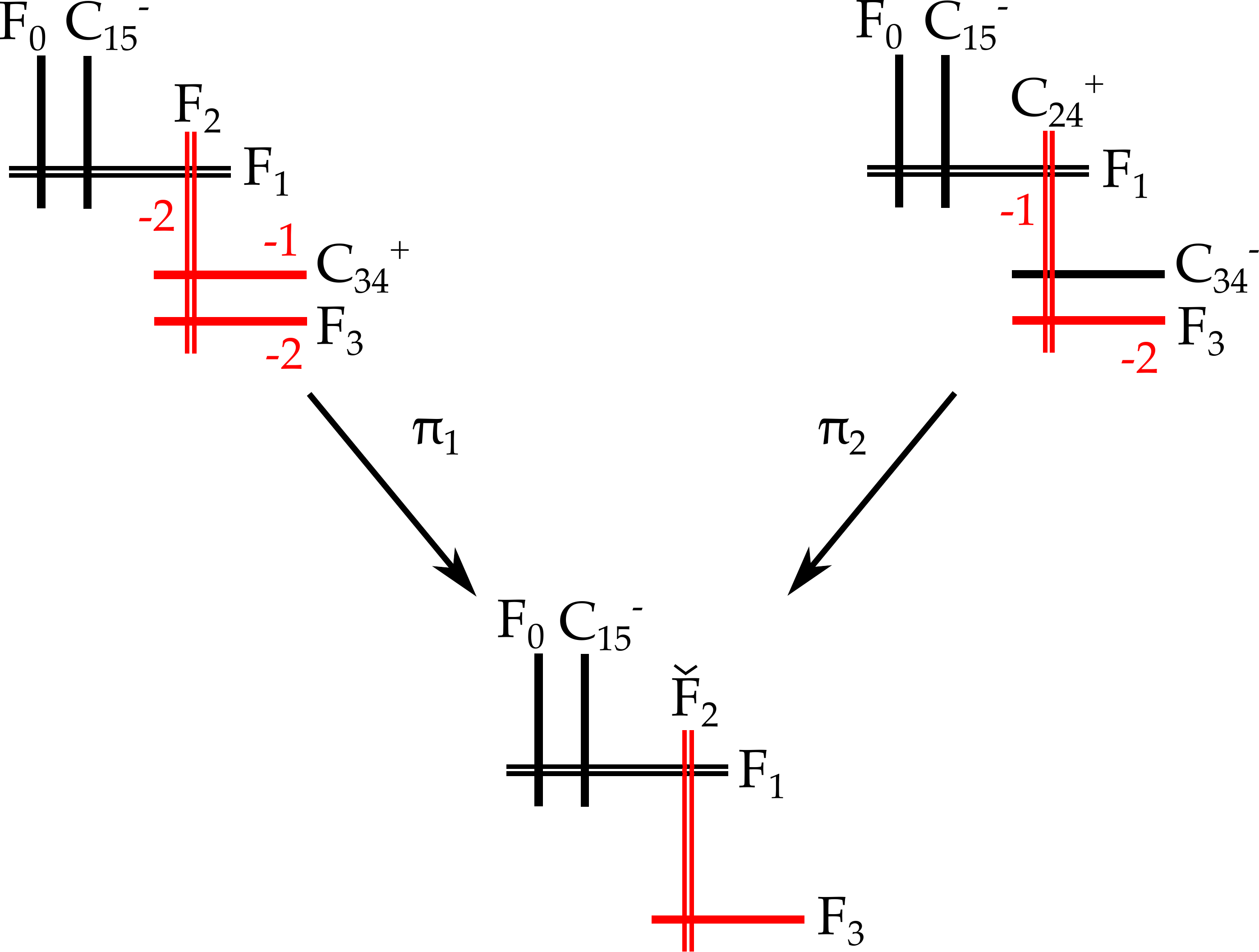} 
  \caption{Flop of a $\sigma_1$ wrapping configuration from phase $6$ (left) to phase $8$ (right) where $\sigma_1 \cdot_Y F_1 =1$. The red numbers denote the self intersections of the curves inside $\sigma_1$. }
 \label{fig:I1sGoodFlop}
\end{figure}

Consider the flop of the curve $C_{3,4}^+$ depicted in figure \ref{fig:I1sGoodFlop}. In this case it is simpler to consider first blowing down this curve, and then blowing up. The starting configuration, shown on the left of figure \ref{fig:I1sGoodFlop}, appears in phase 6 of table \ref{tab:I1sConfigsPhase1to8} where the section intersects $F_1$ in codimension one. The splitting in this phase is given by,
\be 
F_4 \rightarrow C_{3,4}^+ + F_1 + F_2 + C_{1,5}^-
\ee
These curves have the following self-intersections, i.e. normal bundle degrees, inside $D_{F_4}$,
\be \ba 
(C_{3,4}^+)^2_{D_{F_4}} &= -1 \\
(C_{1,5}^-)^2_{D_{F_4}} &= -1 \\
(F_1)^2_{D_{F_4}} &= -2 \\
(F_2)^2_{D_{F_4}} &= -2 \,,\\
\ea \ee
determined by the box graph for this phase. For the curves $F_i$ do not split,
\be
(F_i)^2_{D_{F_i}} = 0 \,.
\ee
In the configuration shown $F_2, C_{3,4}^+, F_3 \subset \sigma_1$ and the self intersections in $\sigma_1$ are given by the red numbers appearing next to these curves in the figure. Now consider the blow down of the curve $C_{3,4}^+$ which changes the canonical class of $D_{F_4}$ and $\sigma_1$, 
\be 
K_{\sigma_1} = \pi_1^*K_{\check{\sigma}_1} + C_{3,4}^+, \qquad K_{D_{F_4}} = \pi_1^*K_{\check{D}_{F_4}} + C_{3,4}^+  \,.
\ee
Under the blow up $\pi_2$ of the singular geometry we reach the $I_1^*$ fiber obtained by the splitting,
\be  \ba
F_4 &\rightarrow  C_{2,4}^+ + F_1 + C_{1,5}^-\\
F_2 &\rightarrow  C_{2,4}^+ + C_{3,4}^-
\ea \ee
The configuration in this phase, phase 8, is shown on the right in figure \ref{fig:I1sGoodFlop}, where the flopped curve $C_{3,4}^- \not\subset \sigma_1$ and the canonical class of the divisor $\check{D}_{F_2}$ is
\be 
K_{\hat{D}_{F_2}} = \pi_2^*K_{\check{D}_{F_2}} + C_{3,4}^-  \,.
\ee
Only the normal bundle of the curve $F_2$, which becomes $C_{2,4}^+$, is altered by this flop as no other curve intersected $C_{3,4}^+$ in the original configuration. As the intermediate stage in this description of the flop is singular the self intersection of the curve $C^+_{2,4}$ in the divisors $\hat{D}_{F_4}$, $\hat{\sigma}_1$ and $\hat{D}_{F_2}$ in phase 8 is computed by always pulling back to one of the resolved geometries,
\be\ba 
(C_{2,4}^+)^2_{\hat{\sigma}_1/\hat{D}_{F_4}} =& - K_{\hat{\sigma}_1/\hat{D}_{F_4}}\cdot_{\hat{\sigma}_1/\hat{D}_{F_4}} C_{2,4}^+ -2\\
=& - K_{\check{\sigma}_1/\check{D}_{F_4}}\cdot_{\check{\sigma}_1/\check{D}_{F_4}} \check{F}_2 -2\\
=& -(K_{\sigma_1/D_{F_4}} - C_{3,4}^+) \cdot_{\sigma_1/D_{F_4}} F_2 -2\\
=& -(0 -1) -2 = -1 \,.
\ea\ee
In the above, the second equality sign holds as the canonical class of $\check{D}_{F_4}$ and $\check{\sigma}_1$ is unchanged by the blow up $\pi_2$.
\be\ba 
(C_{2,4}^+)^2_{\hat{D}_{F_2}} =& - K_{\hat{D}_{F_2}} \cdot_{\check{D}_{F_2}} C_{2,4}^+ -2\\
=& -(\pi_2^*K_{\check{D}_{F_2}} + C_{3,4}^-) \cdot_{\hat{D}_{F_2}} C_{2,4}^+ -2\\
=& -(\pi_2^*K_{D_{F_2}} + C_{3,4}^-) \cdot_{\hat{D}_{F_2}} C_{2,4}^+ -2\\
=& -(-2 + 1) -2 = -1 \,.
\ea\ee
Thus the curve $C_{2,4}^+$ has normal bundle degree $(-1,-1)$ in the flopped geometry which is exactly what we expect from the splitting in phase 8. The flop discussed here exactly reproduces what was claimed in the previous section: a curve contained inside the section is flopped to one which intersects it at a point.

\subsection{Flops to Singular Sections}
\label{sec:FlopToSingularSection}

\begin{figure}
  \centering
  \includegraphics[width=15cm]{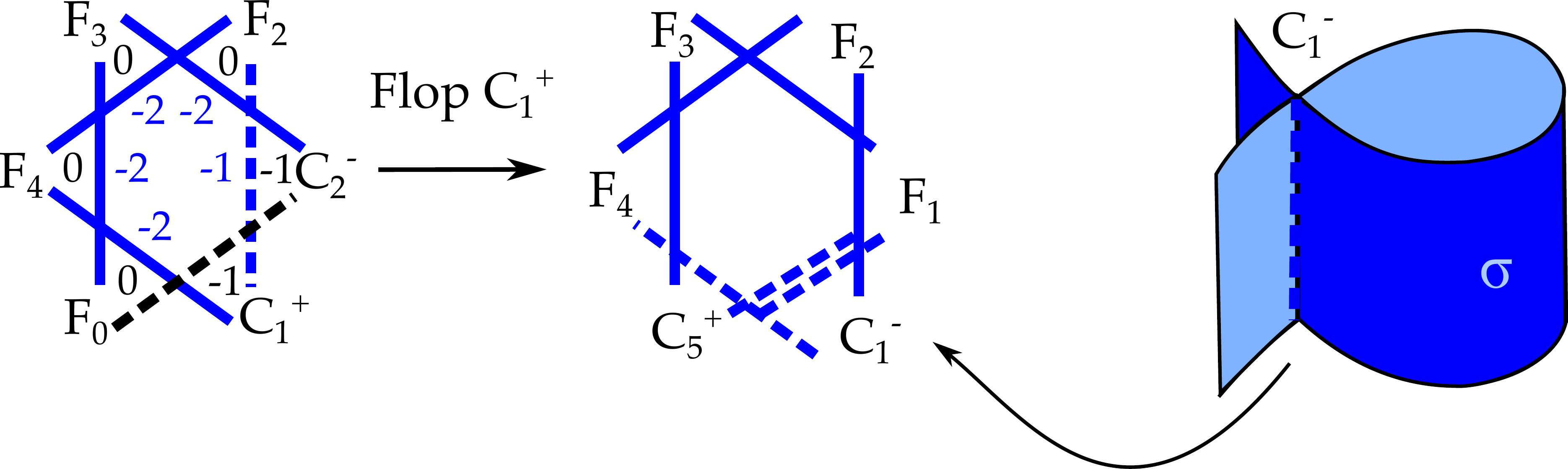} 
  \caption{ 
  {The almost fully wrapped fiber (the rational curves contained in the section $\sigma$ are shown in blue) shown on the left flops via $C_1^+$ to the fiber, which is  fully contained in the section. However the section  
  is now singular along the curve $C_1^-$, along which it self-intersects as shown on the far right.  
   The numbers in black and blue denote the degree of the normal bundle of the curves inside the divisors $D_{F_i}$ and the section $\sigma$, respectively.}
   }
\label{fig:finalflop}
\end{figure}

It was mentioned in section \ref{sec:FlopsIntro} that certain configurations do not flop into configurations within the class of fibers that we considered here. All such fibers are of the type that the entire fiber except for one curve is contained inside the section.  We now briefly comment on this.
Consider for instance flopping the curve $C_1^+$ on the left hand side  of figure \ref{fig:finalflop}. In this configuration the splitting is given by $F_1 \rightarrow C_1^+ + C_2^-$ and the curve $C_1^+$ has normal bundle $(-1,-1)$ inside of $D_{F_1}$. 

Proceeding as described above, we blow up every point along $C_1^+$ and in doing so we obtain the exceptional divisor $E$. The two points at which $C_1^+$ intersected the section become two curves contained inside the section. Under the contraction of the $C_1^+$ ruling of the exceptional divisor $E$,  the two curves contained in the section are identified. Thus we obtain a curve which is contained inside the section twice. The section is now singular as it meets itself along this curve\footnote{We thank Dave Morrison for discussions on this point.}. This configuration is shown on the right hand side of figure \ref{fig:finalflop}. 
In our analysis we assumed throughout that the section is a smooth divisor in the Calabi--Yau. Clearly, after this flop this condition ceases to hold, and it would be interesting to study such configurations, and to determine
whether or not the singular section is consistent from the point of view of the F-theory compactification. We will comment on this further in the discussion section \ref{sec:Disc}.

\section{Singlets}
\label{sec:singlets}

As a final application of our method, we now turn to discuss $U(1)$-charged GUT singlets. Mathematically, this corresponds to analyzing the codimension two fibers with rational section for an $I_1$ to $I_2$ enhancement. 
Apart from the interest in the types of singlet charges that are possible, this has wide-ranging implications for Higgsing the $U(1)$ symmetries to a discrete gauge symmetry, as in e.g. 
\cite{Mayrhofer:2014haa,Mayrhofer:2014laa,Cvetic:2015moa}.
Other phenomenologically interesting implications, in particular when applied to four-folds, concern the possible Yukawa couplings of the type ${\bf R} \overline{\bf R} {\bf 1}$ as well as non-renormalizable couplings, which e.g. could regenerate proton decay operators.  
After some general properties of singlets, we first discuss the situation in three-folds in section \ref{sec:SingThree}, and for four-folds in section \ref{sec:SingletsCY4}.

\subsection{Constraints on Singlet Curves}

Consider a smooth Calabi--Yau three- or four-fold $Y$. 
An $I_1$ fiber consists of a single nodal rational curve $F_0$, with
arithmetic genus $p_a(F_0) = 1$, such that 
\be
D_{F_0} \cdot_Y F_0 = 0
\ee
Above a codimension two locus, the node splits 
\be
F_0 \rightarrow C^+ + C^- \,,
\ee
where $C^\pm$ are smooth rational curves, {which intersect in an $I_2$ Kodaira fiber}. Consistency with codimension
one requires that 
\begin{equation}
  D_{F_0} \cdot_Y C^+ = - D_{F_0} \cdot_Y C^- \,,.
\end{equation}
As both $C^\pm$ are smooth rational curves contained inside $D_{F_0}$, it follows by Corollary \ref{cor:DC} that 
\begin{equation}\label{eqn:I1sum}
  \hbox{deg}(N_{C^+/D_{F_0}}) +\hbox{deg}(N_{C^-/D_{F_0}}) = -4 \,.
\end{equation}
However, as these curves do not arise as complete intersections, their normal bundles in $Y$ are not fixed by the degrees of $N_{C^\pm/D_{F_0}}$. We require one of the curves in the $I_2$ fiber to be contractible. Without loss of generality,  we take $C^-$ to be the contractible curve. 
In Calabi--Yau three-folds this condition is known to have three solutions, as summarized in Theorem \ref{thm:minus2}, which will be discussed in the next section.  For four-folds we are not aware of a similar result, and we will therefore conduct a survey without imposing the additional contractibility condition in section \ref{sec:SingletsCY4}. 


\subsection{Singlets in Three-folds}
\label{sec:SingThree}

In this section, let $Y$ be a smooth Calabi--Yau three-fold. We will first determine the possible section configurations that are consistent from the point of view of normal bundle degrees in a three-fold. 
Following this, we determine the possible singlet charges and fiber types. 

\subsubsection{Normal Bundle Constraints}

We start by considering the possible normal bundle degrees for rational curves in an $I_2$ fiber. We assume $C^-$ to be contractible.
Theorem \ref{thm:minus2} implies that a contractible rational curve  can have the following normal bundles in $Y$:
\begin{itemize}
\item[A)] $N_{C^{-}/Y} = \mathcal{O}(-1)\oplus \mathcal{O}(-1)$
\item[B)] $N_{C^{-}/Y} = \mathcal{O}\oplus \mathcal{O}(-2)$
\item[C)] $N_{C^{-}/Y} = \mathcal{O}(1)\oplus \mathcal{O}(-3)$ \,.
\end{itemize}
We do not constrain $C^+$ to be contractible therefore its normal bundle takes the general form
\be 
N_{C^+/Y} = \mathcal{O}(p) \oplus \mathcal{O}(-2 -p), \quad p \geq -1 \,.
\label{CpNB}
\ee 
We consider a fibration with two rational sections, $\sigma_0$ and $\sigma_1$.
In codimension one both sections intersect $F_0$, therefore it is sufficient to just consider one of the sections to find the possible configurations for the fiber in codimension two. For an $I_1$ local enhancement to $I_2$ the constraint from codimension one is,
\be 
\sigma \cdot_Y (C^+ + C^-) = 1 \,.
\label{I1codim1}
\ee
For each case A$-$C there always exists the solution, where the section intersects transversally either $C^+$ or $C^-$ and does not contain any curves in the fiber. The two cases will differ in the possible wrapping configurations.

As the normal bundle of $C^+$ is the same for cases A$-$C we can first derive some general statements irrespective of the normal bundle of $C^-$. Consider $C^+ \subset \sigma$, 
using Theorem \ref{thm:NormalBundleSES} (iii), there exists an embedding
\be
N_{C^+/\sigma}  \ \hookrightarrow\  N_{C^+/Y}=  \mathcal{O}(p) \oplus \mathcal{O}(-2 -p) \,,\qquad p\geq -1 \,,
\ee
in the following two cases:
\begin{enumerate}
\item[(i)] $\hbox{deg}(N_{C^+/\sigma}) = p$ 
\item[(ii)] $\hbox{deg}(N_{C^+/\sigma}) \leq -p -2 $.
\end{enumerate}
Using Corollary \ref{cor:DC} one finds that for (i)
\be 
\sigma \cdot_Y C^+ = -p -2 \,.
\label{CpLowerBound}
\ee
Combining \eqref{CpLowerBound} with \eqref{I1codim1}, one obtains the intersection of $C^-$ with $\sigma$,
\be 
\sigma \cdot_Y C^- = p+3 \,.
\label{CmUpperBound}
\ee
The intersections of $\sigma$ with $C^+$ (resp. $C^-$) will be bounded from below (resp. above) by \eqref{CpLowerBound} (resp. \eqref{CmUpperBound}).

Now let us consider case A where $C^-$ has normal bundle degree $(-1,-1)$. If $C^- \subset \sigma$ then in order for $N_{C^{-}/\sigma}$ to embed inside $N_{C^{-}/Y}$ we must have,
\be 
\hbox{deg}(N_{C^-/\sigma}) \leq -1 \,.
\ee
This is a consequence of Theorem \ref{thm:NormalBundleSES} part (ii) and as a result the intersections of $\sigma$ with $C^{\pm}$ are 
\be 
(\sigma \cdot_Y C^+, \sigma \cdot_Y C^-) = (2,-1) , \ (1,0) , \ (0,1), \ (-p-2,p+3) \,.
\ee   
The codimension one constraint \eqref{I1codim1} then specifies the upper bound for the intersection of $\sigma$ with $C^+$. The possible configurations which realize these intersections are:
\begin{itemize}
\item[A.1)] $\sigma \cdot_Y C^+ = 2$, $\sigma \cdot_Y C^- = -1$ \\
The lower bound on $\sigma \cdot_Y C^-$ is achieved by $C^- \subset \sigma$, with $\hbox{deg}(N_{C^-/\sigma}) = -1$. To obtain the correct intersection for $C^+$ with the section there are two possibilities:
\begin{itemize}
\item[(i)] $C^+ \not\subset \sigma$ \\
The correct intersections are automatic in this case as in any $I_2$ fiber the curves $C^\pm$ intersect each other in two points, and $C^-$ is contained inside the section.
\item[(ii)] $C^+ \subset \sigma$ \\
The degree of $N_{C^+/\sigma}$ is determined using Corollary \ref{cor:DC}, requiring $\sigma \cdot_Y C^+ = 2$ implies $\hbox{deg}(N_{C^+/\sigma})= -4$.
This solution is only valid when $N_{C^+/Y} = \mathcal{O}(-4)$ can be embedded non-trivially into $N_{C^+/Y}$ which is true for 
\be\label{A1p}
-1 \leq p \leq 2 \,.
\ee
\end{itemize}
\item[A.2)] $\sigma \cdot_Y C^+ = 1$, $\sigma \cdot_Y C^- = 0$ \\
There are two configurations, which realize the above intersections. The first is given by $C^+ \not\subset \sigma$, but $\sigma$ intersects $C^+$ transversally.  In this case the section does not contain any components of the fiber. The second solution is given by $C^+, C^- \subset \sigma$ and $\hbox{deg}(N_{C^+/\sigma}) = -3$ and $\hbox{deg}(N_{C^-/\sigma}) = -2$.
One can check using Corollary \ref{cor:DC} that these values give the correct intersection values for $\sigma \cdot_Y C^{\pm}$. The latter configuration can only be realized  for 
\be\label{A2p}
-1 \leq p \leq 1
\ee
\item[A.3)] $\sigma \cdot_Y C^+ = 0$, $\sigma \cdot_Y C^- = 1$ \\
The solutions in this case can be obtained from the solutions in A.2 by exchanging $C^{\pm}$. The configuration where the entire fiber is contained inside the section is a solution for
\be\label{A3p}
p = -1\  \hbox{ or }\  0 \,.
\ee
\item[A.4)] $\sigma \cdot_Y C^+ = -p-2$, $\sigma \cdot_Y C^- = p+3$ \\
As was detailed above, to achieve a negative intersection with the section, $C^+$ must be contained inside it with $\hbox{deg}(N_{C^+/\sigma}) = p$. There are two possibilities for $C^-$:
\begin{itemize}
\item[(i)] $C^- \not\subset \sigma$ \\
The section, from the containment of $C^+$, intersects $C^-$ in two points necessarily. In order to satisfy \eqref{I1codim1} $C^-$ requires $p+1$ additional intersections with the section.
\item[(ii)] $C^- \subset \sigma$ \\
In this case we require $\hbox{deg}(N_{C^-/\sigma})= -p-5$ to satisfy $\sigma \cdot_Y C^- = p+3$.
This solution is valid for $p\geq -1$ as for these values of $p$ the following embedding always exists
\be 
\mathcal{O}(-p-5) \ \hookrightarrow \ \mathcal{O}(-1) \oplus \mathcal{O}(-1) \,.
\ee
\end{itemize}
\end{itemize}
The full set of configurations for A are summarized below.  The configurations which have been marked $(*)$ are only valid when $p$ falls within the ranges specified in (\ref{A1p}), (\ref{A2p}) and (\ref{A3p}), respectively.
\be 
\begin{array}{c|c||l|l}
 \sigma \cdot_Y C^+ & \sigma \cdot_Y C^- & C^+ \hbox{configuration} & C^- \hbox{configuration} \\ \hline \hline
2 & -1 & C^+ \not\subset \sigma, \sigma \cdot_Y C^+ = 2 & C^- \subset \sigma, \hbox{deg}(N_{C^-/\sigma}) = -1  \\
& & C^+ \subset \sigma, \hbox{deg}(N_{C^+/\sigma}) = -4 & C^- \subset \sigma, \hbox{deg}(N_{C^-/\sigma}) = -1 \ (*)\\ \hline
1 & 0 & C^+ \not\subset \sigma, \sigma \cdot_Y C^+= 1 & C^- \not\subset \sigma, \sigma \cdot_Y C^- = 0 \\
& & C^+ \subset \sigma, \hbox{deg}(N_{C^+/\sigma}) = -3 & C^- \subset \sigma, \hbox{deg}(N_{C^-/\sigma}) = -2 \ (*) \\ \hline
0 & 1 & C^+ \not\subset \sigma, \sigma \cdot_Y C^+ = 0 & C^- \not\subset \sigma, \sigma \cdot_Y C^- = 1 \\
& & C^+ \subset \sigma, \hbox{deg}(N_{C^+/\sigma}) = -2 & C^- \subset \sigma, \hbox{deg}(N_{C^-/\sigma}) = -3 \ (*) \\ \hline
-p -2 & p+3 & C^+ \subset \sigma, \hbox{deg}(N_{C^+/\sigma}) = p & C^- \not\subset \sigma, \sigma \cdot_Y C^- = p+3  \\
& & C^+ \subset \sigma, \hbox{deg}(N_{C^+/\sigma}) = p & C^- \subset \sigma, \hbox{deg}(N_{C^-/\sigma}) = -p-5 
\end{array}
\ee

For case B the curve $C^-$ has normal bundle degree $(0,-2)$. To find the lower bound for the intersection of $C^-$ with the section we need to consider $C^- \subset \sigma$. Requiring $N_{C^-/\sigma}$ to embed inside $N_{C^-/Y}$ gives the constraint 
\be 
\hbox{deg}(N_{C^-/\sigma}) \leq 0 \,,
\ee
where $\hbox{deg}(N_{C^-/\sigma}) \neq -1$. This bounds the intersection of $C^-$ with the section from below,
\be 
\sigma \cdot_Y C^- \geq -2 \ \Rightarrow\ \sigma \cdot_Y C^+\leq 3 \,.
\ee
The possible intersections are given by
\be 
(\sigma \cdot_Y C^+, \sigma \cdot_Y C^-) = (3,-2) , \ (1,0) , \ (0,1), \ (-p-2,p+3)\,.
\ee  
The intersection of $C^+$ with $\sigma$ can not take the value $-1$ due to the constraint $\hbox{deg}(N_{C^-/\sigma}) \neq -1$. The solutions for the last three intersection sets are the same as those given for case A therefore we shall only detail the solutions for the first set here.
\begin{itemize}
\item[B.1)] $\sigma \cdot_Y C^+ = 3, \sigma \cdot_Y C^- = -2$ \\
The two configurations for this set of intersections must have $
C^- \subset \sigma, \hbox{deg}(N_{C^-/\sigma}) = 0$.
This is mandated by the intersection of the section with $C^-$. There are two possibilities for $C^+$:
\begin{itemize}
\item[(i)] $C^+ \not\subset \sigma$ \\
The containment of $C^-$ inside the section means that $C^+$ intersects the section twice through the intersection of $C^{-}$ and $C^+$ in the fiber. Consistency with codimension one requires an additional transverse intersection between $\sigma$ and $C^+$.
\item[(ii)] $C^+ \subset \sigma$ \\
Requiring $\sigma \cdot_Y C^+ = 3$ means that $\hbox{deg}(N_{C^+/\sigma}) = -5$. 
This configuration is a valid solution for 
\be\label{B1p}
-1 \leq p \leq 3 \,.
\ee
\end{itemize}
\end{itemize}
The configurations for case B are ($p$ is constrained in the (*)'ed configurations as in (\ref{B1p}), (\ref{A2p}) and (\ref{A3p}), respectively)
\be 
\begin{array}{c|c||l|l}
 \sigma \cdot_Y C^+ & \sigma \cdot_Y C^- & C^+ \hbox{configuration} & C^- \hbox{configuration} \\ \hline \hline
3 & -2 & C^+ \not\subset \sigma, \sigma \cdot_Y C^+ = 3 & C^- \subset \sigma, \hbox{deg}(N_{C^-/\sigma}) = 0  \\
& & C^+ \subset \sigma, \hbox{deg}(N_{C^+/\sigma}) = -5 & C^- \subset \sigma, \hbox{deg}(N_{C^-/\sigma}) = 0 \ (*) \\ \hline
1 & 0 & C^+ \not\subset \sigma, \sigma \cdot_Y C^+ = 1 & C^- \not\subset \sigma, \sigma \cdot_Y C^- \sigma = 0 \\
& & C^+ \subset \sigma, \hbox{deg}(N_{C^+/\sigma}) = -3 & C^- \subset \sigma, \hbox{deg}(N_{C^-/\sigma}) = -2 \ (*) \\ \hline
0 & 1 & C^+ \not\subset \sigma, \sigma \cdot_Y C^+ = 0 & C^- \not\subset \sigma, \sigma \cdot_Y C^- = 1 \\
& & C^+ \subset \sigma, \hbox{deg}(N_{C^+/\sigma}) = -2 & C^- \subset \sigma, \hbox{deg}(N_{C^-/\sigma}) = -3 \ (*) \\ \hline
-p -2 & p+3 & C^+ \subset \sigma, \hbox{deg}(N_{C^+/\sigma}) = p & C^- \not\subset \sigma, \sigma \cdot_Y C^- = p+3  \\
& & C^+ \subset \sigma, \hbox{deg}(N_{C^+/\sigma}) = p & C^- \subset \sigma, \hbox{deg}(N_{C^-/\sigma}) = -p-5 
\end{array}
\ee
Finally, in case C, the curve $C^-$ has normal bundle $(1,-3)$. If $C^{-} \subset \sigma$ then the only wrapped configuration which gives negative intersections with the section is 
\be 
\hbox{deg}(N_{C^-/\sigma}) = 1 \ \Rightarrow \ C^{-} \cdot_Y \sigma = -3 \,.
\ee
This generates the upper bound $\sigma \cdot_Y C^+ \leq 4$.  The set of possible intersections are
\be 
(\sigma \cdot_Y C^+, \sigma \cdot_Y C^-) = (4,-3) , \ (1,0) , \ (0,1), \ (-p-2,p+3) \,.
\ee  
Once again, the solutions for second and fourth set of intersections are the same as those given in A. Though the third set of intersections has appeared previously the solutions for this normal bundle case are more restricted and we will find only one solution.
\begin{itemize}
\item[C.1)] $\sigma \cdot_Y C^+ = 4, \sigma \cdot_Y C^- = -3$ \\
The two solutions to this set of intersection numbers both require
$C^- \subset \sigma, \hbox{deg}(N_{C^-/\sigma}) = 1$.
To obtain the correct intersection for $C^+$ with the section there are two possibilities:
\begin{itemize}
\item[(i)] $C^+ \not\subset \sigma$ \\
In addition to the two intersections $C^+$ has with the section through the intersection of $C^+$ and $C^-$ two further intersections are required to satisfy the codimension one constraint \eqref{I1codim1}.
\item[(ii)] $C^+ \subset \sigma$ \\
The degree of the normal bundle $N_{C^+/\sigma}$ is fixed by the intersection $\sigma \cdot_Y C^+= 4$ to be $\hbox{deg}(N_{C^+/\sigma})= -6$. 
This is a valid solution for 
\be\label{C4p}
-1 \leq p \leq 4\,.
\ee
\end{itemize}
\item[C.3)] $\sigma \cdot_Y C^+ = 0, \sigma \cdot_Y C^- = 1$ \\
This set of intersections has appeared in A and B however the configuration given by $C^+, C^- \subset \sigma$ and $\hbox{deg}(N_{C^+/\sigma}) = -3, \hbox{deg}(N_{C^-/\sigma}) = -2$ is not a valid solution here as $N_{C^-/\sigma}$ does not embed into $N_{C^-/Y} = \mathcal{O}(1)\oplus \mathcal{O}(-3)$. The only solution is given by $C^+, C^- \not\subset \sigma$ and $\sigma \cdot_Y C^- = 1$. 
\end{itemize}

The full set of solutions for case C are (with ranges of $p$ in the (*)'ed configurations  constrained as in (\ref{C4p}) and (\ref{A3p}))

\be 
\begin{array}{c|c||l|l}
 \sigma \cdot_Y C^+ & \sigma \cdot_Y C^- & C^+ \hbox{configuration} & C^- \hbox{configuration} \\ \hline \hline
4 & -3 & C^+ \not\subset \sigma, \sigma \cdot_Y C^+ = 4 & C^- \subset \sigma, \hbox{deg}(N_{C^-/\sigma}) = 1  \\
& & C^+ \subset \sigma, \hbox{deg}(N_{C^+/\sigma}) = -6 & C^- \subset \sigma, \hbox{deg}(N_{C^-/\sigma}) = 1 \ (*) \\ \hline
1 & 0 & C^+ \not\subset \sigma, \sigma \cdot_Y C^+ = 1 & C^- \not\subset \sigma, \sigma \cdot_Y C^- = 0 \\ \hline
0 & 1 & C^+ \not\subset \sigma, \sigma \cdot_Y C^+ = 0 & C^- \not\subset \sigma, \sigma \cdot_Y C^- = 1 \\
& & C^+ \subset \sigma, \hbox{deg}(N_{C^+/\sigma}) = -2 & C^- \subset \sigma, \hbox{deg}(N_{C^-/\sigma}) = -3 \ (*) \\ \hline
-p -2 & p+3 & C^+ \subset \sigma, \hbox{deg}(N_{C^+/\sigma}) = p & C^- \not\subset \sigma, \sigma \cdot_Y C^- = p+3  \\
& & C^+ \subset \sigma, \hbox{deg}(N_{C^+/\sigma}) = p & C^- \subset \sigma, \hbox{deg}(N_{C^-/\sigma}) = -p-5 
\end{array}
\ee


\subsubsection{Compilation of Fibers and $U(1)$ Charges}

The solutions for each case A$-$C are presented in table \ref{tab:I2fibersCY3} where the intersection sets appear along the horizontal axis and the different normal bundles run along vertically. 
The $I_2$ fibers are labeled as follows:
\begin{itemize}
\item The components of the fiber coloured in red are those contained inside the section and the red numbers appearing next to these components denote the degree of the normal bundle of those components inside $\sigma$. 
\item Red dots on unwrapped fiber components correspond to transverse singlet intersections with $\sigma$. The red numbers next to a sequence of such dots denote the number of such transverse intersection points. 
\end{itemize}
Not every set of $\sigma \cdot_Y C^{\pm}$ intersections can be realized in each case A$-$C. Where an intersection column has been left blank there is no configuration corresponding to that set of intersections with $\sigma$.

\begin{table}
\centering
\begin{tabular}{|c||c|c|c|c|c|c|}
\hline
\multirow{2}{*}{\includegraphics[height=.7cm]{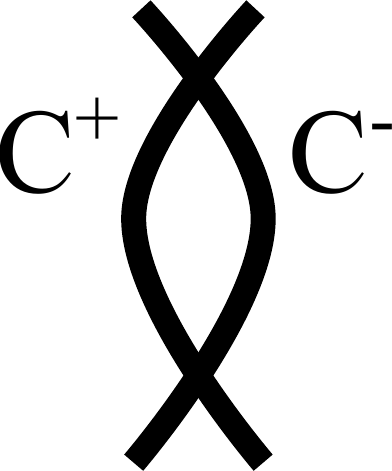}} & $\sigma\cdot C^+ = -p-2$ & $\sigma \cdot C^+ = 0$ & $\sigma \cdot C^+ = 1$ & $\sigma \cdot C^+ = +2$ & $\sigma \cdot C^+ = +3$ & $\sigma \cdot C^+ = +4$ \\
& $\sigma \cdot C^- = p+3$ & $\sigma \cdot C^- = 1$ & $\sigma \cdot C^- = 0$ & $\sigma \cdot C^- = -1$ & $\sigma \cdot C^- = -2$ & $\sigma \cdot C^- = -3$ \\ \hline \hline
\includegraphics[height=1.3cm]{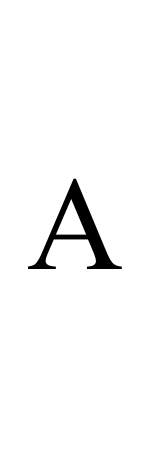} & \includegraphics[height=1.3cm]{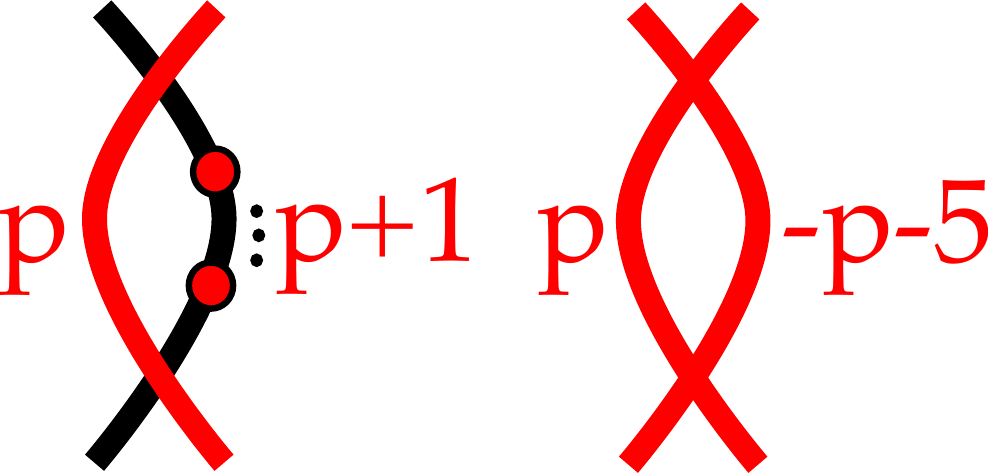} & \includegraphics[height=1.3cm]{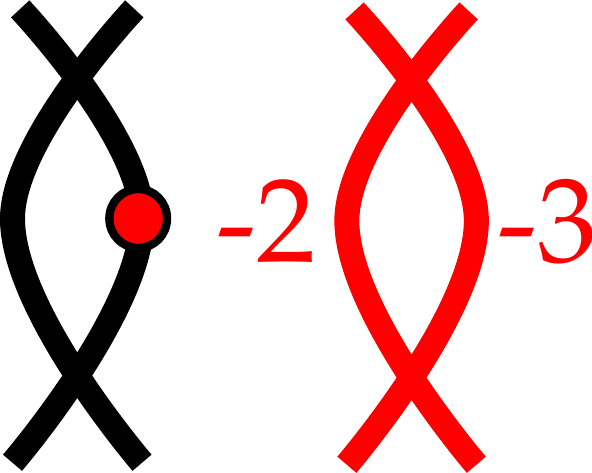} & \includegraphics[height=1.3cm]{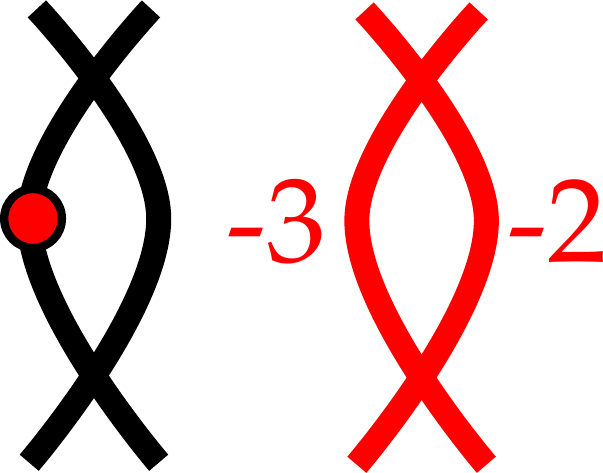} & \includegraphics[height=1.3cm]{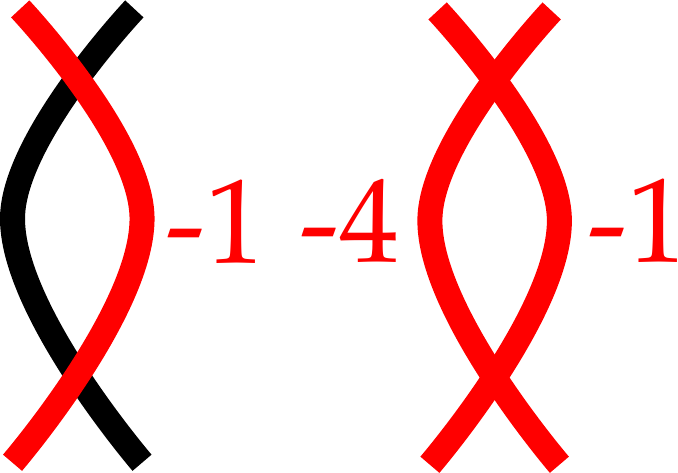}& &  \\ 
\includegraphics[height=1.3cm]{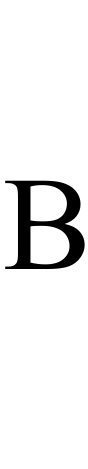} & \includegraphics[height=1.3cm]{I2FibA4.pdf} & \includegraphics[height=1.3cm]{I2FibA3.pdf} & \includegraphics[height=1.3cm]{I2FibA2.pdf} & &  \includegraphics[height=1.3cm]{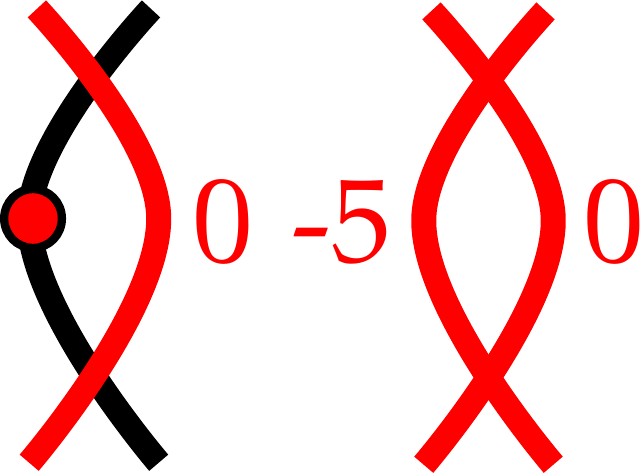} &\\
\includegraphics[height=1.3cm]{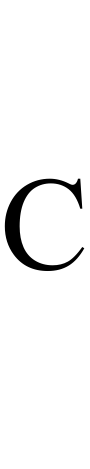} & \includegraphics[height=1.3cm]{I2FibA4.pdf} & \includegraphics[height=1.3cm]{I2FibA3.pdf} & \includegraphics[height=1.3cm]{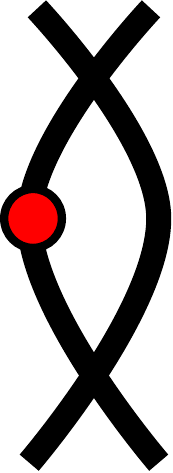} & & & \includegraphics[height=1.3cm]{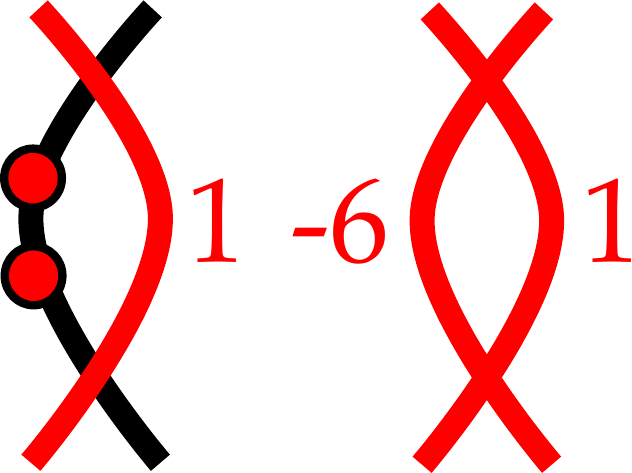} \\\hline
\end{tabular}
\caption{Consistent wrapping configurations for $I_1 \rightarrow I_2$ for normal bundle cases A$-$C. The components in red are those contained inside the section with their normal bundle degrees in $\sigma$ indicated by the red numbers adjacent to the component. Configurations where both components of the $I_2$ fiber are contained inside the section (excluding those appearing in the first column) are only valid for certain ranges of $p$, see main text for more details.}
\label{tab:I2fibersCY3}
\end{table}

\begin{figure}
\centering
 \includegraphics[width=15cm]{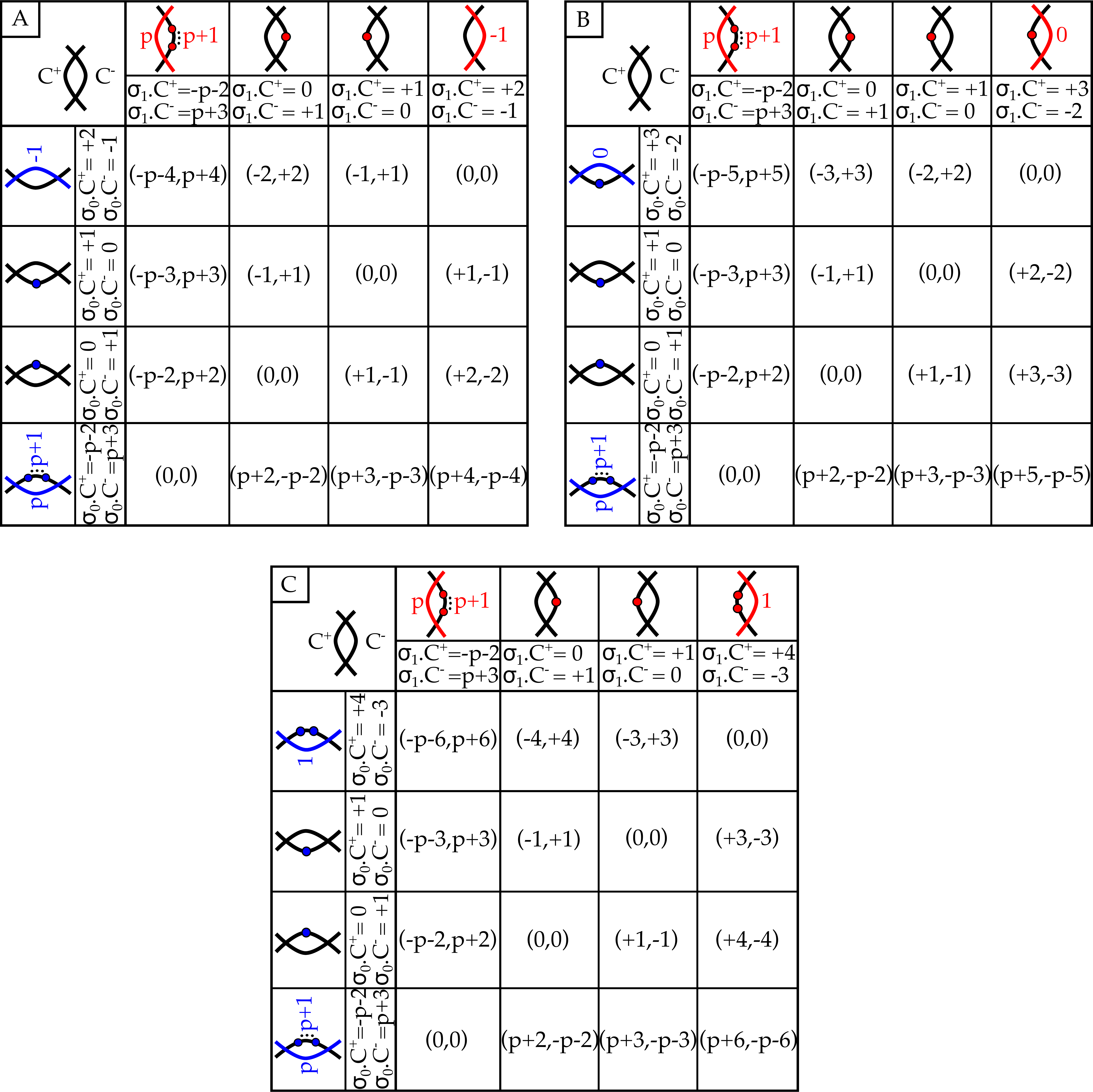} 
 \caption{$U(1)$ charges of singlets for normal bundles cases A$-$C. Configurations for $\sigma_1$ ($\sigma_0$) are along the horizontal (vertical) axis and the charges are the pairs $(a,-a)$ in the grid. Only one representative has been chosen for each distinct set of intersections $\sigma \cdot_Y C^{\pm}$ therefore there are more realizations of each charge than shown.}
 \label{fig:SingletCharges}
\end{figure}

The $U(1)$ charges of singlets can be determined by combining configurations for $\sigma_0$ and $\sigma_1$ in each case A$-$C. As both sections intersect $F_0$ in codimension one the Shioda map, $S(\sigma_1)$, is given by
\be 
S(\sigma_1) = \sigma_1 - \sigma_0 \,.
\ee
Singlet charges are obtained by computing $S(\sigma_1) \cdot_Y C^{\pm}$. The set of possible singlet charges and the associated $I_2$ fibers are shown in figure \ref{fig:SingletCharges}. The fibers along the horizontal (resp. vertical) axis, coloured in red (resp. blue), are for $\sigma_1$ (resp. $\sigma_0$). The entries $(a,-a)$ are the $U(1)$ charges obtained by combining configurations for $\sigma_1$ and $\sigma_0$. Only one representative has been chosen for each distinct set of intersections $\sigma \cdot_Y C^{\pm}$, wherefore there are more realizations of each charge than shown in the figure. The singlet charges which appear in each normal bundle pairing are:
\be\label{SingletCharges}
\hbox{$U(1)$ charges of singlets in} \quad 
\left\{ 
\ba
\hbox{A} &\in  \left\{0, \pm 1, \pm 2, \pm (p+2), \pm (p+3), \pm (p+4) \right\}  \cr
\hbox{B} &\in \left\{0, \pm 1, \pm 2, \pm 3, \pm (p+2), \pm (p+3), \pm (p+5) \right\} \cr
\hbox{C} &\in \left\{0, \pm 1, \pm 3, \pm 4, \pm (p+2), \pm (p+3), \pm (p+6)  \right\} \,.
\ea
\right.  
\ee
The charges are dependent on $p$, appearing in \eqref{CpNB}, which defines the normal bundle of $C^+$.

Singlet configurations ($I_2$ fibers in the presence of an one additional rational section) with charges  
\be 
S(\sigma_1) \cdot_Y C^- = \{-1, +1,+ 2 \} \,,
\ee
have appeared in \cite{Morrison:2012ei,Mayrhofer:2012zy, Braun:2013yti}.  The zero section in these configurations is holomorphic i.e. $\sigma_0$ does not contain curves in the fiber over codimension two. The range of possible singlet charges was extended in \cite{Klevers:2014bqa} where a singlet configuration with charge $+3$ was found. Comparing these fibers to those in figure \ref{fig:SingletCharges}, we find the same configurations in the following normal bundle cases:
\be
\begin{array}{c|c|c}
\hbox{Charge }S(\sigma_1) \cdot_Y C^-  & I_2 \hbox{ fiber} & \hbox{Realization} \\ \hline
\multirow{2}{*}{$-1$} & \multirow{2}{*}{\includegraphics[height=1cm]{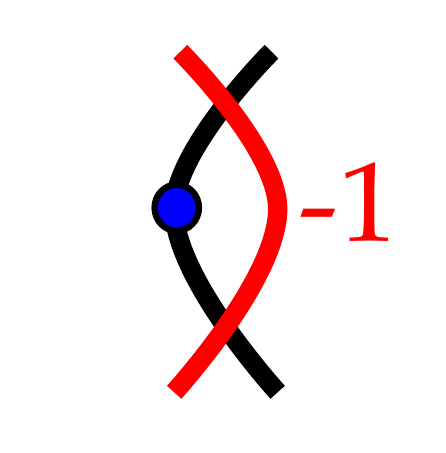}} & \multirow{2}{*}{\hbox{A} }\\
&&  \\
\multirow{2}{*}{$+1$} & \multirow{2}{*}{\includegraphics[height=1cm]{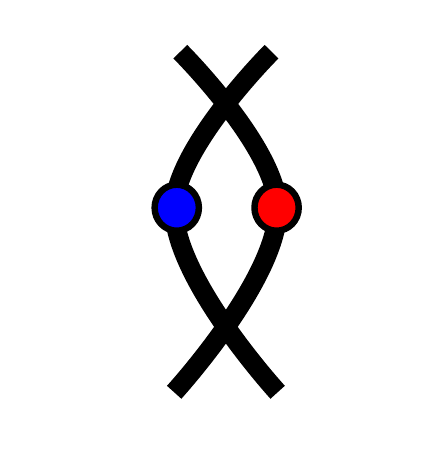}} & \multirow{2}{*}{\hbox{A}-\hbox{C}} \\
&&\\
\multirow{2}{*}{$+2$} & \multirow{2}{*}{\includegraphics[height=1cm]{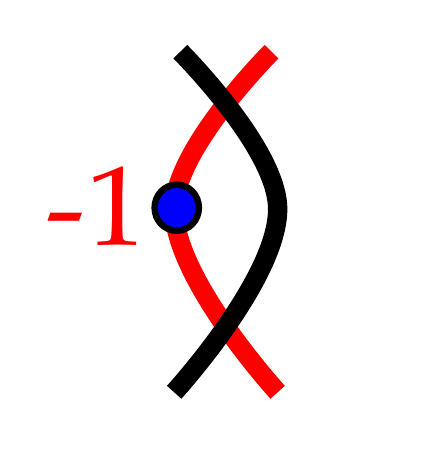}} &  \multirow{2}{*}{\hbox{A-C when $p=-1$}} \\
 &&\\
\multirow{2}{*}{$+3$} & \multirow{2}{*}{\includegraphics[height=1cm]{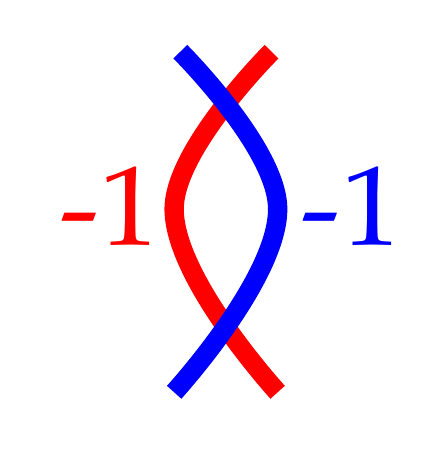}} & \multirow{2}{*}{\hbox{A when} $p = -1$} \\
&&
\end{array}
\ee

Finally, we compare the singlet charges found above with those required for every ${\bf \bar{5}}_{q_1}$ and  ${\bf 5}_{q_2}$ in \eqref{FundamentalCharges} to form a Yukawa coupling
\be
{\bf 5}_{q_1} {\bf \bar{5}}_{q_2} {\bf 1}_{-q_1-q_2} \,.
\label{551coupling}
\ee
Generically, in the geometry all such couplings will be present for base varieties of dimension $\geq 3$ and correspond to codimension three enhancements to $SU(7)$, which will be discussed in detail in section \ref{sec:CodimThree}. Using the set of $\bf \bar{5}$ charges in \eqref{FundamentalCharges}, the set of singlets, ${\bf 1}_{-q_1-q_2}$, for each codimension one fiber in (\ref{I5fibertypes}) is
\be\label{551SingletCharges}
\hbox{$U(1)$ charges of  GUT singlets in} \quad 
\left\{ 
\ba
I_5^{(01)} &\in  \left\{0, \pm 1, \pm 2, \pm 3, \pm 4, \pm 5, \pm 6 \right\} \cr
I_5^{(0|1)} &\in \left\{0, \pm 5, \pm 10, \pm 15, \pm 20, \pm 25 \right\} \cr
I_5^{(0||1)} &\in \left\{0, \pm 5, \pm 10, \pm 15, \pm 20, \pm 25\right\} \,.
\ea
\right.  
\ee
Comparison (after multiplication by five) yields, that the singlet charges in \eqref{551SingletCharges} fall within the charges derived from analysing $I_1\rightarrow I_2$ enhancements in \eqref{SingletCharges}. It would be interesting to analyze this further from the point of view of four-fold normal bundle consistencies at the Yukawa points. 

\begin{sidewaystable}
\begin{tabular}{|c||c|c|c|c|c|c|c|c|}
\hline
{\includegraphics[height=1.3cm]{I2Fib.pdf}} 
& \includegraphics[height=1.5cm]{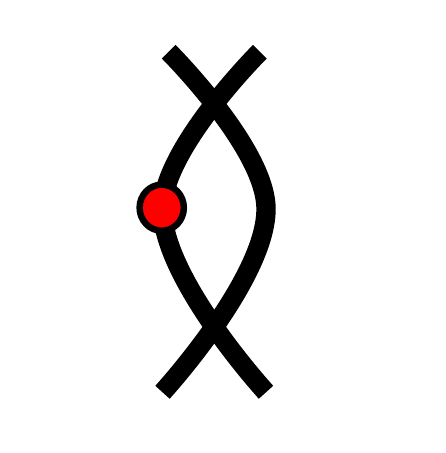} &\includegraphics[height=1.5cm]{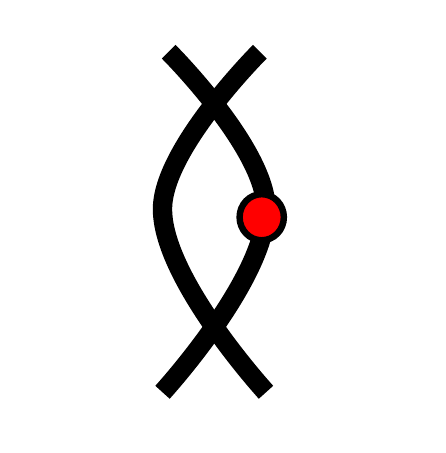} &\includegraphics[height=1.5cm]{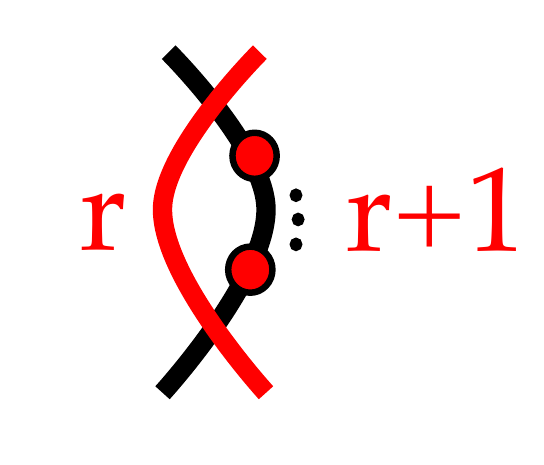} &\includegraphics[height=1.5cm]{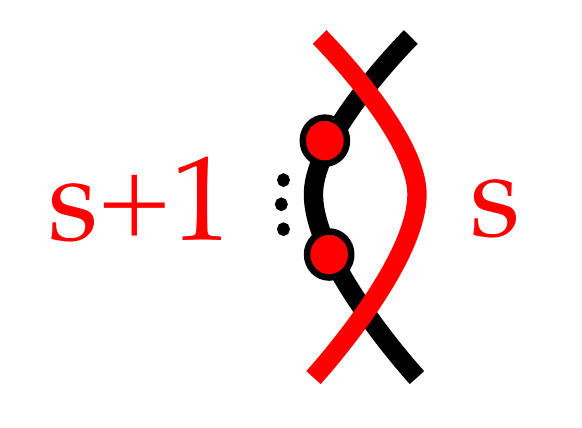} &\includegraphics[height=1.5cm]{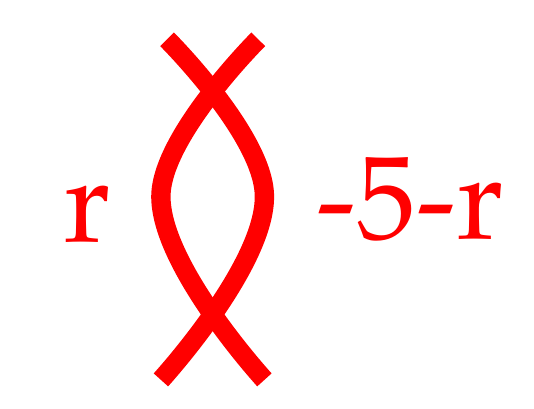}  \cr
$\sigma_0 ,  \sigma_1 \cdot_Y (C^+, C^-)$ 
& $(1,0)$ & $ (0,1)$ & $ (-r-2,r+3)$ & $ (s+3,-s-2)$ & $(-r-2, r+3)$ \cr \hline\hline
 \includegraphics[height=1.5cm]{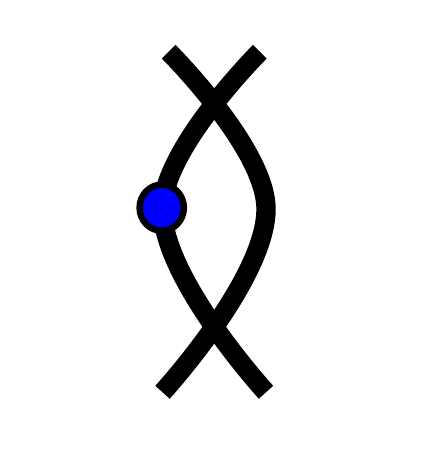} & $(0,0)$ & $(-1,+1 )$& $(-r-3, r+3)$& $(s+2,-s-2 )$& $(-r-3, r+3 )$ \cr
 $(1,0)$  &&&&&
 \cr \hline
\includegraphics[height=1.5cm]{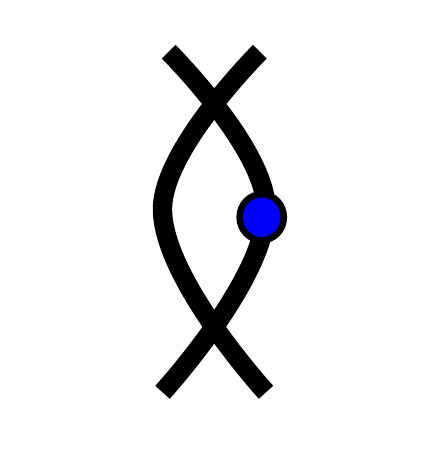}  & $(+1,-1)$ & $(0,0)$& $(-r-2, r+2)$& $(s+3, -s-3 )$& $(-r-2,r+2)$ \cr
$(0,1)$ &&&&&
\cr\hline
\includegraphics[height=1.5cm]{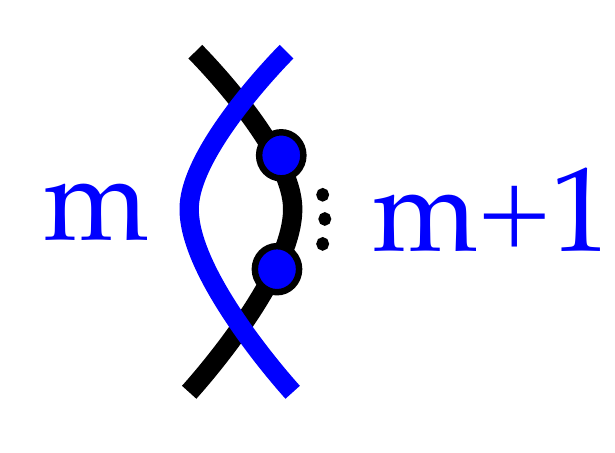}  & $(m+3,-m-3)$ & $(m+2,-m-2)$& $(-r+m,r-m)$& $(s+m+5,-s-m-5)$& $(-r+m,r-m)$ \cr
$(-m-2,m+3)$ &&&&&
\cr\hline
\includegraphics[height=1.5cm]{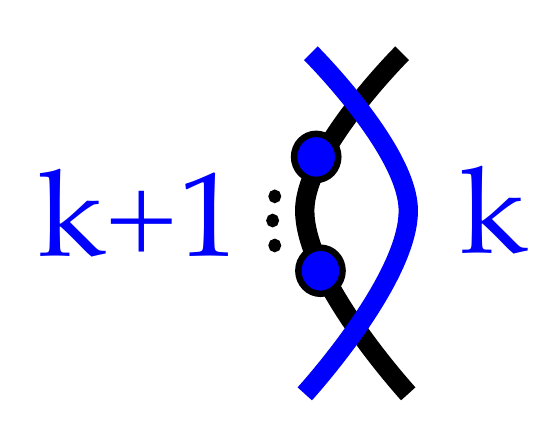}  & $(-k-2,k+2)$ & $(-k-3, k+3)$& $(-r-k-5,r+k+5)$& $(s-k,-s+k)$& $(-r-k-5,r+k+5)$ \cr
$(k+3,-k-2)$&&&&&
\cr\hline
\includegraphics[height=1.5cm]{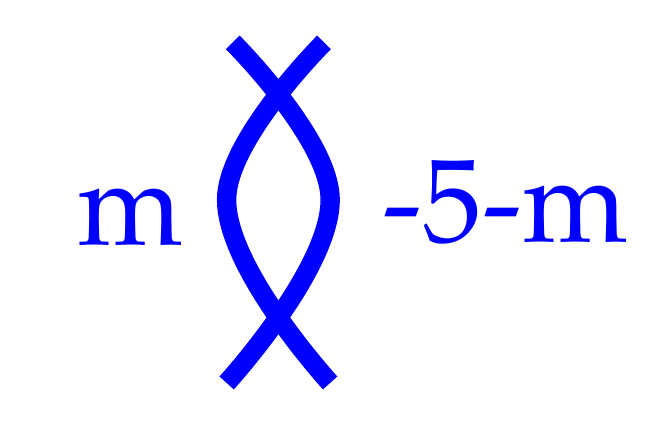}    & $(m+3, -m-3)$ & $(m+2,-m-2)$& $(-r+m, r-m)$& $(s+m+5, -s-m-5 )$& $(-r+m, r-m )$ \cr
$(-m-2,m+3 )$&&&&&
 \cr\hline
\end{tabular}
\caption{Consistent wrapping configurations for $I_1 \rightarrow I_2$ in four-folds and $U(1)$ charges. Configurations for $\sigma_1$ ($\sigma_0$) are along the horizontal (vertical) axis and the charges are the pairs $(a,-a)$ in the grid. The components in red (blue) are those contained inside $\sigma_1$ ($\sigma_0$) with their normal bundle degrees in $\sigma_{\alpha}$ indicated by the red (blue) numbers adjacent to the component.}
\label{tab:I2fibers}
\end{sidewaystable}


\subsection{Singlets in Four-folds}
\label{sec:SingletsCY4}

One of the criteria for the codimension two  $I_2$ fiber is that one of the curves needs to be contractible. In the case of three-folds discussed in the last section, the relevant criterion goes back to 
 Theorem \ref{thm:minus2}.  
A similar result, 
which constrains the normal bundle of contractible curves in four-folds, to our knowledge, is not known. 
Nevertheless, we can consider a general types of $I_2$ fiber, and without imposing contractibility, determine the consistent section configurations and corresponding charges.

The result of this analysis is summarized in table \ref{tab:I2fibers}. The normal bundle degrees $\hbox{deg}(N_{C^{\pm}/\sigma_\alpha})$ of  curves $C^\pm$  that are wrapped by the sections in the $I_2$ fiber,
 represented by $r,s,m$ and $k$ in the table, have been left un-constrained, i.e. we do not impose that one of the curves in the $I_2$ fiber is contractible. 
  The intersections of $C^{\pm}$ with the 
 section are calculated using Corollary \ref{cor:DC}, the only input being the values of $r,s,m$ and $k$. 
 In the table, these  intersections with $\sigma_0$ and $\sigma_1$ are shown below each fiber type, and the $U(1)$ charge is again computed using the Shioda map $S(\sigma_1) = \sigma_1 - \sigma_0$. 
It would be interesting to generalize the results of \cite{Reid, Laufer} to four-folds in order to further constraint the normal bundles and thereby the $U(1)$ charge values in four-folds.


\section{Codimension three Fibers and Yukawa Couplings}
\label{sec:CodimThree}

In elliptic Calabi--Yau four-folds there are codimension three points in the
base of the fibration, above which the codimension two fibers can enhance
further, i.e. again some of the rational curves become reducible. From an
F-theory point of view, the fibers above such points in the base are of
interest as they give rise to coupling of matter fields in Yukawa
interactions.


\subsection{Codimension three Fibers and Phases}

The codimension three fibers for $SU(5)$ with $\overline{\bf 5}$ and ${\bf 10}$ matter were determined from the box graphs using mutual compatibility of the 
relative cones of effective curves in \cite{Hayashi:2014kca}. The Yukawa couplings ${\bf 10} \times {\bf 10} \times {\bf 5}$ and ${\bf \bar{5}} \times {\bf \bar{5}} \times {\bf 10}$ occur at codimension three loci, where the fiber enhances from the $I_6$ and $I_1^*$ fibers, that realize the fundamental and anti-symmetric matter, to   monodromy-reduced $IV^*$ or $I_2^*$  fibers, which correspond to a local enhancement of the symmetry to $E_6$ and $SO(12)$, respectively.  Physically, the Yukawas can be thought of as generated by the splitting of matter curves into other matter curves, plus, potentially, roots \cite{MS}.

Here we will focus on the coupling between singlets and two fundamentals: ${\bf
5} \times \bar{\bf 5}\times  {\bf1}$. These are realized above codimension
three loci with an $SU(7)$ enhancement.  This is the simplest instance in
which the fibers (without the presence of additional sections) are not
standard Kodaira fibers in codimension three, but are monodromy-reduced, i.e.
the fiber is not $I_7$, but remains $I_6$. However, if there is a suitable
additional section, there is an enhancement to a full $I_7$ fiber \cite{EY,
Lawrie:2012gg}. 

We will now explain how the box graphs can be used to determine the consistent codimension three fibers. The analysis works for general types of fibers, but we will concentrate here on $SU(5)$ with $\overline{\bf 5}$ matter, i.e. the phases and fibers shown in figure \ref{fig:SU5BoxGraphs5}. 
As before, $F_i$ are the rational curves associated to the simple roots of $SU(5)$. First consider two codimension two $I_6$ fibers, which are characterized by the splitting 
\be
F_i \rightarrow C^{T +}_i + C^{T-}_i \,,\qquad 
F_j \rightarrow C^{B+}_j + C^{B-}_j \,.
\ee
The superscripts $T$op and $B$ottom label the curves in the two $I_6$ fibers in codimension two.
The combined phase is obtained by stacking the box graphs for each $I_6$ fiber on top of each other.  
Representation theoretically we are looking at the decomposition
\be
\ba
\mathfrak{su}(7) \quad \rightarrow & \quad \mathfrak{su}(5) \oplus \mathfrak{su}(2) \oplus \mathfrak{u}(1) \,.
\ea
\ee
Denote by $\widetilde{F}$ the  curve associated to the simple root
$\widetilde{\alpha}$ of the $\mathfrak{su}(2)$. Then in the combined box graph
this acts between the two layers, from the bottom to the top layer, e.g. 
\be
\includegraphics[width=4cm]{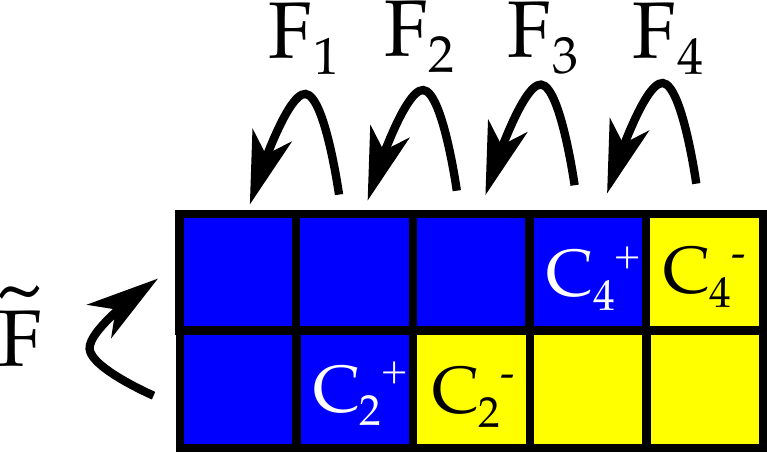} \,.
\ee
The combined box graphs need to satisfy both the flow rules for the $SU(5)$, as well as compatibility with the action of this additional root.

\begin{figure}
\centering
\includegraphics[width=16cm]{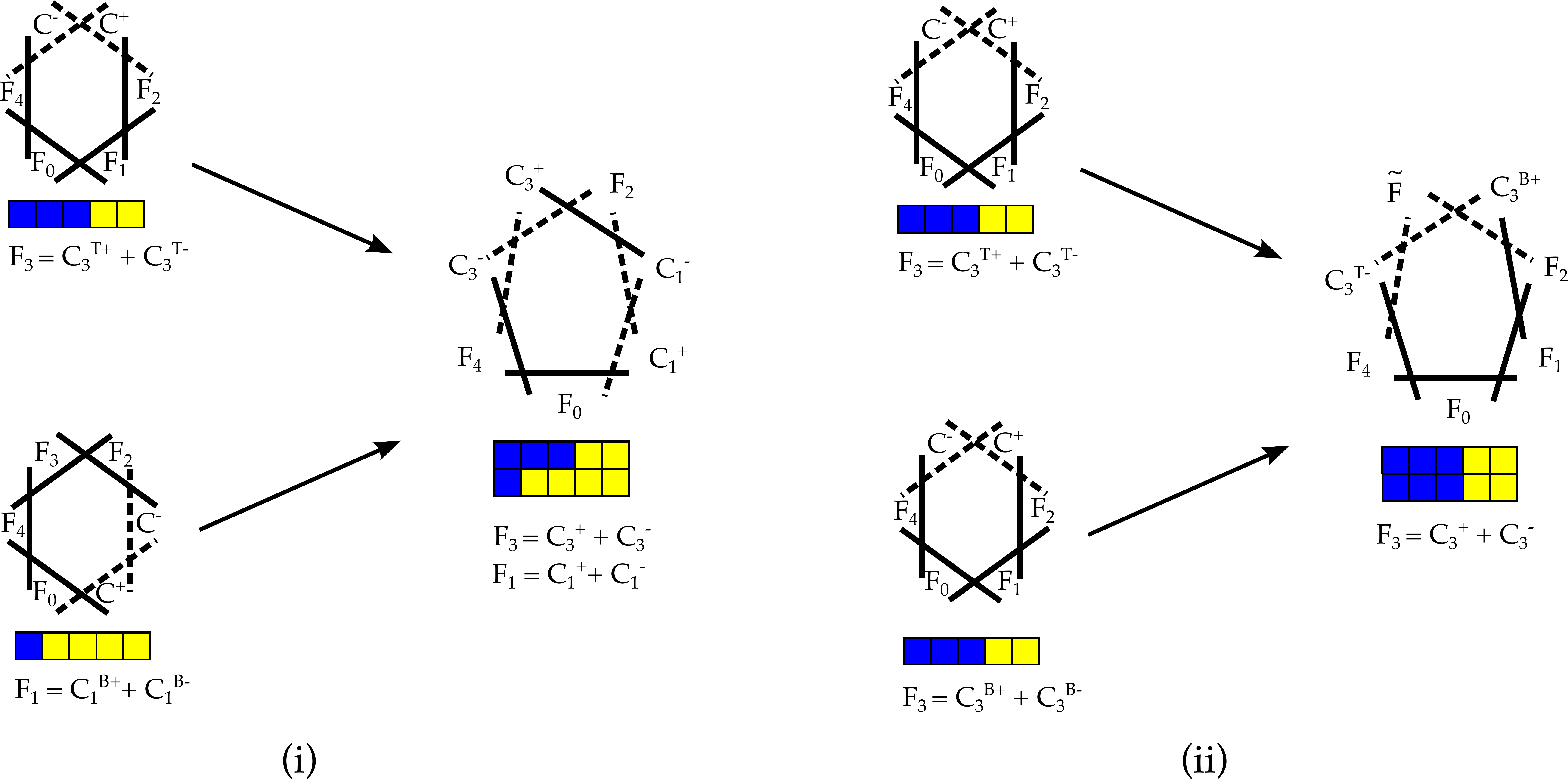}
\caption{Construction of the fiber in codimension three, where two codimension
  two $I_6$ fibers in the phases/box graphs shown on the left, collide to give
  a fiber of type $I_7$ in codimension three. The box graph for the $I_7$ is
  shown on the right of each figure. Figure (i) shows the codimension three
  enhancement when the two $I_6$ fibers are in different phases/box graphs,
  whereas in (ii) they are in the same phase. 
Note that for each of these enhancements it is necessary to have at least one extra rational section.}
\label{fig:I7Box}
\end{figure}

Let us first assume $i\not=j$. In this case, e.g. shown in figure \ref{fig:I7Box}, both $F_i$ and $F_j$ are reducible, and the extremal generators of the relative
cone of curves are 
\be\label{SU7Cone}
C_i^{T+}, \ C_i^{T-} , \ C_j^{B+}, \ C_j^{B-}, \  F_k,\ k\not= i,j \,. 
\ee
In particular $\widetilde{F}$ is not extremal. The resulting fiber is obtained applying similar rules to the standard box graph analysis, summarized in section \ref{sec:Box} (for more details on how the fiber is determined from the graph we refer the reader to \cite{Hayashi:2014kca, Braun:2014kla, ABSSN}) and exemplified in part (i) of figure \ref{fig:I7Box}. 

For $i=j$, the phases of the two $I_6$ fibers agree, and the extremal generators are
\be
C_i^{T-},\ C_j^{B+},\  \widetilde{F}, \   F_k,\ k\not= i,j  \,,
\ee
where $\widetilde{F}$ remains irreducible, and the curves in the $I_6$ fibers, which became reducible, split as follows 
\be\ba\label{SamePhaseSplit}
C_i^{T+} \quad &\rightarrow \quad C_i^{B+} + \widetilde{F}\cr
C_j^{B-} \quad &\rightarrow \quad C_j^{T-} + \widetilde{F}\,.
\ea\ee
Note that this is the splitting from the $I_6$ Top and Bottom codimension two
fibers respectively. 
The rational curves in the fiber in codimension three  intersect again in an $I_7$ fiber, which is shown in part (ii) of figure \ref{fig:I7Box}. 

Let us reemphasize that in both these cases, it is paramount that the fiber has an additional rational section, as otherwise there is a monodromy reduction from $I_7$ to $I_6$.

\subsection{Codimension three Fibers with Rational Sections}
\label{sec:CodimThreeConsts}

Like in the splitting from codimension one to two that we analyzed in section \ref{sec:Codim2Analysis}, we require various conditions on the intersection numbers of the section $\sigma$ with the fiber components to be retained, when passing from codimension two to three: 
\begin{enumerate}
\item The section $\sigma$ intersects the fiber as $\sigma \cdot_Y \hbox {Fiber} =1$.

\item  Let $C$ be a rational curve in the fiber, which remains irreducible when passing from codimension two to codimension three, and let $S_C \not\subset \sigma$, i.e. matter surface obtained by fibering 
$C$  over the matter locus is not contained in the section, but let $C$ be
contained in $\sigma$ in codimension three.
Then $\sigma \cdot_Y C$ needs to be preserved in codimension three.


\item If $S_C\subset\sigma$ in codimension two, and $C \rightarrow C^+ + C^-$ then   by Corollary \ref{cor:DC} 
\be
\sigma \cdot_Y C = -4 - \hbox{deg} (N_{C^+/\sigma} )- \hbox{deg} (N_{C^-/\sigma} )\,.
\ee 
\end{enumerate}
Note that, obviously, a curve that is contained in the codimension two fiber
continues to be contained in the codimension three fiber to which the
codimension two fiber degenerates.  The compatibility between codimension
two and three has to be imposed for {\it every} codimension two fiber whose
codimension two locus in the base passes through the codimension three point
in question (i.e. all the codimension two fibers that correspond to matter
that participates in the Yukawa coupling).

Note also, that the constraints on  the normal bundle derived  for four-folds  $Y$ in section \ref{sec:four-folds}   need to be respected. The normal bundle of the rational curves in the fiber have to be such that they embed into the normal bundle $N_{C/Y}$. From Theorem \ref{thm:four-foldNB} observe that the normal bundles of $F_i$ in the four-fold $Y$ are 
\be
N_{F_i/Y} = \mathcal{O} \oplus \mathcal{O} \oplus \mathcal{O}(-2) \,, 
\ee 
and the normal bundles of the curves $C_i^\pm$, obtained from the splitting $F_i \rightarrow C^+_i + C^-_i$, which correspond to weights of the fundamental representation, are 
\be
N_{C_i^\pm/Y} = \mathcal{O} \oplus \mathcal{O}(-1) \oplus \mathcal{O}(-1) \,. 
\ee


\subsection{Charged Singlet Yukawas}

\begin{figure}
\centering
\includegraphics[width=16cm]{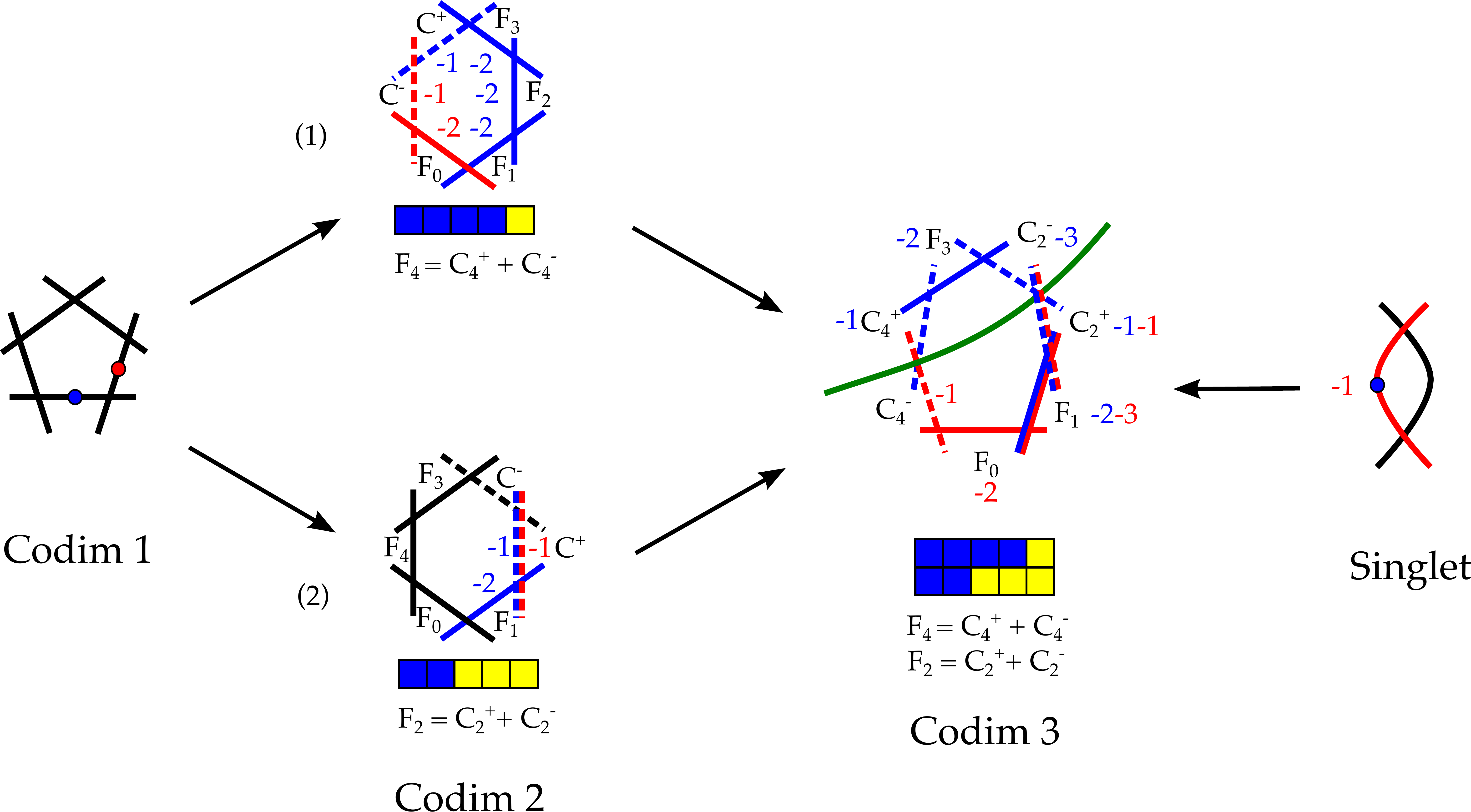}
\caption{Example of a codimension three fiber with one additional rational
  section where the codimension two fibers are in different phases. 
Codimension one:  $I_5$ fiber with two sections, $\sigma_0$ (blue) and $\sigma_1$ (red). 
Codimension two:  $I_6$ fiber with sections as indicated (the configuration is
described in (\ref{Config12})), corresponding to $\overline{\bf 5}$ matter,
with charge $11$ and charge $1$, respectively. 
Here the two $I_6$ fibers are in {\it different} phases. The curves, $C^\pm$,
into which the $F_i$ that become reducible in codimension two have split are
shown by dotted lines.  Colored fiber components correspond to
rational curves that are contained in the respective sections. The numbers
next to these indicate the degree of the normal bundle of these curves in the
section.  Codimension three: $I_7$ fiber with sections, as well as the
corresponding box graph, obtained by stacking the box graphs associated to the
codimension two fibers. Again, fiber components that are contained in the
sections $\sigma_{0/1}$ are colored accordingly. The green line indicates
where the $I_7$ fiber needs to be ``cut" to determine the singlet that couples
to the two fundamental matter multiplets. On the far right the $I_2$  fiber
that realizes this singlet is shown.  \label{fig:SU7WithSection}} 
\end{figure}

We now consider the Yukawa couplings that are realized by codimension three
enhancements to $I_7$ involving charged singlets, i.e. $\overline{\bf 5}
\times {\bf 5} \times {\bf 1}$ couplings.
First consider the case of the two $I_6$ fibers in different phases. 
An example is shown in figure \ref{fig:SU7WithSection}. Starting with an $I_5^{(0|1)}$ model at the far left in codimension one, the next two entries correspond to the codimension two fibers. The blue/red colored fibers  indicate  the rational curves that are contained in the sections $\sigma_0$ and $\sigma_1$, respectively. 
From figure \ref{fig:0s1Charges} the configurations in codimension two, labeled (1) and (2), correspond to fundamental matter with $U(1)$ charges
\be
q(\overline{\bf 5}^{(1)})  =  +11 \,,\qquad 
q(\overline{\bf 5}^{(2)})  =  +1 \,.
\ee
The codimension three fiber when these two collide can be determined by imposing the requirements in section \ref{sec:CodimThreeConsts}. 
The compatibility conditions have to be satisfied for both of the two  $I_6$ fibers enhancing to the $I_7$ fiber. 
For instance, consider the $I_6$ fiber (1). We can characterize the configuration by 
For instance, the configurations of the  $I_6$ fibers (1) and (2) can be
characterized by
\be\label{Config12}
\begin{array}{lll}
(1)  :  & F_1, F_2, F_3 \subset \sigma_0  &\hbox{deg}(N_{F_i/\sigma_0}) =-2  \cr
& C_4^+ \subset \sigma_0 & \hbox{deg}(N_{C_4^+/\sigma_0}) =-1 \cr
& C_4^{-}, F_0 \not\subset \sigma_0 & \sigma_0 \cdot_{Y} C_4^{-} = \sigma_0 \cdot_Y F_0 =1 \cr
& F_0\subset \sigma_1 & \hbox{deg}(N_{F_0/\sigma_1}) =-2 \cr
& C_4^{-}\subset \sigma_1   & \hbox{deg}(N_{C_4^-/\sigma_1}) =-1 \cr
& C_4^+, F_1  \not\subset \sigma_1 & \sigma_1 \cdot_Y C_4^+ = \sigma_1 \cdot_Y F_1 =1 \cr
&&\cr
(2)  :  & F_1 \subset \sigma_0  &\hbox{deg}(N_{F_1/\sigma_0}) =-2  \cr
& C_2^+ \subset \sigma_0 & \hbox{deg}(N_{C_2^+/\sigma_0}) =-1 \cr
& C_2^{-}, F_0 \not\subset \sigma_0 & \sigma_0 \cdot_{Y} C_2^{-} = \sigma_0 \cdot_Y F_0 =1 \cr
& C_2^{+}\subset \sigma_1   & \hbox{deg}(N_{C_2^+/\sigma_1}) =-1 \cr
& C_2^-, F_1  \not\subset \sigma_1 & \sigma_1 \cdot_Y C_2^- = \sigma_1 \cdot_Y F_1 =1 \,.
\end{array}
\ee
The fibers split as determined by the box graphs, and applying the
compatibility conditions on the sections in codimension three determines the
fibers \footnote{Note that the codimension three fiber is not unique, but
  only unique in terms of the intersection numbers. This is similar to the
  codimension two fibers, where, for example, $\sigma \cdot_Y F=1$ can be
  either realized in terms of a transverse intersection, or in terms of
  $F\subset \sigma$ with $\hbox{deg}(N_{F/\sigma}) =-3$. These ambiguities
however do not change the charges or, in the case of codimension three, the
possible Yukawa couplings.}, e.g. it is clear that all the components that are
contained in either of the codimension two fibers have to continue to be
contained in the sections. Furthermore, imposing that the intersection numbers
and normal bundles are consistent, results in the configuration shown in
figure \ref{fig:SU7WithSection}.

From the $I_7$ we can obtain the $I_2$ fiber and thereby the singlet that participates in the Yukawa coupling. 
As we consider two $I_6$ fibers in different phases $\widetilde{F}$ is not extremal, see (\ref{SU7Cone}) for the configuration in figure \ref{fig:SU7WithSection}, but is given in terms of 
\be\label{Ftilde}
\widetilde{F} \quad \rightarrow \quad  C_4^+ + F_3 + C_2^-  \,,
\ee
which can be read off from the box graph 
or directly from the fiber. 
In figure \ref{SU7Cone} the component $\widetilde{F}$ is shown, separated from
its conjugate component, by the green cut through the $I_7$ fiber. 
The combination in equation (\ref{Ftilde})
are uncharged under the GUT group $SU(5)$, i.e. geometrically 
\be
D_{F_i} \cdot_Y \widetilde{F} =0 \,,\qquad i=0, \cdots, 4 \,,
\ee
as required for a singlet, but intersects the sections as 
\be\ba
\sigma_0 \cdot_Y \widetilde{F}&= \sigma_0\cdot_Y (C_4^+ + F_3 + C_2^- ) = -1 + 0 + 1 =0 \cr
\sigma_1 \cdot_Y \widetilde{F}&= \sigma_1\cdot_Y (C_4^+ + F_3 + C_2^- ) = 1 + 0 + 1 =2 \,. 
\ea\ee
Likewise we can consider the conjugate field, given by the curve (so to speak the other half of the cut $I_7$ fiber)
\be
\overline{\widetilde{F}} \quad \rightarrow \quad   C_4^- + F_0 + F_1 + C_2^+\,,
\ee
which intersects the sections as 
\be\ba
\sigma_0 \cdot_Y \overline{\widetilde{F}}&= \sigma_0\cdot_Y (C_4^- + F_0 + F_1 + C_2^+) = 1+ 1+ 0 -1=1 \cr
\sigma_1 \cdot_Y \overline{\widetilde{F}}&= \sigma_1\cdot_Y (C_4^- + F_0 + F_1 + C_2^+) = -1+ 0+ 1+ -1 =-1 \,. 
\ea\ee
Applying Shioda (and multiplying by 5 for the $SU(5)$ normalization) we obtain that the charges of these singlets are indeed $\mp10$, as required for the coupling to the matter of charge $\pm 11$ and $\mp1$, i.e. $\overline{\bf 5}_{11} \times {\bf 5}_{-1} \times {\bf 1}_{-10}$.

\begin{figure}
\centering
\includegraphics[width=16cm]{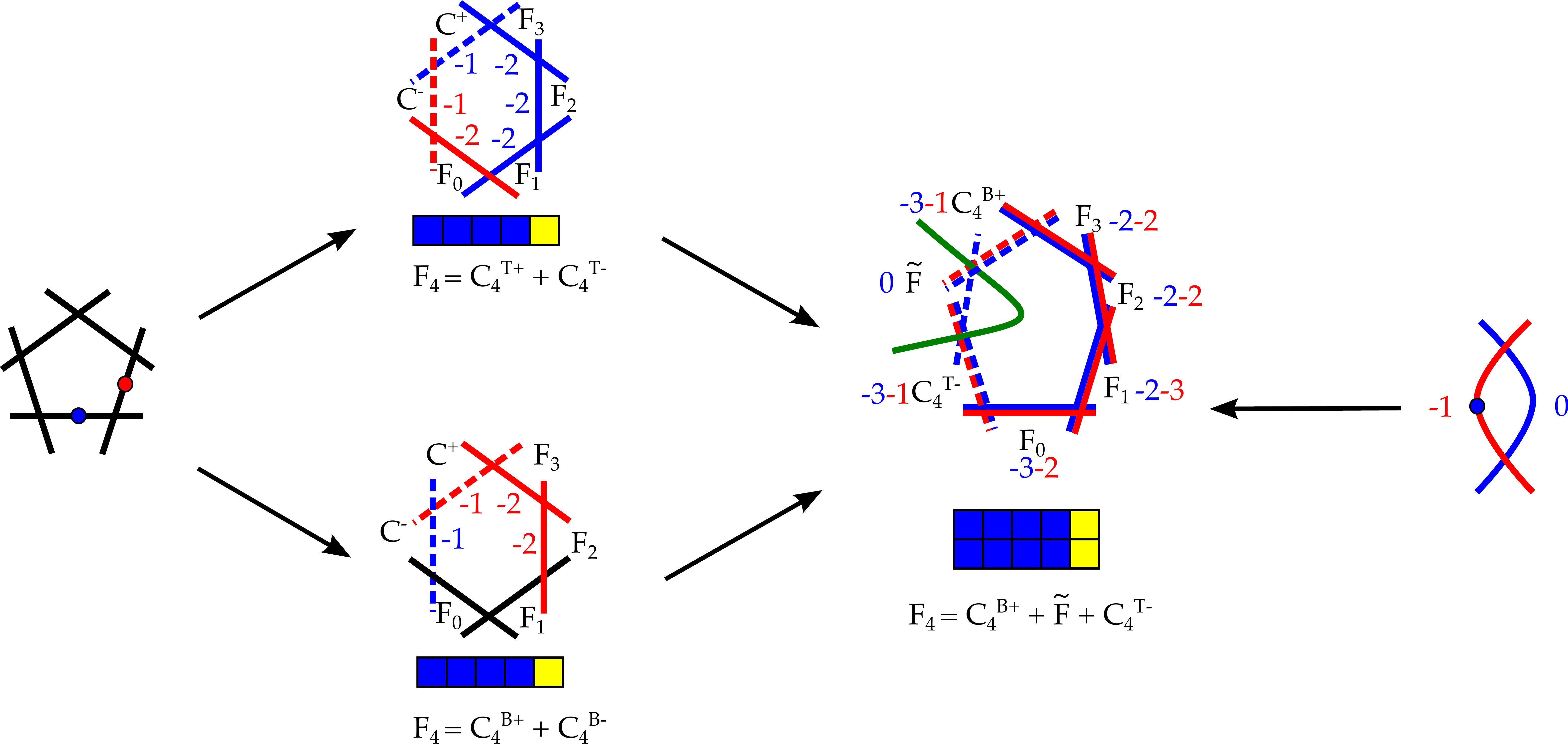}
\caption{
Example of a codimension three fiber with one additional rational section,
where the codimension two fibers are in the same phase. 
The matter corresponds to charge $+11$ (T) and charge $-9$ (B) $\overline{\bf
5}$ matter and a singlet of charge $20$.   
The notation is as in figure \ref{fig:SU7WithSection}. 
\label{fig:SU7WithSectionSameP}}
\end{figure}

Finally, let us briefly comment on the case when the two $I_6$ fibers are in the same phase, an example is shown in figure \ref{fig:SU7WithSectionSameP}. 
The charges are 
\be\label{qsamePCharges}
q(\overline{\bf 5}^T) = +11 \qquad 
q(\overline{\bf 5}^B) = -9 \,. 
\ee
The splitting from codimension two to codimension three of the fiber components is that
in (\ref{SamePhaseSplit}) and part (ii) in figure \ref{fig:I7Box}, and
$\tilde{F}$ is an irreducible, new fiber component.  Again we impose
ompatibility with the section configurations in codimensions two and three,
as well as consistent normal bundle configurations.  The resulting
codimension three fiber is shown in  figure \ref{fig:SU7WithSectionSameP}.
The singlet charge is obtained by intersecting $\widetilde{F}$ with the
sections.  Note, that $\widetilde{F} \cdot_Y D_{F_i}=0$, which is consistent
with this being the singlet, and 
\be\ba
\sigma_0 \cdot_Y \widetilde{F}&= -2  \cr
\sigma_1 \cdot_Y \widetilde{F}& =2 \,. 
\ea\ee
Likewise, the conjugate field is
\be
\overline{\widetilde{F}} \quad \rightarrow\quad  C_4^{B+} + F_3 + F_2 + F_1 + F_0 + C_4^{T-}
\ee
and 
\be
\ba
\sigma_0 \cdot_Y \overline{\widetilde{F}}&= 3  \cr
\sigma_1 \cdot_Y \overline{\widetilde{F}}& =-1 \,. 
\ea
\ee
The associated $I_2$ fiber, which realizes these intersections, is shown in
figure \ref{fig:SU7WithSectionSameP}, and matches the required charge of $20$
from (\ref{qsamePCharges}), such that the coupling $\overline{\bf 5}_{-9}{\bf 5}_{-11} {\bf
1}_{20}$ is uncharged.

\section{Multiple $U(1)$s and Higgsing}
\label{sec:multiU1s}
The analysis shown in the preceding sections has been for a single additional
rational section of the elliptic fibration, which generates one $U(1)$
symmetry. This can be extended to the case of elliptic fibrations with
multiple rational sections, which generates multiple $U(1)$ symmetries. 
Furthermore, based on the classification of singlets, we can consider
the possible Higgsings of the abelian symmetry to discrete subgroups. The case
of partial Higgsing of multiple $U(1)$s is left for future work. 

\subsection{Multiple $U(1)$s  and Rational Sections}

The set of rational sections, $\sigma_i$, in an elliptic fibration generate
the Mordell--Weil group, which is a finitely generated abelian group
\be 
\mathbb{Z}^n \oplus \Gamma \,,
\ee
where $n$ is the number of rational sections in the fibration and $\Gamma$ is
the discrete part of the Mordell--Weil group,  which we do not consider here.
The zero-section $\sigma_0$ is the origin of the Mordell--Weil group, and
$\sigma_i$, $i=1, \cdots, n$, are the generators of the free part. 

\begin{figure}
\centering
 \includegraphics[width=13cm]{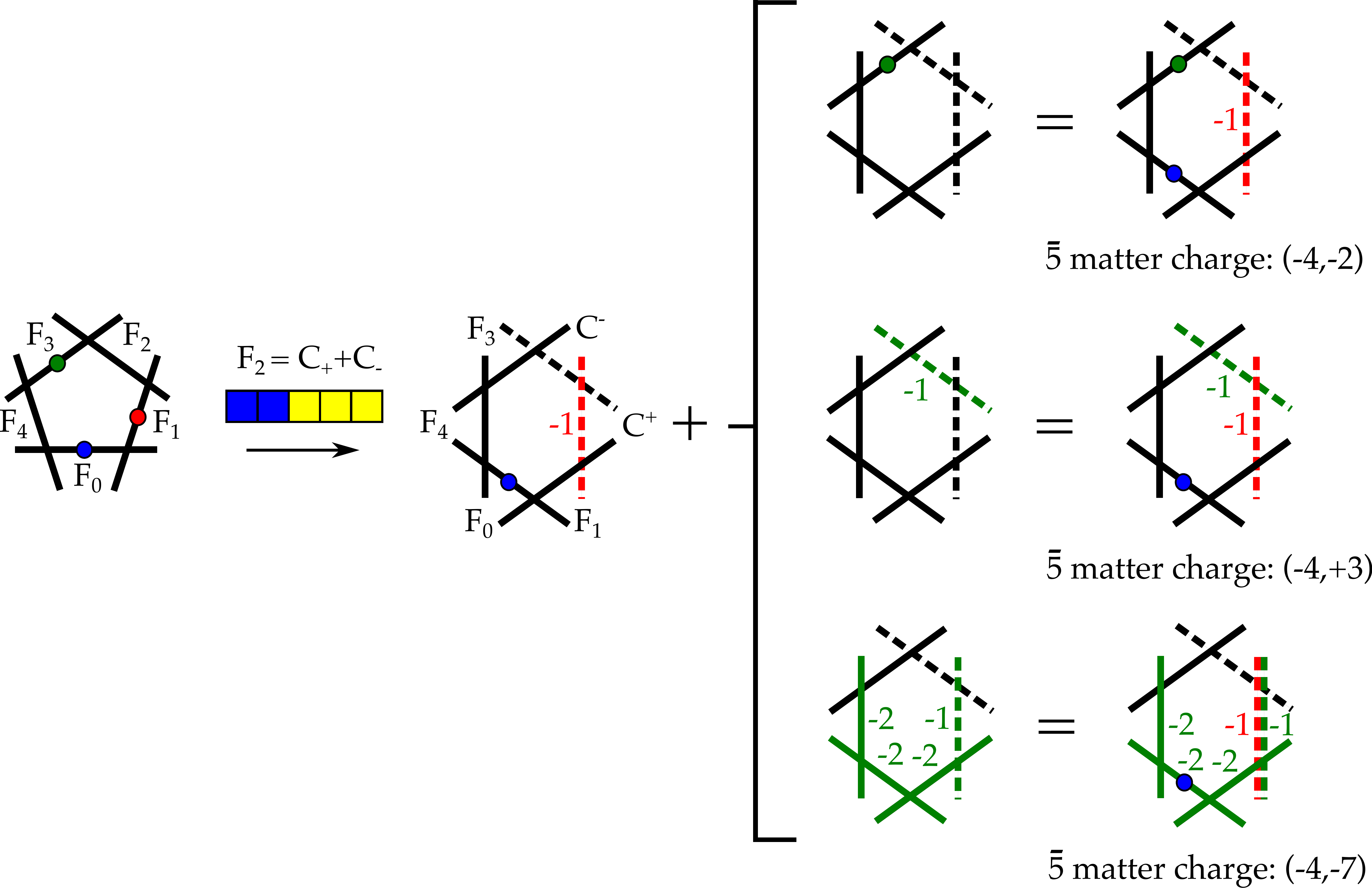} 
 \caption{Example set of $\bf \bar{5}$ charges for an $I_5^{(0|1||2)}$ model in the phase where $F_2$ splits. 
 The sections $\sigma_0/ \sigma_1/\sigma_2$ are colored blue/red/green.
 The configurations for $\sigma_0$ and $\sigma_1$ are fixed to give charge $-4$ under $U(1)_1$. Combining this with the possible configurations for $\sigma_2$ gives the set of charges under $U(1)_2$.}
 \label{fig:MultiU1s}
\end{figure}

The key point to note is that our analysis for one rational section applies
independently to each generator of the free part of the Mordell--Weil group.
The set of configurations for each section in an $I_k \rightarrow I_{k+1}$
enhancement is therefore just given by those in figures \ref{fig:InFib1} and
\ref{fig:InFib2}, where the section, $\sigma_i$, is taken to intersect
$F_{m_i}$ in codimension
one.  One can then construct the Shioda map, $S(\sigma_i)$ for each section,
which defines the generator of the abelian gauge factor $U(1)_i$.  Let us
consider an example with two additional rational sections, $\sigma_1$ and
$\sigma_2$, where the codimension one fiber type is $I_5^{(0|1||2)}$, as
depicted in figure \ref{fig:MultiU1s}. For each phase, the possible charges
for $\bf \bar{5}$ matter under $U(1)_1$, are given in figure
\ref{fig:0s1Charges} (modulo the fully wrapped configurations). To each of
these one can overlay a configuration for $\sigma_2$ in the same phase and
compute the charge under $U(1)_2$ by  intersection $C^{\pm}$ with 
\begin{equation}
S(\sigma_2) = 5 \sigma_2 - 5 \sigma_0 + 2D_{F_1} + 4 D_{F_2} + 6 D_{F_3} + 3 D_{F_4} \,.
\end{equation}
Further, consider $\sigma_1$ such that $q_{\bf \bar{5}} = -4$ in the
phase where $F_2$ splits. This is shown in figure \ref{fig:MultiU1s}. This
configuration can be combined with any one of the three possible
configurations for $\sigma_2$, each of which gives a different charge under
$U(1)_2$. Repeating this for every configuration in all phases gives the full
set of charges for this codimension one fiber.  Following this procedure we
determine all possible combinations, and it can be shown that all known
explicit realizations of models with multiple $U(1)$ factors form a subclass
of the models obtained here. 


\subsection{Higgsing and Discrete Symmetries}

In section \ref{sec:singlets} the set of possible codimension two $I_2$ fibers with rational sections were determined along with the corresponding singlet charges.
One application of this result is to use such $U(1)$ charged singlets to Higgs the $U(1)$ symmetry to a discrete subgroup $\mathbb{Z}_q$. 
Examples of such Higgsing have recently been considered in \cite{Mayrhofer:2014haa,Mayrhofer:2014laa,Cvetic:2015moa}\footnote{Other discussions of discrete symmetries in F-theory compactifications without section (i.e. genus one fibrations) have appeared in \cite{Klevers:2014bqa, Morrison:2014era,Braun:2014oya,Anderson:2014yva,Garcia-Etxebarria:2014qua}} 
for $q = 2,3$. Though Higgsing different singlet configurations of the same
charge leads to the same discrete symmetry in the F-theory compactification, this was shown not to be the case upon the circle reduction to M-theory. 
This can be seen field theoretically by reducing F-theory in 6d along an $S^1$ to M-theory in 5d \cite{Mayrhofer:2014haa,Mayrhofer:2014laa, Cvetic:2015moa, Anderson:2014yva,Garcia-Etxebarria:2014qua}.
Turning on a vacuum expectation value for the Higgs field, $S_q$, of charge $q$ breaks the $U(1)$ in F-theory to $\mathbb{Z}_{q}$. 
Starting in 6d, and compactifying to 5d on a circle, the masses of the Kaluza-Klein modes are labeled by the charge $q$, the mode number (or KK-charge) $n$ and the Wilson line $\xi$ along the circle 
\be 
m_n^q = | q \xi + n| \,.
\ee
The massless spectrum depends on the value of $\xi$ and for $\xi = k/q$ with
integral $k$ the KK-charge  $n=-k$ becomes massless.
There are  $q$ distinct values for the Wilson line, modulo the action of $SL_2\mathbb{Z}$, which correspond to distinct M-theory vacua, between which the Tate-Shafarevich group acts \cite{Braun:2014oya}.

Equipped with the set of $I_2$ fibers and their corresponding charges, given in figure \ref{fig:SingletCharges}, we can now consider the Higgsing with more general singlet configurations, with charges beyond $q=2,3$.
Furthermore, it is possible to determine for a fixed singlet charge $q$, the fibers which realize the $q$ different choices of 5d Higgs fields. Note that the KK-charge $n$ is computed by intersecting with the zero-section
\be
\sigma_0 \cdot_Y C^\pm = n^{\pm} \,.
\ee
That is, we look for configurations where $C^+$, or $C^-$, has intersections with $\sigma_0$ within the set
\be 
n^{\pm}= \sigma_0 \cdot_Y C^{\pm} \in \{0,\cdots, q-1 \} \hbox{ mod } q \,.
\ee
 The result is that for charges up to $q = 9$ it is always possible, by tuning the degree of the normal bundle of the curve $C^+$ in (\ref{CpNB}), to obtain curves in the $I_2$ fiber with the desired intersections with $\sigma_0$. 
 It would be interesting to study how these configurations are related via flop transitions such as in the case of $q=3$ studied in \cite{Cvetic:2015moa}. For charges $q \geq 10$ the set of KK-charges, which do not have a realization grows with $q$ and it would be interesting to explore how the other configurations could be realized.


\section{Discussion and Outlook}
\label{sec:Disc}

In this paper we determined the possible $U(1)$ charges of matter in F-theory
compactifications to four and six dimensions, by classifying the possible
configurations of  rational sections in codimension two fibers.  Our analysis
for charged matter in the fundamental and anti-symmetric representations of
$SU(n)$ in sections \ref{sec:SU5F} and \ref{sec:SU5A} holds for both Calabi--Yau
three- and four-folds.  The main inputs were the classification result of
codimension two fibers in \cite{Hayashi:2014kca} as well as constraints on
rational curves and their normal bundles in Calabi--Yau varieties, as discussed
in section \ref{sec:NormalBundles}.  There are various exciting directions for
future research.  
\begin{itemize}
\item Building complete models: \\
In our analysis we did not discuss constraints from charged matter Yukawa
couplings, only couplings between fundamental matter and singlets.  It would
be interesting to see whether codimension three constraints will provide
further conditions as to how various codimension two fiber types can co-exist
in a given model. The codimension three fibers and possible Coulomb phases
without additional sections were derived already in \cite{Hayashi:2013lra,
Hayashi:2014kca} and it would be interesting to generalize this to models with
rational sections.  Clearly further constraints that would select subsets of
compatible codimension two fibers would also be of interest for model
building, and could play an important role for a  systematic study of the
phenomenology similar to \cite{Dolan:2011iu, Dolan:2011aq,
Krippendorf:2014xba}.

\item Explicit realizations: \\
The charges and fibers in explicitly known fibrations with various numbers of
abelian factors \cite{Morrison:2012ei, Borchmann:2013jwa, Cvetic:2013nia,
Borchmann:2013hta, Cvetic:2013jta, Cvetic:2013qsa, Braun:2013yti,
Braun:2013nqa, Braun:2014qka, Klevers:2014bqa, Grimm:2010ez, Braun:2011zm,
Mayrhofer:2012zy, Morrison:2014era, Kuntzler:2014ila, Lawrie:2014uya},
as well as the matter charges in the singlet-extended $E_8$ model \cite{Baume:2015wia}, form a strict subset of the fibers that we have found in the present paper. 
It would be extremely interesting to determine realizations for the new fiber types, including the singlets that we classified in section \ref{sec:singlets}.

\item Flops: \\
Our classification assumes that the section, which is a divisor in the
Calabi--Yau variety, is smooth. We have observed in section \ref{sec:flops}
that, by flopping codimension two fibers with certain section configurations,
the resulting fiber has a section which self-intersects in a curve in
the fiber, and is thus no longer smooth. It would be very interesting to
study such flops concretely, to determine the complete flop chain when the
allowed configurations include such singular sections. It would also be
interesting to study the flops for the $I_2$ fibers realizing different
KK-charges for the singlets, generalizing the analysis for charge 3 singlets
in \cite{Cvetic:2015moa}.

\item Singlets: \\
Unlike the charged matter, the analysis for the  classification of singlets in
section \ref{sec:singlets} is comprehensive only for Calabi--Yau three-folds,
as we impose that one of the curves in the $I_2$ fiber should be contractible.
A similar criterion for contractibility for higher-dimensional Calabi--Yau
varieties is not known to us, however  we have determined all possible
codimension two $I_2$ fibers with rational section, without necessarily
requiring contractibility of the curves, in table \ref{tab:I2fibers}.  It
would be interesting to determine a contractibility criterion on the normal
bundle of rational curves in four-folds and to thereby  constrain the singlet
configurations in table \ref{tab:I2fibers} to the allowed set in four-folds.
Note that no such disclaimer holds for the charged matter in sections
\ref{sec:SU5F} and \ref{sec:SU5A}, which do not rely on imposing any
contractibility on the curves, and our  results hold for codimension two in
three- and four-folds alike. 

\item Higgsing and discrete groups:\\
We determined the singlet fibers for $U(1)$ charges up until $q=9$, including
realizations for each KK-charge.  This allows a comprehensive study of
discrete symmetries by giving vacuum expectation values to these singlets, and
it would be interesting to determine the effects on the low energy theories,
for instance like in \cite{Ibanez:1991pr}.

\end{itemize}


\subsection*{Acknowledgements}

We thank Andreas Braun, Herb Clemens, Thomas Grimm, Denis Klevers, Dave Morrison, Eran Palti, Damiano Sacco and in particular Dima Panov for numerous discussions. 
This work is supported in part by the STFC grant ST/J002798/1. 

\newpage


\appendix
\section{Details for Anti-Symmetric Matter}
\label{app:I1sSplittings}

In this appendix the various details of the enhancements from $I_5$ to
$I_1^*$, which gives rise to matter in the ${\bf 10}$ representation of
$SU(5)$, are collected. Tables \ref{tab:10SplitPart1} and
\ref{tab:10SplitPart2} list the sixteen different enhancements that can
occur, as determined in \cite{Hayashi:2014kca}, and represented by the appropriate
box graph. The possible $U(1)$ charges listed in section \ref{sec:SU5A} are
determined by studying each of these sixteen enhancements and asking in what
ways fiber curves, or collections of fiber curves, can be contained inside
the section, whilst remaining consistent with the intersection data in
codimension one. There are eleven qualitatively different ``splitting
types'', which were previously listed in section \ref{sec:SU5A}, and for each of these
it is determined what the possible configurations of curves in any rational
section for that particular splitting type are. 

\begin{figure}
  \centering
  \begin{subfigure}[b]{0.4\textwidth}
    \centering
    \includegraphics[scale=2.5]{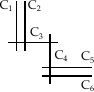}
    \caption{}
    \label{fig:I1sSchema}
  \end{subfigure}
  \begin{subfigure}[b]{0.4\textwidth}
    \centering
    \includegraphics[scale=2.5]{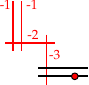}
    \caption{}
    \label{fig:I1sSchemaWrapped}
  \end{subfigure}
  \caption{(i) is a schematic depiction of an $I_1^*$ fiber and (ii) is this
  $I_1^*$ fiber in the configuration (1123--x). As usual if a component is
  colored red then it is contained inside the section, and the red integer
adjacent to the component is the degree of the normal bundle to that component
in the section. A red node indicates an additional transverse intersection
with the section.}
\end{figure}

\subsection{Codimension two $I_1^*$ Fibers}\label{app:wrapnotation}

For the purpose of this
appendix a new notation will need to be introduced to concisely summarize all
of the different configurations as there are many configurations that
realize the same intersection numbers of the curves with the section. Each
fiber will be displayed as in figure \ref{fig:I1sSchema}.
As such there is an obvious choice of ordering $C_1,\cdots,C_6$, where
these curves can be curves associated to either roots or weights.
If a curve $C_i$ is contained within the section it is such that
$\text{deg}(N_{C_i/\sigma}) \leq -1$ by Theorems \ref{thm:NormalBundleSES}
and \ref{thm:abNB}, and by the analysis it is also
known that this value always happens to be in the (negative) single digits.
The notation is then given by the string $(n_1n_2n_3n_4n_5n_6)$ where the
$n_i$ are
\begin{enumerate}
  \item[(i)] If $C_i$ is contained inside the section then $n_i = -
    \text{deg}(N_{C_i/\sigma})$.
  \item[(ii)] If $C_i$ is uncontained in the section and has an additional
    transverse intersection with the section then the $n_i$ is replaced by
    an ``x''. Additional here means that there is a transverse intersection
    that does not come from the intersection(s) of $C_i$ with another curve
    $C_j$ which is contained inside the section.
  \item[(iii)] If the curve $C_i$ is otherwise then the $n_i$ is replaced
    with an en-dash ``--''.
\end{enumerate}

Such a string completely determines the configuration, for example consider
the configuration (1123--x) on the fiber presented in figure
\ref{fig:I1sSchema}. Such a configuration is represented in figure
\ref{fig:I1sSchemaWrapped}. The string fixes that
\begin{itemize}
  \item $C_1, C_2, C_3, C_4 \subset \sigma$ with $\text{deg}(N_{C_1/\sigma}) =
    \text{deg}(N_{C_2/\sigma}) = -1$, $\text{deg}(N_{C_3/\sigma}) = -2$, and
    $\text{deg}(N_{C_4/\sigma}) = -3$.
  \item $C_5 \not\subset \sigma$ and $\sigma \cdot_Y C_5 = 1$ from the
    single intersection point between $C_5$ and the contained curve $C_4$.
  \item $C_6 \not\subset \sigma$ and $\sigma \cdot_Y C_6 = 2$ with one
    contribution from the intersection point of $C_6$ and $C_4$, and an
    additional contribution from the extra transverse intersection of the
    section with $C_6$.
\end{itemize}

\begin{table}
$
\begin{array}{|c|c|c|c||c|c|}\hline
\# &\hbox{Box Graph} & \hbox{Splitting} &  \hbox{Intersections} & I_5^{(0|1)} \, S_f \hbox{ values} & I_5^{(0||1)} \, S_f \hbox{ values} \cr\hline
&&&&&\cr
1 &  \multirow{2}{*}{\includegraphics[width=2cm]{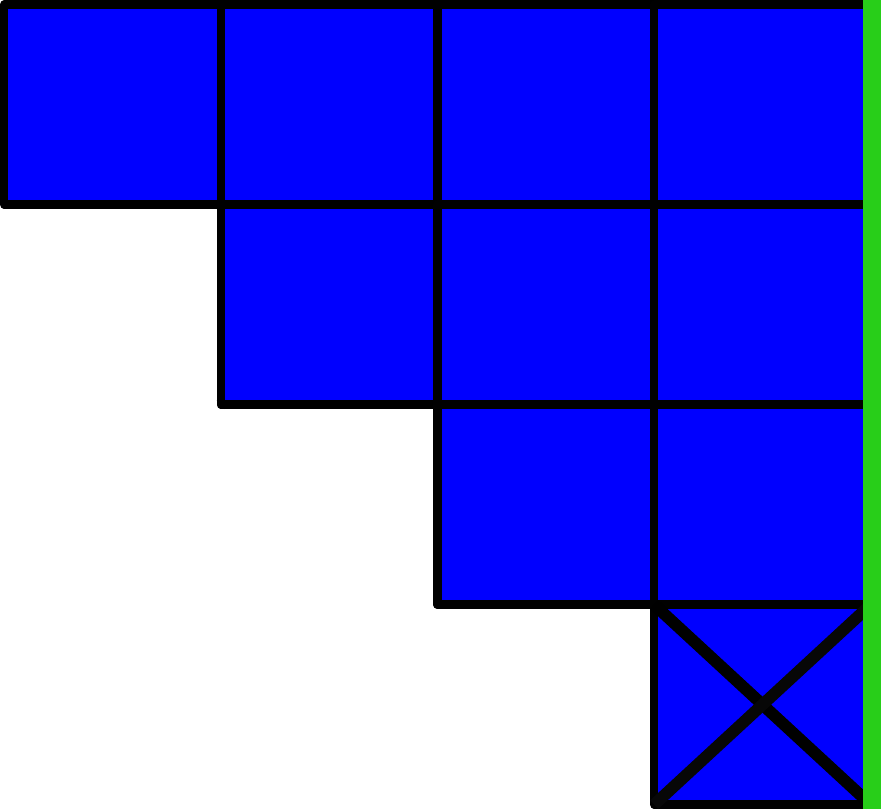}}
& F_0 \rightarrow   C_{4,5}^+ + F_2 + F_3 + \tilde{F}_0 
& \multirow{2}{*}{\includegraphics[height=2cm]{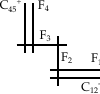}}
& S_f \cdot_Y C_{4,5}^+ = +2 & S_f \cdot_Y C_{4,5}^+ = +4
 \cr
& & \tilde{F}_0 = C_{1,2}^-& 
& S_f \cdot_Y C_{1,2}^-=+3 & S_f \cdot_Y C_{1,2}^-=+6 \cr
&&&&&\cr
&&&&&\cr\hline
&&&&&\cr
2&  \multirow{2}{*}{\includegraphics[width=2cm]{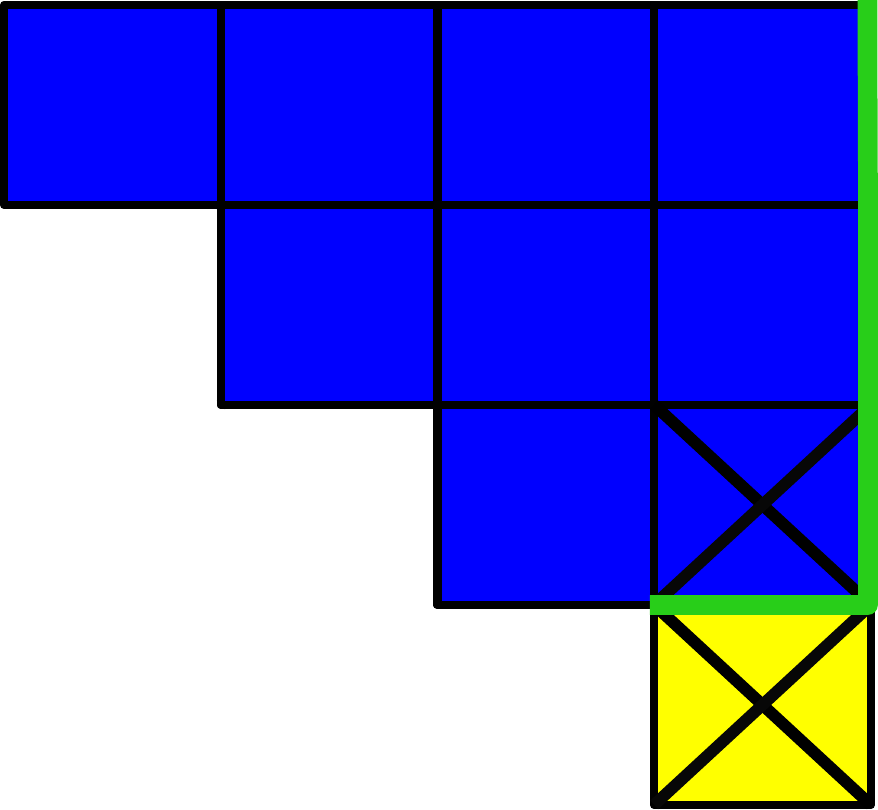} }
&  F_3 \rightarrow C_{3,5}^+ +  C_{4,5}^-  
& \multirow{2}{*}{\includegraphics[height=2cm]{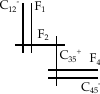}}
& S_f \cdot_Y C_{3,5}^+ = +2 & S_f \cdot_Y C_{3,5}^+ = +4\cr
&& F_0 \rightarrow C_{3,5}^+ +F_2 + \tilde{F}_0 
&& S_f \cdot_Y C_{4,5}^- = -2 & S_f \cdot_Y C_{4,5}^- = -4 \cr
&&\tilde{F}_0= C_{12}^- 
&& S_f \cdot_Y C_{1,2}^- = +3 & S_f \cdot_Y C_{1,2}^- = +6 \\
&&&&&\cr\hline
&&&&&\cr
 3 & \multirow{2}{*}{\includegraphics[width=2cm]{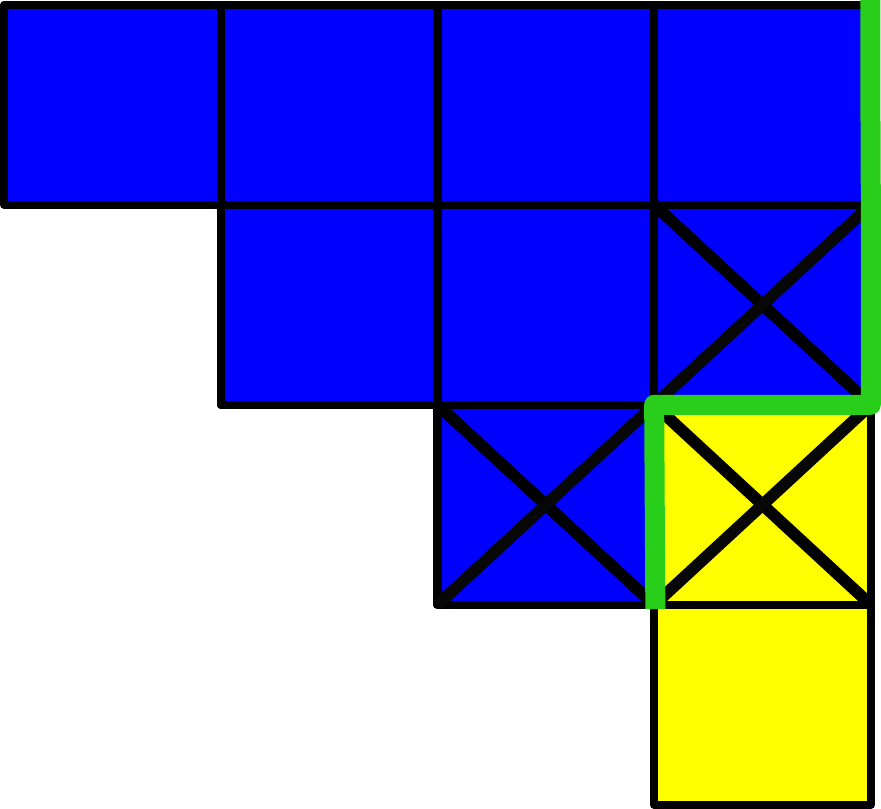} }
&F_2 \rightarrow C_{2,5}^+ +  C_{3,5}^-  
& \multirow{2}{*}{\includegraphics[height=2cm]{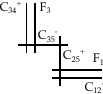}}
& S_f \cdot_Y C_{2,5}^+ = +2 & S_f \cdot_Y C_{2,5}^+ = -1 \cr
&& F_4 \rightarrow C_{3,4}^+ +  C_{3,5}^-  
&& S_f \cdot_Y C_{3,5}^- = -2 & S_f \cdot_Y C_{3,5}^- = -4 \cr
&& F_0 \rightarrow C_{2,5}^+ +  \tilde{F}_0  
&& S_f \cdot_Y C_{3,4}^+ = +2 & S_f \cdot_Y C_{3,4}^+ = +4\cr
&&\tilde{F}_0= C_{1,2}^- 
&& S_f \cdot_Y C_{1,2}^- = +3 & S_f \cdot_Y C_{1,2}^- = +6 \cr\hline
&&&&&\cr
 4& \multirow{2}{*}{\includegraphics[width=2cm]{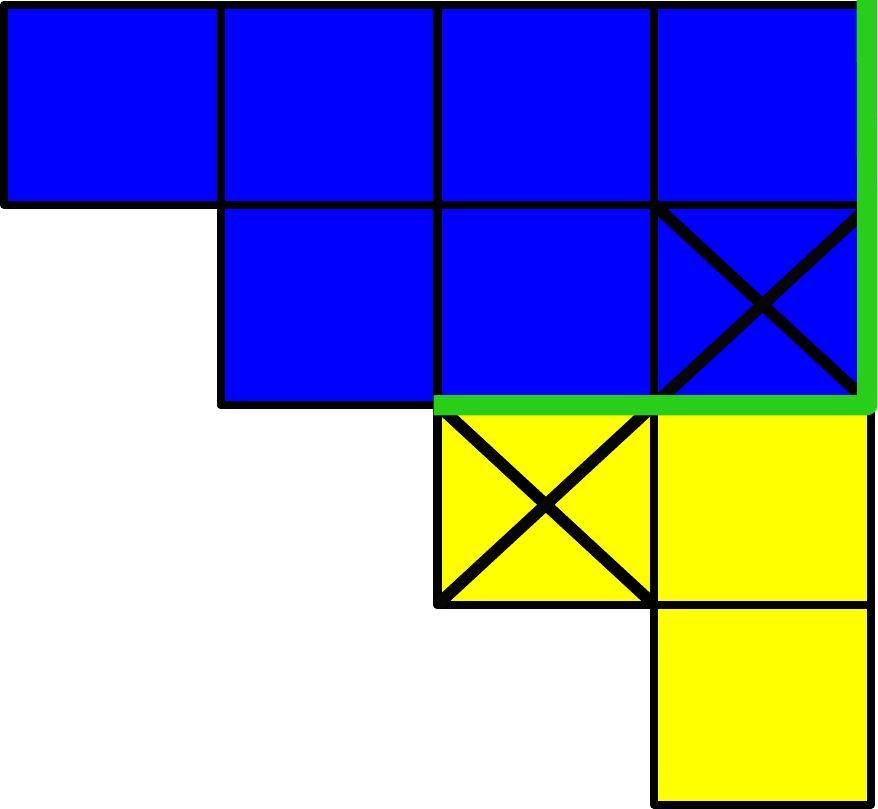} }
&  F_2 \rightarrow C_{2,5}^+ +  C_{3,4}^- + F_4 
& \multirow{2}{*}{\includegraphics[height=2cm]{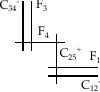}}
& S_f \cdot_Y C_{2,5}^+ = +2 & S_f \cdot_Y C_{2,5}^+ = -1 \cr
&& F_0 \rightarrow C_{2,5}^+ + \tilde{F}_0 
&& S_f \cdot_Y C_{3,4}^-=-2 & S_f \cdot_Y C_{3,4}^-=-4 \cr
&&\tilde{F}_0= C_{12}^- &&&\cr
&&&&& \cr\hline
&&&&& \cr
 5& \multirow{2}{*}{\includegraphics[width=2cm]{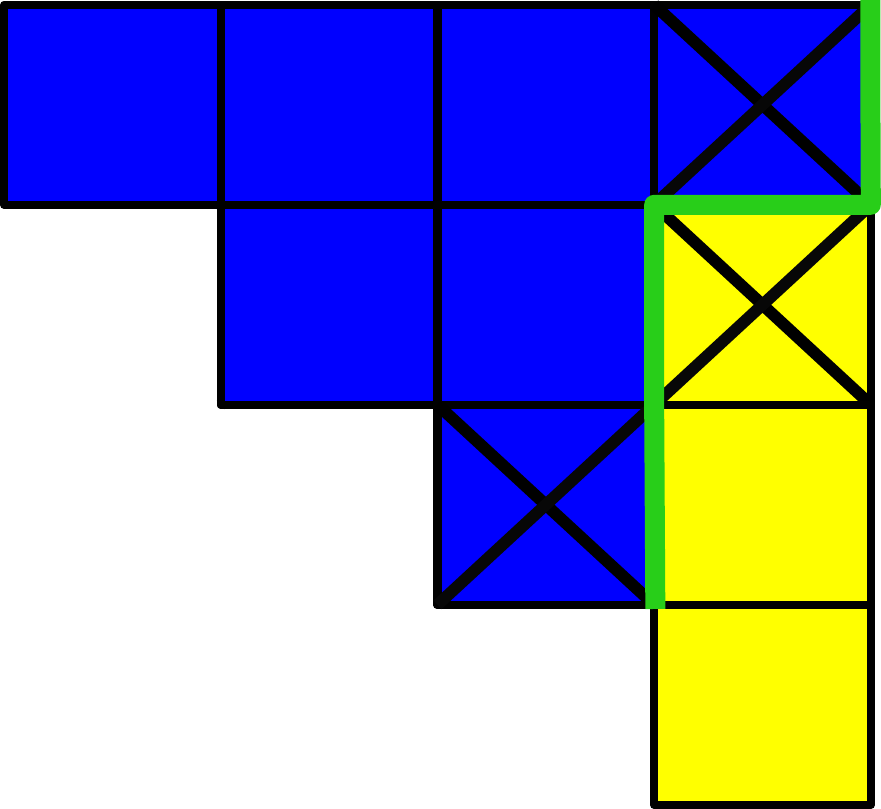} }
&F_1 \rightarrow   C_{1,5}^+ + C_{2,5}^- 
& \multirow{2}{*}{\includegraphics[height=2cm]{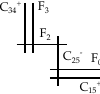}}
& S_f \cdot_Y C_{1,5}^+ = -3 & S_f \cdot_Y C_{1,5}^+ = -1 \cr
&& F_4 \rightarrow C_{3,4}^+ +F_2+ C_{2,5}^-  
&& S_f \cdot_Y C_{2,5}^- = -2 & S_f \cdot_Y C_{2,5}^- = +1\cr
&&&& S_f \cdot_Y C_{3,4}^+ = +2 & S_f \cdot_Y C_{3,4}^+ = +4 \cr
&&&&&\cr\hline
&&&&&\cr
 6& \multirow{2}{*}{\includegraphics[width=2cm]{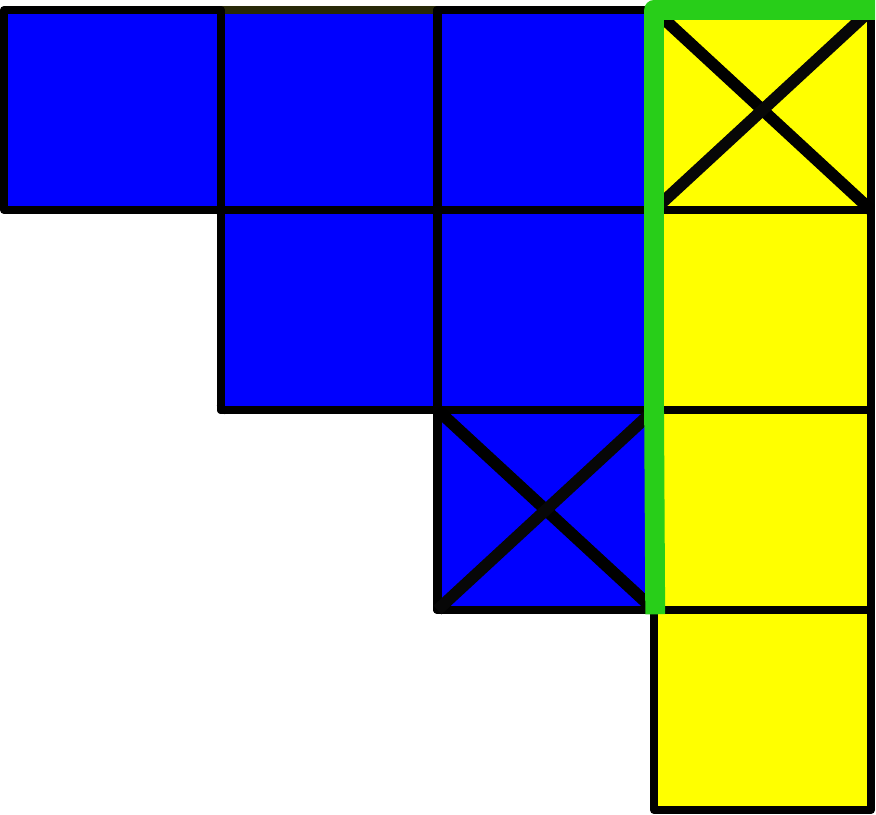} }
&F_4 \rightarrow   C_{3,4}^+ + F_1 + F_2 + C_{1,5}^- 
& \multirow{2}{*}{\includegraphics[height=2cm]{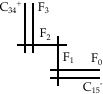}}
& S_f \cdot_Y C_{3,4}^+ = +2 & S_f \cdot_Y C_{3,4}^+ = +4 \cr
&&&& S_f \cdot_Y C_{1,5}^- = +3 & S_f \cdot_Y C_{1,5}^- = +1\cr
&&&&&\cr
&&&&&\cr\hline
&&&&&\cr
 7& \multirow{2}{*}{\includegraphics[width=2cm]{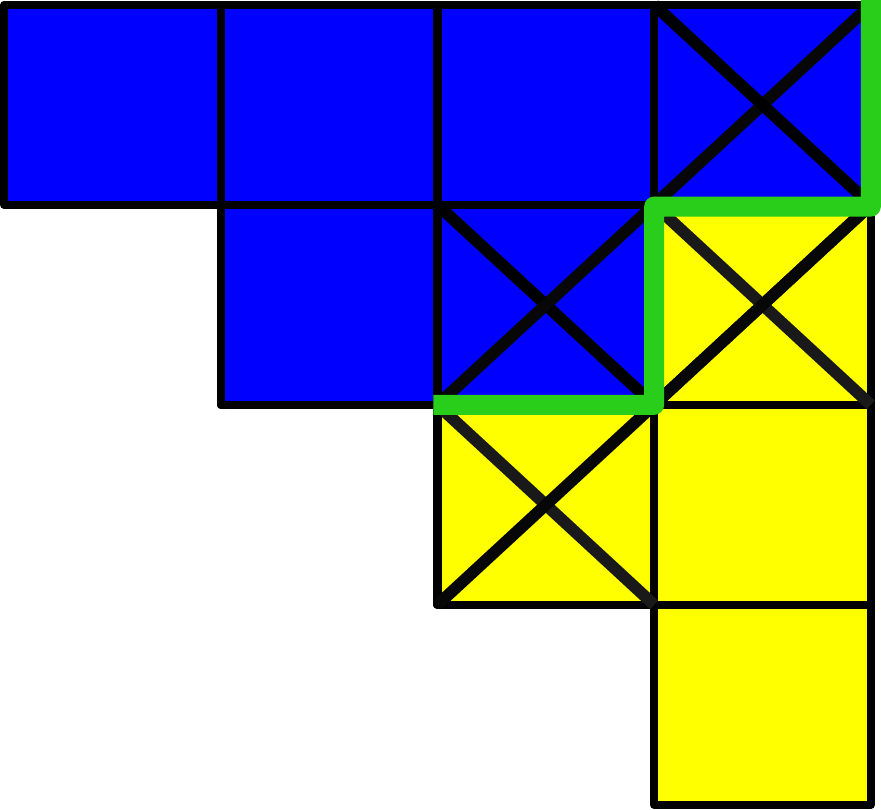} }
&F_1 \rightarrow C_{1,5}^+ +  C_{2,5}^-  
& \multirow{2}{*}{\includegraphics[height=2cm]{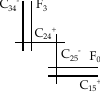}}
& S_f \cdot_Y C_{1,5}^+ = -3 & S_f \cdot_Y C_{1,5}^+ = -1 \cr
&& F_2 \rightarrow C_{2,4}^+ +  C_{3,4}^-  && S_f \cdot_Y C_{2,5}^- = -2 & S_f \cdot_Y C_{2,5}^- = +1 \cr
&& F_4 \rightarrow C_{2,4}^+ +  C_{2,5}^-  && S_f \cdot_Y C_{2,4}^+ = +2 & S_f \cdot_Y C_{2,4}^+ = -1 \cr
&&&& S_f \cdot_Y C_{3,4}^- = -2 & S_f \cdot_Y C_{3,4}^- = -4 \cr\hline
&&&&&\cr
 8& \multirow{2}{*}{\includegraphics[width=2cm]{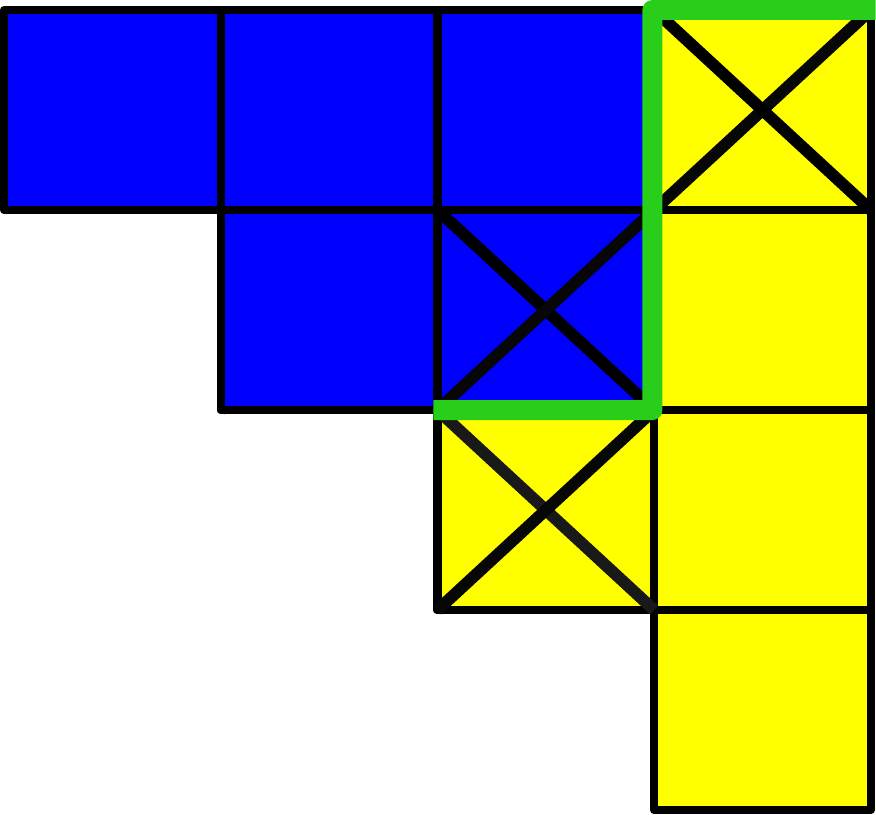} }
&F_2 \rightarrow   C_{2,4}^+ + C_{3,5}^- 
& \multirow{2}{*}{\includegraphics[height=2cm]{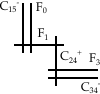}}
& S_f \cdot_Y C_{2,4}^+ = +2 & S_f \cdot_Y C_{2,4}^+ = -1 \cr
&& F_4 \rightarrow C_{2,4}^+ +F_1+ C_{1,5}^-  && S_f \cdot_Y C_{3,4}^- = -2 & S_f \cdot_Y C_{3,4}^- = -4\cr
&&&& S_f \cdot_Y C_{1,5}^- = +3 & S_f \cdot_Y C_{1,5}^- = +1 \cr
&&&&&\cr\hline
\end{array}
$
\caption{Splitting rules for $SU(5)\times U(1)$ with ${\bf 10}$ and Shioda map details $S_f$ for $I_5^{(0|1)}$ and $I_5^{(0||1)}$ for phases $1-8$.
\label{tab:10SplitPart1}}
\end{table}
 

\begin{table}
$
\begin{array}{|c|c|c|c||c|c|}\hline
\# &\hbox{Box Graph} & \hbox{Splitting} & \hbox{Intersections} & I_5^{(0|1)} \, S_f \hbox{ values} & I_5^{(0||1)} \, S_f \hbox{ values} \cr\hline
&&&&&\cr
9 &  \multirow{2}{*}{\includegraphics[width=2cm]{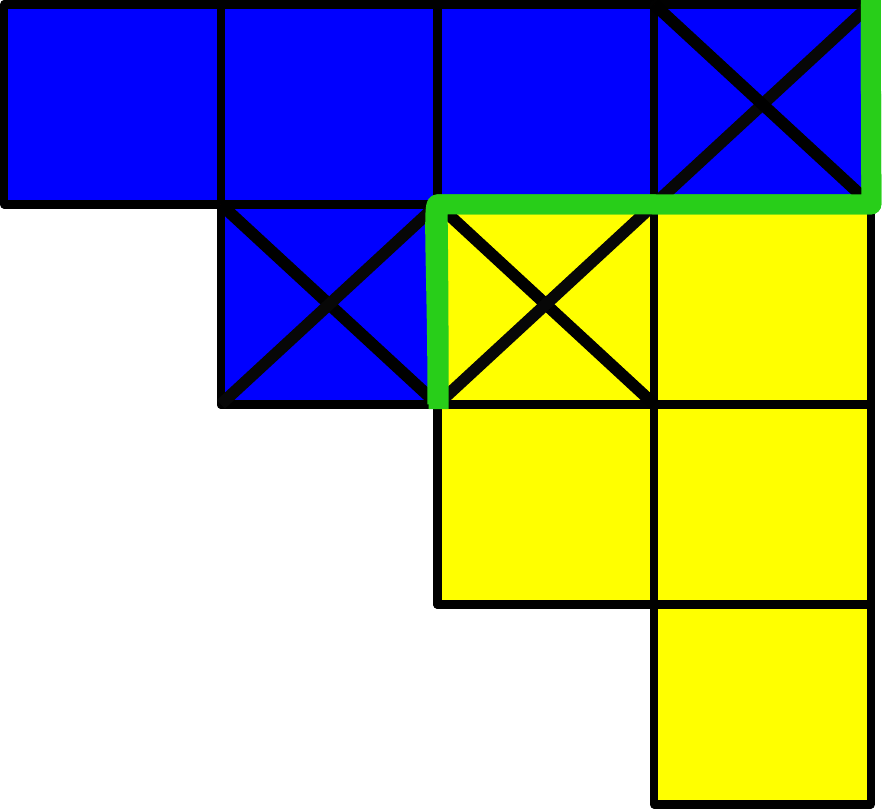}}
&F_1 \rightarrow C_{1,5}^+ + F_4 + C_{2,4}^- 
&\multirow{2}{*}{\includegraphics[height=2cm]{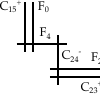}}
& S_f \cdot_Y C_{1,5}^+ = -3 & S_f \cdot_Y C_{1,5}^+ = -1\cr
&& F_3 \rightarrow C_{2,3}^+ + C_{2,4}^-  && S_f \cdot_Y C_{2,4}^- = -2 & S_f \cdot_Y C_{2,4}^- = +1 \cr
&&&& S_f \cdot_Y C_{2,3}^+ = +2 & S_f \cdot_Y C_{2,3}^+ = -1 \cr
&&&&&\cr\hline
&&&&&\cr
10&  \multirow{2}{*}{\includegraphics[width=2cm]{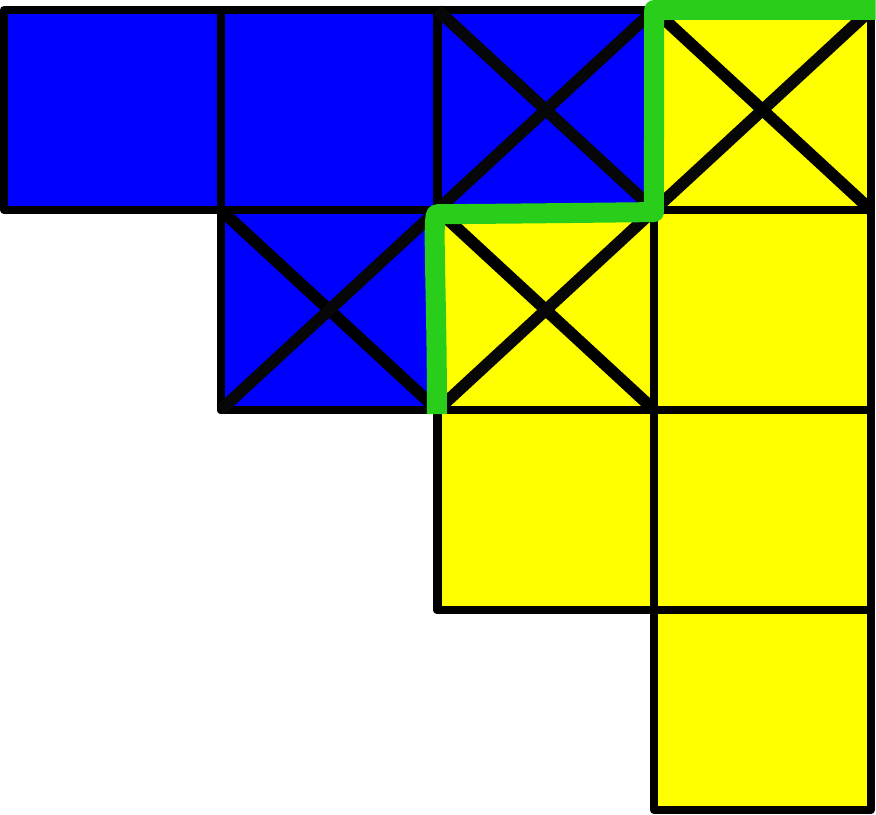} }
&  F_1 \rightarrow C_{1,4}^+ +  C_{2,4}^-  
& \multirow{2}{*}{\includegraphics[height=2cm]{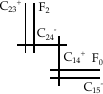}}
& S_f \cdot_Y C_{1,4}^+ = -3 & S_f \cdot_Y C_{1,4}^+ = -1 \cr
&& F_3 \rightarrow C_{2,3}^+ + C_{2,4}^- && S_f \cdot_Y C_{2,4}^- = -2 & S_f \cdot_Y C_{2,4}^- = +1 \cr
&& F_4 \rightarrow C_{1,4}^+ + C_{1,5}^- && S_f \cdot_Y C_{2,3}^+ = +2 & S_f \cdot_Y C_{2,3}^+ = -1 \cr
&&&& S_f \cdot_Y C_{1,5}^- = +3 & S_f \cdot_Y C_{1,5}^- = +1 \cr\hline
&&&&&\cr
11& \multirow{2}{*}{\includegraphics[width=2cm]{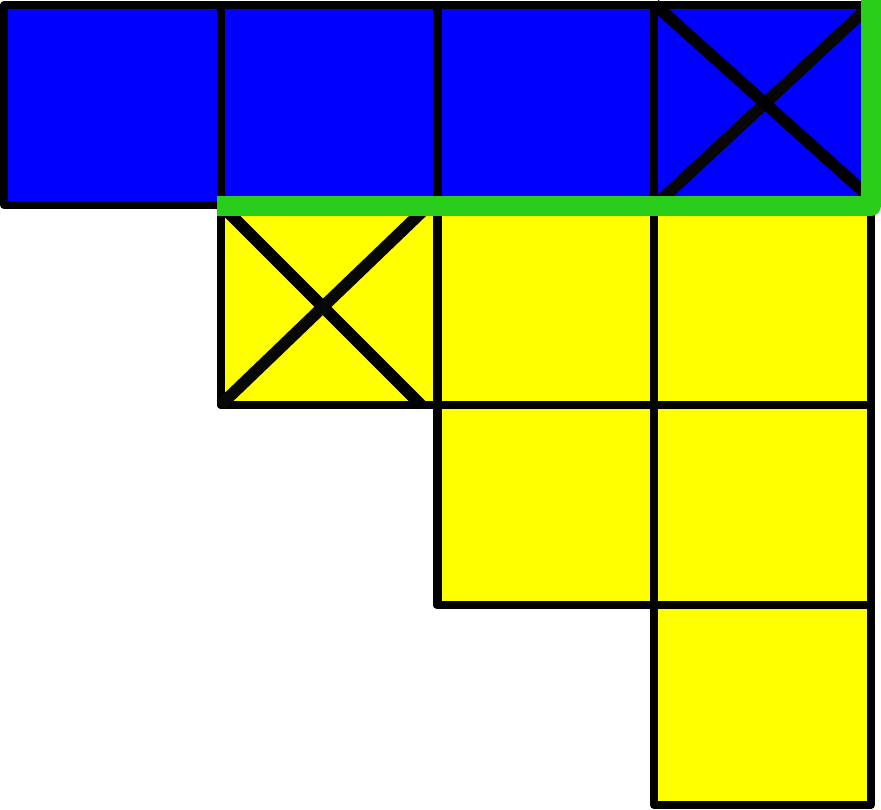} }
&F_1 \rightarrow   C_{1,5}^+ + F_4 + F_3 + C_{2,3}^- 
& \multirow{2}{*}{\includegraphics[height=2cm]{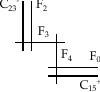}}
& S_f \cdot_Y C_{1,5}^+ = -3 & S_f \cdot_Y C_{1,5}^+ = -1 \cr
&&&& S_f \cdot_Y C_{2,3}^- = -2 & S_f \cdot_Y C_{2,3}^- = +1 \cr
&&&&&\cr
&&&&&\cr\hline
&&&&&\cr
12& \multirow{2}{*}{\includegraphics[width=2cm]{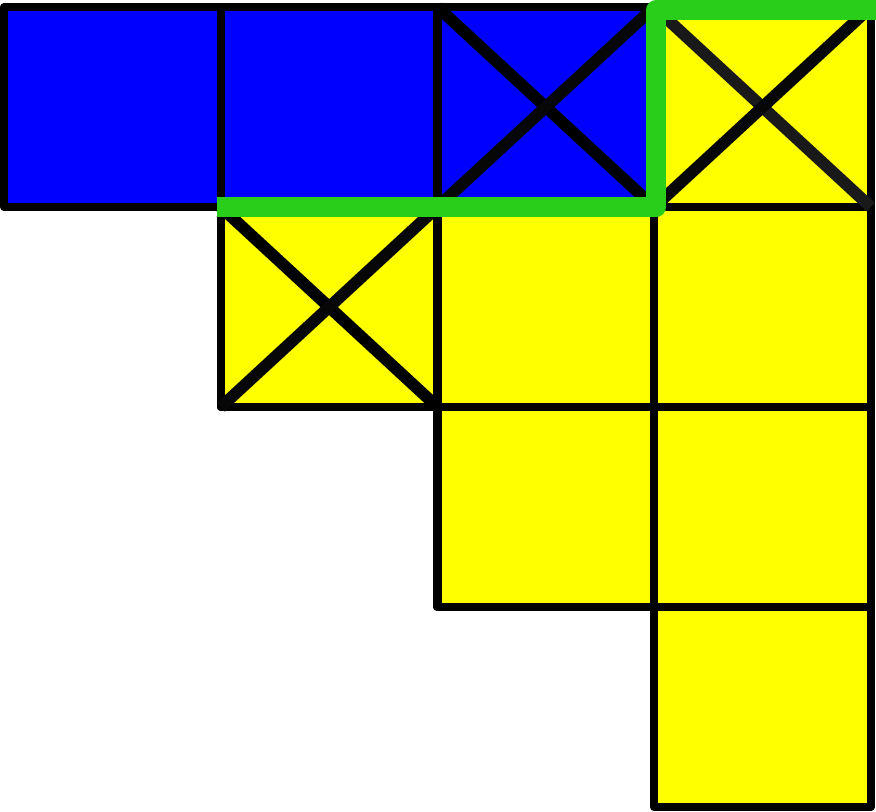} }
&F_1 \rightarrow   C_{1,4}^+ + F_3 + C_{2,3}^- 
& \multirow{2}{*}{\includegraphics[height=2cm]{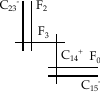}}
& S_f \cdot_Y C_{1,4}^+ = -3 & S_f \cdot_Y C_{1,4}^+ = -1 \cr
&& F_4 \rightarrow C_{1,4}^+ + C_{1,5}^- 
&& S_f \cdot_Y C_{2,3}^- = -2 & S_f \cdot_Y C_{2,3}^- = +1 \cr
&&&& S_f \cdot_Y C_{1,5}^- = +3 & S_f \cdot_Y C_{1,5}^- = +1 \cr
&&&&&\cr\hline
&&&&&\cr
13& \multirow{2}{*}{\includegraphics[width=2cm]{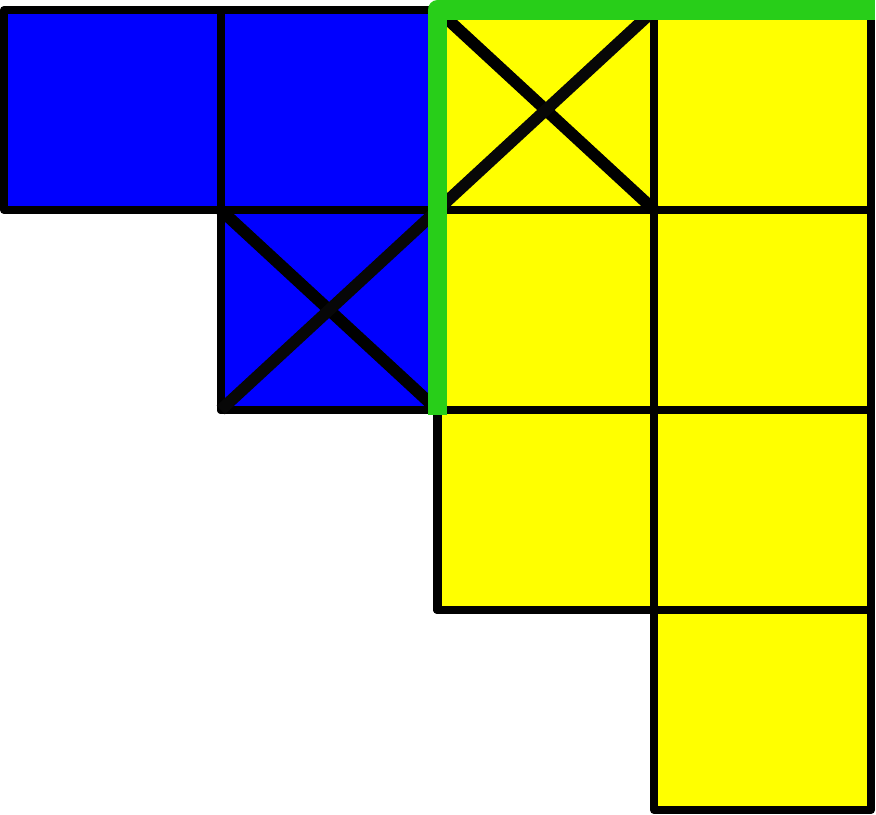} }
&F_3 \rightarrow   C_{2,3}^+ + F_1 + C_{1,4}^- 
& \multirow{2}{*}{\includegraphics[height=2cm]{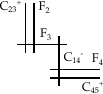}}
& S_f \cdot_Y C_{2,3}^+ = +2 & S_f \cdot_Y C_{2,3}^+ = -1 \cr
&&
F_0 \rightarrow C_{1,4}^- + \tilde{F}_0 
&& S_f \cdot_Y C_{1,4}^- = +3 & S_f \cdot_Y C_{1,4}^- = +1 \cr
&&\tilde{F}_0= C_{4,5}^+
&& S_f \cdot_Y C_{4,5}^+ = +2 & S_f \cdot_Y C_{4,5}^+ = -4\cr
&&&&&\cr\hline
&&&&&\cr
14& \multirow{2}{*}{\includegraphics[width=2cm]{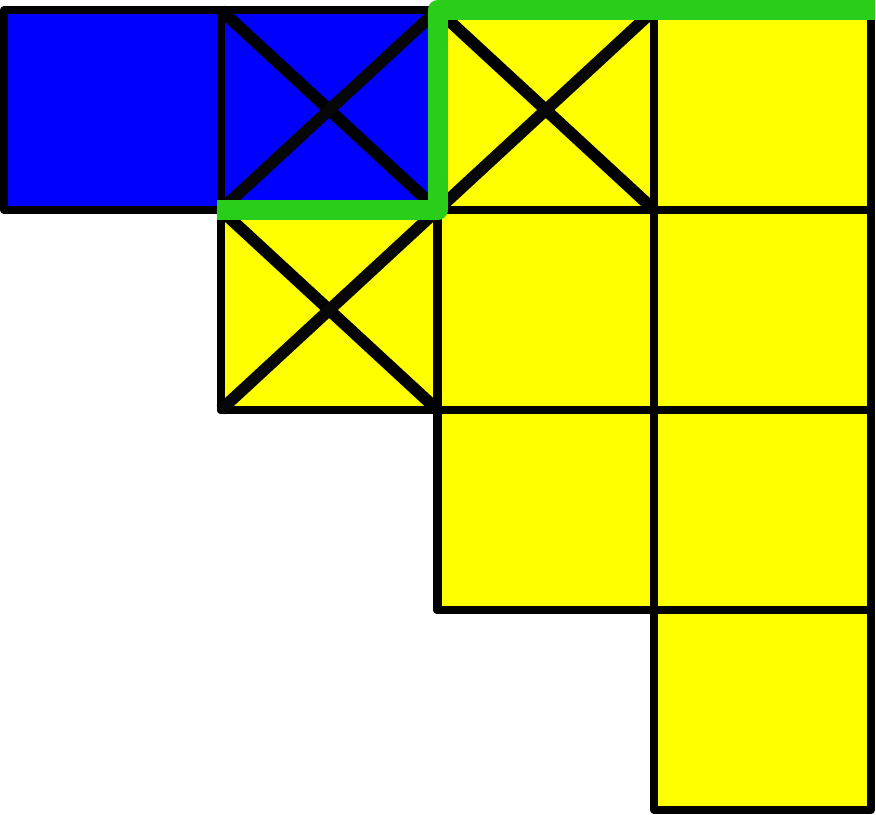} }
&F_1 \rightarrow   C_{1,3}^+ + C_{2,3}^- 
& \multirow{2}{*}{\includegraphics[height=2cm]{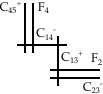}}
& S_f \cdot_Y C_{1,3}^+ = -3 & S_f \cdot_Y C_{1,3}^+ = -1 \cr
&& F_3 \rightarrow C_{1,3}^+ +  C_{1,4}^-  
&& S_f \cdot_Y C_{2,3}^- = -2 & S_f \cdot_Y C_{2,3}^- = +1 \cr
&& F_0 \rightarrow C_{1,4}^- +  \tilde{F}_0  
&& S_f \cdot_Y C_{1,4}^- = +3 & S_f \cdot_Y C_{1,4}^- = +1 \cr
&&\tilde{F}_0= C_{4,5}^+
&& S_f \cdot_Y C_{4,5}^+ = +2 & S_f \cdot_Y C_{4,5}^+ = +4 \cr\hline
&&&&&\cr
15& \multirow{2}{*}{\includegraphics[width=2cm]{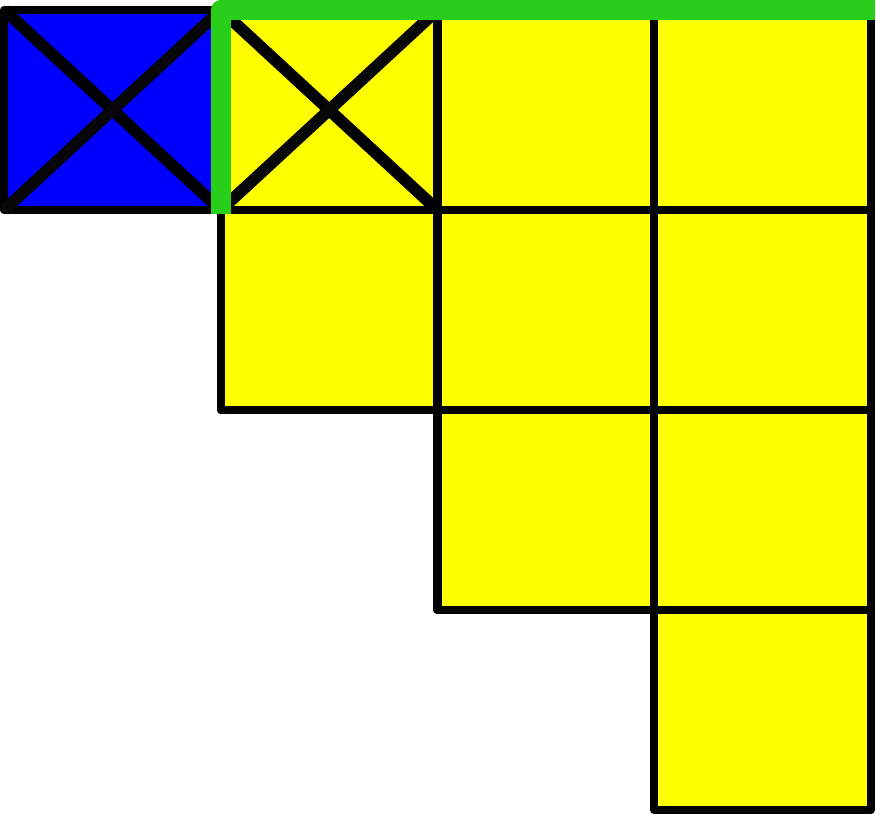} }
&F_2 \rightarrow   C_{1,2}^+ + C_{1,3}^- 
& \multirow{2}{*}{\includegraphics[height=2cm]{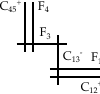}}
& S_f \cdot_Y C_{1,2}^+ = -3 & S_f \cdot_Y C_{1,2}^+ = -6 \cr
&&
F_0 \rightarrow \tilde{F}_0 + F_3 + C_{1,3}^- 
&& S_f \cdot_Y C_{1,3}^- = +3 & S_f \cdot_Y C_{1,3}^- = +1 \cr
&& F_0 = C_{4,5}^+
&&&\cr
&&&&&\cr\hline
&&&&&\cr
16& \multirow{2}{*}{\includegraphics[width=2cm]{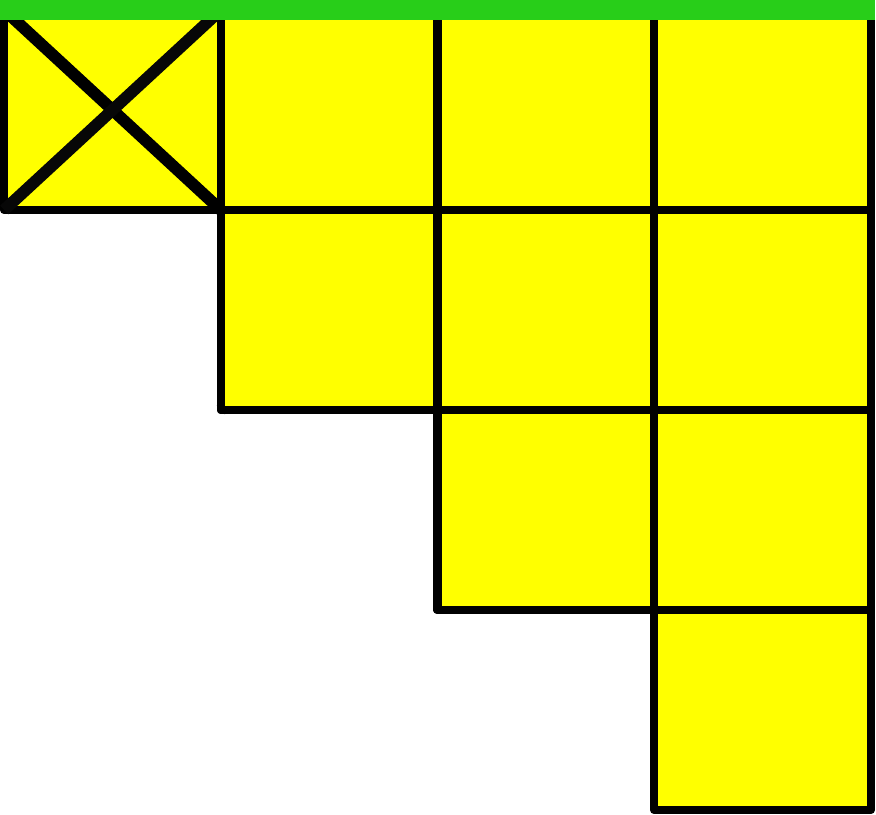} }
&F_0 \rightarrow   C_{1,2}^- + F_2 + F_3 + \tilde{F}_0 
& \multirow{2}{*}{\includegraphics[height=2cm]{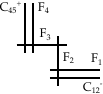}}
& S_f \cdot_Y C_{4,5}^+ = +2 & S_f \cdot_Y C_{4,5}^+ = +4 \cr
&& F_0 = C_{4,5}^+
&& S_f \cdot_Y C_{1,2}^- = +3 & S_f \cdot_Y C_{1,2}^- = +6 \cr
&&&&&\cr
&&&&&\cr\hline
\end{array}
$
\caption{Splitting rules for $SU(5)\times U(1)$ with ${\bf 10}$ and Shioda map details $S_f$ for $I_5^{(0|1)}$ and $I_5^{(0||1)}$ for phases $9-16$.
\label{tab:10SplitPart2}}
\end{table}

\subsection{Compilation of Codimension two Fibers}

\begin{figure}
  \centering
  \includegraphics[scale=2.5]{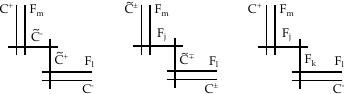}
  \caption{The structure and ordering of the $I_1^*$ fibers of A-type,
  B-type, and C-type, respectively.}
  \label{fig:I1sTypes}
\end{figure}

In this section the different sets of intersection numbers and the possible
realizations as configurations of the fiber curves contained within the
section are enumerated for each splitting type introduced in section
\ref{sec:I1sconfigs}. Figure \ref{fig:I1sTypes} demonstrates the ordering of
the fiber components for each of the three major types, and fixes the
ordering of the notation $(n_1\cdots n_6)$. All the configurations,
determined by a similar procedure to that used in section \ref{sec:SU5A}
for the A.2 splitting types, are listed in table \ref{tbl:I1sSTconfigs}.

For each splitting type there are many more configurations than there are
possible sets of intersections numbers between the split curves and the
section. Multiple configurations correspond to the same intersection
numbers, the same $U(1)$ charges. In table \ref{tbl:I1sSTconfigs} the
intersection numbers are listed for each set of configurations with common
intersection numbers. The intersection numbers $\sigma \cdot_Y C$ are given
as a tuple of integers in the same ordering as the strings describing the
configurations. The intersections of the section with curves that do not
split are not included in such a listing as they are always determined by
codimension one: they are either zero or one depending on whether the section
intersects that component in codimension one.

\begin{table}
  \centering
  \begin{tabular}{|c|c|c|}
    \hline
    \text{Splitting type} & \text{Intersection numbers} &
    \text{Configurations} \cr\hline
    \multirow{3}{*}{A.1} & (-1,1,-1,1) & 
      (1231-- --), (1231--3), (12313--), (123133) \cr\cline{2-3}
    & (0,0,0,0) & 
      (-- -- -- --x--), (2222--2), (222232) \cr\cline{2-3}
    & (1,-1,1,-1) & 
      (--213--1), (3213--1), (--21331), (321331) \cr\hline
    \multirow{3}{*}{A.2} & (-1,1,-1,2) & 
      (12312x), (123124) \cr\cline{2-3} 
    & (0,0,0,1) & 
      (-- -- -- -- --x), (22222--), (222223) \cr\cline{2-3}
    & (1,-1,1,0) &
      (--21-- -- --), (321-- -- --), (--21322), (321322) \cr\hline
    \multirow{2}{*}{A.3} & (-1,1,0,0) & 
      (1-- -- -- -- --), (123222) \cr\cline{2-3}
    & (0,0,1,-1) & 
      (-- -- -- -- --1), (222321) \cr\hline
    \multirow{4}{*}{B.1} & (-1,1,-1) & 
      (1223--1), (122331) \cr\cline{2-3}
    & (0,0,0) & 
      (-- -- -- --x--), (2222--2), (222232) \cr\cline{2-3}
    & \multirow{2}{*}{(1,-1,1)} &
      (--221-- --), (3221-- --), (--2213--), (--221--3), \cr
    & & (32213--), (3221--3), (--22133), (322133) \cr\hline
    \multirow{3}{*}{B.2} & (-1,1,0) & 
      (122-- -- --), (122322) \cr\cline{2-3}
    & (0,0,1) & 
      (-- -- -- -- --x), (22222--), (222223) \cr\cline{2-3}
    & (1,-1,2) & 
      (--2212x), (32212x), (--22124), (322124) \cr\hline
    \multirow{2}{*}{B.3} & (-1,0,0) & 
      (1-- -- -- -- --), (123222) \cr\cline{2-3}
    & (0,-1,1) & 
      (-- -- --12--), (22312--), (-- -- --123), (223123) \cr\hline
    \multirow{4}{*}{B.4} & (-1,1,-1) & 
      (1--2321), (132321) \cr\cline{2-3}
    & (0,0,0) & 
      (--x-- -- -- --), (2--2222), (232222) \cr\cline{2-3}
    & \multirow{2}{*}{(1,-1,1)} &
      (-- --212--), (3--212--), (--3212--), (-- --2123) \cr
    & & (33212--), (3--2123), (--32123), (332123) \cr\hline
    \multirow{3}{*}{B.5} & (0,1,-1) & 
      (-- -- -- -- --1), (222321) \cr\cline{2-3}
    & (1,0,0) &
      (x-- -- -- -- --), (--22222), (322222) \cr\cline{2-3}
    & (2,-1,1) &
      (x2212--), (x22123), (42212--), (422123) \cr\hline
    \multirow{3}{*}{C.1} & (1,-1) & 
      (--222--1), (3222--1), (--22231), (322231) \cr\cline{2-3}
    & (0,0) &
      (-- -- -- --x--), (2222--2), (222232) \cr\cline{2-3}
    & (-1,1) & 
      (1222-- --), (12223--), (1222--3), (122233) \cr\hline
    \multirow{2}{*}{C.2} & (-1,0) & 
      (122-- -- --), (122322) \cr\cline{2-3}
    & (0,-1) &
      (-- -- -- -- --1), (222321) \cr\hline
    \multirow{4}{*}{C.3} & (2,-1) & 
      (x22221), (422221) \cr\cline{2-3}
      & (1,0) & 
        (x-- -- -- -- --), (--22222), (322222) \cr\cline{2-3}
      & (0,1) & 
        (-- -- -- -- --x), (22222--), (222223) \cr\cline{2-3}
      & (-1,2) &
        (12222x), (122224) \cr\hline
    \end{tabular}
    \caption{
      For each of the different splitting types, listed in section
      \ref{sec:I1sconfigs}, for the enhancements from an $I_5$ fiber to an
      $I_1^*$, including the information of which fiber component the section
      intersects in codimension one, all the possible consistent
      configurations of the $I_1^*$ fiber components with the section are
      listed in the third column, using the notation described in section
      \ref{app:wrapnotation}. There are multiple configurations of the curves
      inside the section where all of the fiber curves have the same
      intersection numbers with the section, these are collected and the
      intersection numbers particular to those configurations are listed in
      the second column. These intersection numbers are the relevant datum
      for the computation of the $U(1)$ charges. The tuples of intersection
      numbers do not include the curves which do not split as their intersection
      numbers are always uniquely fixed by codimension one.
    }
    \label{tbl:I1sSTconfigs}
\end{table}

Each of the concrete enhancements from the $I_5$ fiber into an $I_1^*$ fiber,
listed in tables \ref{tab:10SplitPart1} and \ref{tab:10SplitPart2}, are
realizations of one of the splitting types just analyzed. Determining the
splitting type depends on the phase (which fixes whether it is of type A, B, or C), and the
codimension one configuration, which determines the subcase. The
configurations of $I_1^*$ curves in the section can then be determined for
each phase and codimension one configuration of the section. All of the
configurations for each of the sixteen phases are listed in tables 
\ref{tab:I1sConfigsPhase1to8} and \ref{tab:I1sConfigsPhase9to16}.

\begin{sidewaystable}
  \footnotesize
  \centering
  \begin{tabular}{|c|c|c|c|}
    \hline
    \multirow{2}{*}{Phase} & \multicolumn{3}{|c|}{Configurations}
    \cr\cline{2-4}
    & $\sigma \cdot_Y F_0 = 1$ & $\sigma \cdot_Y F_1 = 1$ & $\sigma \cdot_Y
  F_2 = 1$ \cr\hline
  \multirow{4}{*}{1} & 
  (x22221), (422221) & 
  (--222--1), (3222--1), (--22231), (322231) & 
  (122-- -- --), (122322) \cr\cline{2-4}
  & 
  (x-- -- -- -- --), (--22222), (322222) &
  (-- -- -- --x--), (2222--2), (222232) &
  (-- -- -- -- --1), (222321) \cr\cline{2-4}
  &
  (-- -- -- -- --x), (22222--), (222223) &
  (1222-- --), (12223--), (1222--3), (122233) & 
  \cr\cline{2-4}
  &
  (12222x), (122224) &
  & \cr\hline
  \multirow{4}{*}{2} & 
  (-- -- -- -- --1), (222321) &
  (1--2321), (132321) &
  (1-- -- -- -- --), (123222) \cr\cline{2-4}
  &
  (x-- -- -- -- --), (--22222), (322222) &
  (--x-- -- -- --), (2--2222), (232222) &
  (-- -- --12--), (22312--), (-- -- --123), (223123) \cr\cline{2-4}
  &
  \multirow{2}{*}{(x2212--), (x22123), (42212--), (422123)} &
  (-- --212--), (3--212--), (--3212--), (-- --2123)
  & \cr
  & & (33212--), (3--2123), (--32123), (332123) & \cr\hline
  \multirow{3}{*}{3} &
  (12312x), (123124) &
  (1231-- --), (12313--), (1231--3), (123133) &
  (1-- -- -- -- --), (123222) \cr\cline{2-4}
  &
  (-- -- -- -- --x), (22222--), (222223) &
  (-- -- -- --x--), (2222--2), (222232) &
  (-- -- -- -- --1), (222321) \cr\cline{2-4}
  &
  (--21-- -- --), (321-- -- --), (--21322), (321322) &
  (--213--1), (3213--1), (--21331), (321331) & \cr\hline
  \multirow{4}{*}{4} &
  (122-- -- --), (122322) &
  (1223--1), (122331) &
  (-- -- -- -- --1), (222321) \cr\cline{2-4}
  &
  (-- -- -- -- --x), (22222--), (222223) &
  (-- -- -- --x--), (2222--2), (222232) &
  (x-- -- -- -- --), (--22222), (322222) \cr\cline{2-4}
  &
  \multirow{2}{*}{(--2212x), (32212x), (--22124), (322124)} &
  (--221-- --), (3221-- --), (--2213--), (--221--3) &
  \multirow{2}{*}{(x2212--), (x22123), (42212--), (422123)} \cr
  & & (32213--), (3221--3), (--22133), (322133) & \cr\hline
  \multirow{4}{*}{5} &
  (1223--1), (122331) &
  (122-- -- --), (122322) &
  (1-- -- -- -- --), (123222) \cr\cline{2-4}
  &
  (-- -- -- --x--), (2222--2), (222232) &
  (-- -- -- -- --x), (22222--), (222223) &
  (-- -- --12--), (22312--), (-- -- --123), (223123) \cr\cline{2-4}
  &
  (--221-- --), (3221-- --), (--2213--), (--221--3) &
  \multirow{2}{*}{(--2212x), (32212x), (--22124), (322124)} & \cr
  & (32213--), (3221--3), (--22133), (322133) & & \cr\hline
  \multirow{3}{*}{6} &
  (--222--1), (3222--1), (--22231), (322231) &
  (122-- -- --), (122322) &
  (-- -- --221), (232221) \cr\cline{2-4}
  &
  (-- -- -- --x--), (2222--2), (222232) &
  (-- -- -- -- --1), (222321) &
  (1-- -- -- -- --), (123222) \cr\cline{2-4}
  &
  (1222-- --), (12223--), (1222--3), (122233) & & \cr\hline
  \multirow{3}{*}{7} &
  (1231-- --), (12313--), (1231--3), (123133) &
  (12312x), (123124) &
  (x21321), (421321) \cr\cline{2-4}
  &
  (-- -- -- --x--), (2222--2), (222232) &
  (-- -- -- -- --x), (22222--), (222223) &
  (x-- -- -- -- --), (--22222), (322222) \cr\cline{2-4}
  &
  (--213--1), (3213--1), (--21331), (321331) &
  (--21-- -- --), (321-- -- --), (--21322), (321322) &
  (-- -- --12--), (-- -- --123), (22312--), (223123) \cr\hline
  \multirow{4}{*}{8} &
  (1--2321), (132321) &
  (1-- -- -- -- --), (123222) &
  (122-- -- --), (122322) \cr\cline{2-4}
  &
  (--x-- -- -- --), (2--2222), (232222) &
  (-- -- --12--), (22312--), (-- -- --123), (223123) &
  (-- -- -- -- --x), (22222--), (222223) \cr\cline{2-4}
  &
  (-- --212--), (3--212--), (--3212--), (-- --2123) &
  &
  \multirow{2}{*}{(--2212x), (32212x), (--22124), (322124)} \cr
  & 
  (33212--), (3--2123), (--32123), (332123) & & \cr\hline
  \end{tabular}
  \caption{For each of the three configurations of a
    section, $\sigma$, with the codimension one components it is listed for the phases 1
    -- 8, which are the distinct enhancements from $I_5$ to
    $I_1^*$ listed in table \ref{tab:10SplitPart1}, the possible
    consistent configurations of the curves of the $I_1^*$ fiber with the
    section, in the notation of section \ref{app:wrapnotation}. 
    Configurations of the same phase and codimension one configuration that
    are not separated by a horizontal divider have the same intersection
    numbers between all the curves of the $I_1^*$ fiber and the section.
  }
    \label{tab:I1sConfigsPhase1to8}
\end{sidewaystable}

\begin{sidewaystable}
  \footnotesize
  \centering
  \begin{tabular}{|c|c|c|c|}
    \hline
    \multirow{2}{*}{Phase} & \multicolumn{3}{|c|}{Configurations}
    \cr\cline{2-4}
    & $\sigma \cdot_Y F_0 = 1$ & $\sigma \cdot_Y F_1 = 1$ & $\sigma \cdot_Y
  F_2 = 1$ \cr\hline
  \multirow{4}{*}{9} & 
  (1--2321), (132321) &
  (-- -- -- -- --1), (222321) &
  (1223--1), (122331) \cr\cline{2-4}
  &
  (--x-- -- -- --), (2--2222), (232222) &
  (x-- -- -- -- --), (--22222), (322222) &
  (-- -- -- --x--), (2222--2), (222232) \cr\cline{2-4}
  &
  (-- --212--), (3--212--), (--3212--), (-- --2123) &
  \multirow{2}{*}{(x2212--), (x22123), (42212--), (422123)} &
  (--221-- --), (3221-- --), (--2213--), (--221--3) \cr
  &
  (3--2123), (33212--), (--32123), (332123) &
  &
  (32213--), (3221--3), (--22133), (322133) \cr\hline
  \multirow{3}{*}{10} & 
  (1231-- --), (12313--), (1231--3), (123133) &
  (1-- -- -- -- --), (123222) &
  (-- --1321), (3--1321), (--31231), 331321) \cr\cline{2-4}
  &
  (-- -- -- --x--), (2222--2), (222232) &
  (-- -- -- -- --1), (222321) &
  (--x-- -- -- --), (2--2222), (232222) \cr\cline{2-4}
  &
  (--213--1), (3213--1), (--21331), (321331) &
  &
  (1--312--), (13312--), (1--3123), (133123) \cr\hline
  \multirow{4}{*}{11} & 
  (--222--1), (3222--1), (--22231), (322231) &
  (x22221), (422221) &
  (1--222--), (13222--), (1--2223), (132223) \cr\cline{2-4}
  &
  (-- -- -- --x--), (2222--2), (222232) &
  (x-- -- -- -- --), (--22222), (322222) &
  (--x-- -- -- --), (2--2222), (232222) \cr\cline{2-4}
  &
  (1222-- --), (12223--), (1222--3), (122233) &
  (-- -- -- -- --x), (22222--), (222223) &
  (-- --2221), (3--2221), (--32221), (332221) \cr\cline{2-4}
  &
  &
  (12222x), (122224) &
  \cr\hline
  \multirow{4}{*}{12} & 
  (1223--1), (122331) &
  (-- -- -- -- --1), (222321) &
  (1--2321), (132321) \cr\cline{2-4}
  &
  (-- -- -- --x--), (2222--2), (222232) &
  (x-- -- -- -- --), (--22222), (322222) &
  (--x-- -- -- --), (2--2222), (232222) \cr\cline{2-4}
  &
  (--221-- --), (3221-- --), (--2213--), (--221--3) &
  \multirow{2}{*}{(x22212--), (x22123), (42212--), (422123)} &
  (-- --212--), (3--212--), (--3212--), (-- --2123) \cr
  &
  (32213--), (3221--3), (--22133), (322133) &
  &
  (33212--), (3--2123), (--32123), (332123) \cr\hline
  \multirow{4}{*}{13} &
  (122-- -- --), (122322) &
  (1-- -- -- -- --), (123222) &
  (1--2321), (132321) \cr\cline{2-4}
  &
  (-- -- -- -- --x), (22222--), (222223) &
  (-- -- --12--), (22312--), (-- -- --123), (223123) &
  (--x-- -- -- --), (2--2222), (232222) \cr\cline{2-4}
  &
  \multirow{2}{*}{(--2212x), (32212x), (--22124), (322124)} &
  &
  (-- --212--), (3--212--), (--3212--), (-- --2123) \cr
  &
  &
  &
  (33212--), (3--2123), (--32123), (332123) \cr\hline
  \multirow{3}{*}{14} &
  (x21321), (421321) &
  (12312x), (123124) &
  (1231-- --), (12313--), (1231--3), (123133) \cr\cline{2-4}
  &
  (x-- -- -- -- --), (--22222), (322222) &
  (-- -- -- -- --x), (22222--), (222223) &
  (-- -- -- --x--), (2222--2), (222232) \cr\cline{2-4}
  &
  (-- -- --12--), (22312--), (-- -- --123), (223123) &
  (--21-- -- --), (321-- -- --), (--21322), (321322) &
  (--213--1), (3213--1), (--21331), (321331) \cr\hline
  \multirow{4}{*}{15} &
  (-- -- -- -- --1), (222321) &
  (1223--1), (122331) &
  (122-- -- --), (122322) \cr\cline{2-4}
  &
  (x-- -- -- -- --), (--22222), (322222) &
  (-- -- -- --x--), (2222--2), (222232) &
  (-- -- -- -- --x), (22222--), (222223) \cr\cline{2-4}
  &
  \multirow{2}{*}{(x2212--), (x22123), (42212--), (422123)} &
  (--221-- --), (3221-- --), (--2213--), (--221--3) &
  \multirow{2}{*}{(--2212x), (32212x), (--22124), (322124)} \cr
  &
  &
  (32213--), (3221--3), (--22133), (322133) &
  \cr\hline
  \multirow{4}{*}{16} &
  (x22221), (422221) &
  (--222--1), (3222--1), (--22231), (322231) &
  (122-- -- --), (122322) \cr\cline{2-4}
  &
  (x-- -- -- -- --), (--22222), (322222) &
  (-- -- -- --x--), (2222--2), (222232) &
  (-- -- -- -- --1), (222321) \cr\cline{2-4}
  &
  (-- -- -- -- --x), (22222--), (222223) &
  (1222-- --), (12223--), (1222--3), (122233) &
  \cr\cline{2-4}
  &
  (12222x), (122224) &
  &
  \cr\hline
  \end{tabular}
  \caption{
    Similar to table \ref{tab:I1sConfigsPhase1to8}, all the configurations of
    the curves of the $I_1^*$ fiber with the section $\sigma$ are listed for
    the codimension one configurations of the section and the phases 9 -- 16,
    which were listed previously in table \ref{tab:10SplitPart2}. The
    configurations are again listed with the notation of section
    \ref{app:wrapnotation}.
    There can be
    multiple configurations of the curves with the section which have the same
    intersection numbers, and thus the same $U(1)$ charges, and these are
    collected together inside each phase and codimension one configuration.
  }
  \label{tab:I1sConfigsPhase9to16}
\end{sidewaystable}



\section{Charge Comparison to Singlet-Extended $E_8$}
\label{app:E8}

In \cite{Baume:2015wia} $U(1)$ charges for $SU(5)$ models that come from a
Higgsing of $E_8$, extended by non-$E_8$ singlets, are determined. What is
considered is the decomposition of the adjoint of $E_8 \rightarrow SU(5)
\times U(1)^4$, which is then augmented by additional singlets carrying
different charge under the abelian $U(1)^4$ such that for every pair of ${\bf 5}$ and
${\bf \overline{5}}$ representations of $SU(5)$ coming from the decomposition of
$E_8$ there exists a singlet such that the coupling ${\bf 1 5 \overline{5}}$
is uncharged under the $U(1)^4$. Various singlets can be Higgsed to produce
models with fewer abelian symmetries, and determine the tree of possible
theories arising from this singlet-extension of $E_8$. In this appendix the
charges found from this analysis, listed in tables 2.1 and 2.2 of 
\cite{Baume:2015wia}, are compared to the possible $U(1)$ charges determined in the
main body of this paper. In summary, it is found that the charges appearing in
descendants of the singlet-extended $E_8$ form a strict subset of the charges
found herein. 

Consider first the single $U(1)$ models from the singlet-extended $E_8$. There
are eleven such models listed in \cite{Baume:2015wia}, which all have $U(1)$
charges\footnote{Some models have an additional discrete symmetry from the
  Higgsing of the $U(1)$. This is not relevant for this comparison and will be
ignored at this point.} that are subsets of one of the following three classes
of charges
\be
\begin{array}{llllll}
  &\quad & &{\bf 10} &  &{\bf \overline{5}} \cr
  &(1): & &\{-2,-1,0,1,2\} & &\{-3,-2,-1,0,1,2,3\} \cr
  &(2): & &\{-8, -3, 2, 7\} & &\{-11, -6, -1, 4, 9\} \cr
  &(3): & &\{-4, 1, 6\} & &\{-8, -3, 2, 7\} \,.
\end{array}
\ee
For each of the three classes there is at least one model which realizes
matter representations with all of the charges in that class. These three
classes have charges which are subsets of the charges\footnote{
  There is  an 
  overall sign between the charges of class $(2)$ and
  the $I_5^{(0|1)}$ codimension one configurations which were listed in figure
  \ref{fig:0s1Charges}.} from the three
codimension one fiber types, $I_5^{(01)}$, $I_5^{(0|1)}$, and $I_5^{(0||1)}$
respectively, as determined in sections \ref{sec:SU5F} and \ref{sec:SU5A} for the
${\bf \overline{5}}$ and ${\bf 10}$ matter. There are some $U(1)$ charges which come from
the analysis of configurations of the fiber curves with the section which do
not appear to arise from the singlet-extended $E_8$.
The missing charges are
\begin{itemize}
  \item In class $(1)$ the charges $\pm 3$ for the ${\bf 10}$ representation.
  \item In class $(2)$ the charges $-13$ and $+12$ for the ${\bf 10}$ and $14$ for
    the ${\bf \overline{5}}$.
  \item In class $(3)$ the charges $-9$ and $+11$ for the ${\bf 10}$ and $-13$ and
    $+12$ for the ${\bf \overline{5}}$. 
\end{itemize}

The significance of $E_8$ is not entirely clear so that 
this mismatch in the charges of the ${\bf 10}$ and ${\bf \overline{5}}$
matter is perhaps not too surprising. However all the single $U(1)$ models from the
singlet-extended $E_8$ have charges which come from the analysis of the
possible configurations of the section in the present paper, as expected. This
includes also the singlet charges which appear in \cite{Baume:2015wia} as, from
the analysis in section \ref{sec:singlets}, the range of singlet charges depends
on an integer $p$, which specifies the normal bundle of one of the curves in
the $I_2$ fiber.  As we do not know of any constraint on the 
possible values of $p$ it is possible to tune $p$ such that one realizes 
the charges in the singlet-extended $E_8$ analysis.

Moving on to the models with two or more remaining $U(1)$ symmetries after the
further Higgsing of the $U(1)^4$ it appears that there are models which have
charges that are not neatly pairs of charges that would be possible for single
$U(1)$s. As discussed in section \ref{sec:multiU1s}, when there are multiple
$U(1)$s one can consider any linear combination of the $U(1)$ generators and
thus produce another $U(1)$ generator, under which the matter will have
different charges. To be concrete, consider the model labelled $\{4,6,8\}$
from table 2.1 of \cite{Baume:2015wia}. This model has ${\bf \overline{5}}$
matter with $U(1)$ charges $(-4,-4)$ and $(-2,-1)$, among other ${\bf
  \overline{5}}$ matter. Recall that for a single $U(1)$ it was only possible
  to realize a ${\bf \overline{5}}$ matter curve with charge $-4$ in an
  $I_5^{(0|1)}$ model, and thus all the ${\bf \overline{5}}$ matter should
  have charge, under that $U(1)$, which take values in $-14$, $-9$, $-4$, $1$, $6$, and  $11$.
  The model in question also has ${\bf \overline{5}}$ matter with charge $-2$
  (or $-1$ if one studies the second $U(1)$) which is not one of the possible
  charges. However, if one designates the two $U(1)$ generators as $U_1$ and
  $U_2$ respectively then one can define two new $U(1)$s by linear
  combinations of these, as
\begin{equation}
  \begin{aligned}
    U_1^\prime &= U_1 - U_2 \cr
    U_2^\prime &= 2U_1 - 3U_2 \,.
  \end{aligned}
\end{equation}
Under this new pair of $U(1)$ generators the charges of the ${\bf 10}$ and
${\bf \overline{5}}$ curves in the model $\{4,6,8\}$ transform as
\begin{equation}
  \begin{array}{c|c}
    {\bf 10} & {\bf \overline{5}} \cr\hline
    (-2,-2) & (-4,-4) \cr
    (0,1) & (-2,-1) \cr
    (1,0) & (-1,-2) \cr
    (3,3) & (1,1) \cr
    & (3,4) \cr
    & (4,3)
  \end{array}
  \quad\leftrightarrow\quad
  \begin{array}{c|c}
    {\bf 10} & {\bf \overline{5}} \cr\hline
    (2,0) & (4,0) \cr
    (-3,-1) & (-1,-1) \cr
    (2,1) & (4,1) \cr
    (-3,0) & (-1,0) \cr
    & (-6,-1) \cr
    & (-1,1) \,.
  \end{array}
\end{equation}
Now it can be seen that the sets of charges are consistent with the charges
listed in the main text for each additional $U(1)$. Indeed with respect to the
first new generator $U_1^\prime$ the section $\sigma_2$ to which it is
associated seems to be an $I_5^{(0|2)}$ fiber in codimension one, and the
section of the second generator, $\sigma_1$, seems to intersect the
codimension one fiber as $I_5^{(01)}$. The $\{4,6,8\}$ model can be seen to
come from an enhancement of an $I_5^{(01|2)}$ model. 

The remaining multiple $U(1)$ models in table 2.1 of \cite{Baume:2015wia}
which have charges that do not immediately match the charges found in the main
body of this paper can all be brought into the form listed here by taking the
appropriate linear combination of the $U(1)$ generators, and thus all the
$U(1)$ charges found therein can be seen to be $U(1)$ charges that also come
from the analysis of how the section can contain curves in the codimension two
fiber that has been the focus of this paper.


\newpage



\providecommand{\href}[2]{#2}\begingroup\raggedright\endgroup

\end{document}